%% file: eqb3d.tex
\documentclass[11pt]{article}
\usepackage{epsfig,amssymb,amstext}

 \newcommand{\OPBIL}{} 

\input{macros}
\newtheorem{platz}{{\bf Fig.}} 
\rem{
\newcommand{\Capts}[1]{#1}
\newcommand{\FIGo}[3]{%
\marginpar{ \begin{platz} \label{#1} ~ \end{platz} \vspace*{1.5ex} }}

} 
\newcommand{\Capts}[1]{}
\newcommand{\FIGo}[3]{\begin{figure}%
#3%
\caption[]{\capsty #2}%
\label{#1}%
\end{figure}}
\rem{
\includeonly{
  intro
 ,classic
 ,qm
 ,miller
 ,bt
 ,invprob
}
} 

\newcommand{\bstr}{1}

\begin{document}

\OPBIL

\thispagestyle{plain}
\begin{center}
{\huge Triaxial Ellipsoidal Quantum Billiards} \\[7ex]
{\large Holger Waalkens, Jan Wiersig$^\ast$, Holger R.~Dullin$^\dagger$}\\[4ex]
{\large Institut f\"ur Theoretische Physik and}\\[1ex]
{\large Institut f\"ur Dynamische Systeme}\\[1ex]
{\large University of Bremen}\\[1ex]
{\large Bremen, Germany}\\[2ex]
{\large $^\ast$present address:}\\[1ex] 
{\large School of Mathematical Sciences}\\[1ex]
{\large Queen Mary and Westfield College}\\[1ex]
{\large University of London}\\[1ex]
{\large London, UK}\\[2ex]
{\large $^\dagger$present address:}\\[1ex]
{\large Department of Applied Mathematics}\\[1ex]
{\large University of Colorado}\\[1ex]
{\large Boulder, USA}\\[4ex]
\today
\end{center}

\vspace*{1ex}

\renewcommand{\baselinestretch}{\bstr} \normalsize

\section*{Abstract}
The classical mechanics, exact quantum mechanics and semiclassical
quantum mechanics of the billiard in the triaxial ellipsoid is
investigated. 
The system is separable in ellipsoidal coordinates.
A smooth description of the motion is given 
in terms of a geodesic flow on a solid torus, which is
a fourfold cover of the interior of the ellipsoid. 
Two crossing separatrices lead to four generic types of motion. 
The action variables of the system are integrals of a single Abelian
differential of second kind on a hyperelliptic curve of genus 2. 
The classical separability carries
over to quantum mechanics giving two versions of generalized
Lam{\'e} equations according to the two sets of classical
coordinates. 
The quantum eigenvalues define a lattice
when transformed to classical action space. Away from the separatrix
surfaces the lattice is given by {\sl EBK} quantization rules
for the four types of classical motion. 
The transition between the four lattices is described by a 
uniform semiclassical quantization scheme based on a {\sl WKB} ansatz.
The tunneling between tori 
is given by penetration integrals which again are integrals 
of the same Abelian differential that gives the classical action variables.
It turns out that the quantum mechanics of ellipsoidal billiards
is semiclassically most elegantly explained by the investigation of
its hyperelliptic curve and the real and purely imaginary periods of a
single Abelian differential. 

\vspace{3ex}

\noindent
PACS: 03.20.+i, 03.65.Ge, 03.65.Sq
\rem{ 
\vfill
\noindent
41 pages, 17 Figures, V Tables
\clearpage
{\large
\noindent
\vspace*{3ex}\\[3ex]
Holger Waalkens\\[1.5ex]
Institut f\"ur Theoretische Physik and\\[1ex]
Institut f\"ur Dynamische Systeme\\[1ex]
University of Bremen\\[1ex]
Postfach 330~440\\[1ex]
28334 Bremen, Germany\\
\begin{tabbing}
phone: \= \kill 
phone: \> +49 421 218-4566 \\
fax:   \> +49 421 218-4869 \\
email: \> waalkens@physik.uni-bremen.de\\
\end{tabbing}
}
} 
\clearpage


\input{intro}
\input{classic}

\input{regularization}
\input{actionintegrals}
\input{qm}

\input{semiqm}
\input{degenerate}

\input{conclu}

\section*{Acknowledgements}

We thank P.H.~Richter for illuminating discussions and helpful
comments. 
H.D. was supported by the DFG under contract number Du 302.


\bibliographystyle{unsrt}
\bibliography{fg4,extern}

\end{document}

%% file: macros.tex
   \def\bege{\begin{equation}}
   \def\ende{\end{equation}}
   \def\bega{\begin{eqnarray}}
   \def\enda{\end{eqnarray}}
   \def\began{\begin{eqnarray*}}
   \def\endan{\end{eqnarray*}}

\newcommand{\tabstart}{\renewcommand{\baselinestretch}{1.0} \normalsize }
\newcommand{\tabend}{\renewcommand{\baselinestretch}{\bstr} \normalsize }

\newcommand{\bm}[1]{\mbox{\boldmath$#1$\unboldmath}}
\newcommand{\N}{{\mathbb N}} 
\newcommand{\C}{{\mathbb C}}
\newcommand{\R}{{\mathbb R}}
\newcommand{\Z}{{\mathbb Z}}

\newcommand{\rem}[1]{}

\newcommand{\todo}[1]{}
   
\newcommand{\Fell}{{\cal F}}
\newcommand{\Kell}{{\cal K}}

\newcommand{\sn}{{\,\mbox{sn}}}
\newcommand{\cn}{{\,\mbox{cn}}}
\newcommand{\dn}{{\,\mbox{dn}}} 
\def\tO{$O$}
\def\tP{$P$}
\def\tOA{$OA$}
\def\tOB{$OB$}

\newcommand{\psixi}{\psi_\xi}
\newcommand{\psieta}{\psi_\eta}
\newcommand{\psizeta}{\psi_\zeta}
\newcommand{\BId}{{\bm {\tilde I}}}
\newcommand{\Ids}{\tilde{I}_s}

\newcommand{\BIf}{{\bm I}}  

\newcommand{\Iflambda}{I_{\lambda}}
\newcommand{\Ifmu}{I_{\mu}}
\newcommand{\Ifnu}{I_{\nu}}
\newcommand{\BJf}{{\bm J}}  
\newcommand{\Jlambda}{J_\lambda}
\newcommand{\Jmu}{J_\mu}
\newcommand{\Jnu}{J_\nu}

\newcommand{\trace}{\text{tr}\,}

\newcommand{\capsty}{\footnotesize}
\newcommand{\hono}[1]{}
\newcommand{\nono}[1]{}

\newcommand{\nootnote}[1]{}
\newcommand{\EBK}{{\em EBK}}
\newcommand{\WKB}{{\em WKB}}

\newcommand{\equ}{Eq.}
\newcommand{\fig}{Fig.}

\newcommand{\h}{E}

\newcommand{\Bmaslovd}{\tilde{{\bm \alpha}}}

\newcommand{\Gapp}{{\bm G}_{\rm app}}
\newcommand{\Iapp}{\BId_{\rm app}}
\newcommand{\maslovapp}{\Bmaslovd_{\rm app}}

%% file: intro.tex
\section{Introduction}
\label{sec:intro}

Almost exactly 160 years ago, Carl Gustav Jacobi was able to separate
the geodesic flow 
on ellipsoidal surfaces \cite{Jacobi1866}. 
In his letter from December 28, 1838 to his colleague Friedrich
Wilhelm Bessel he wrote:\\ \\
``{\sl Ich habe vorgestern die geod\"atische Linie f\"ur ein
  Ellipsoid mit drei ungleichen Achsen auf Quadraturen
  zur\"uckgef\"uhrt. Es sind die einfachsten Formeln von der Welt,
  Abelsche Integrale, die sich in die bekannten elliptischen
  verwandeln, wenn man zwei Achsen gleich setzt.}''\footnote{English
  translation: The day before yesterday, I reduced the geodesic line
  of an ellipsoid with three unequal axes to quadratures. The
  formulas are the simplest in the world, Abelian integrals,
  transforming into the known elliptical ones if two axes are made equal.}\\ \\
The billiard motion inside an 
$n$-dimensional ellipsoid appears as the singular limit
of the geodesic flow on an $(n+1)$-dimensional ellipsoidal surface with one
semiaxis approaching zero. The starting point in Jacobi's treatment is
what nowadays is called Hamilton-Jacobi ansatz. By introducing
ellipsoidal coordinates Jacobi has shown that the integration of the
Hamilton-Jacobi generating function leads to Abelian integrals. From
Jacobi's point of view this insight essentially solves the
separation problem. Jacobi's integrals represent the Abel
transform of the ellipsoidal coordinates for which the time evolution is
trivial. To give explicit expression for the time evolution of the
ellipsoidal coordinates themselves it is necessary to invert the Abel
map. The solution of this problem, the so-called Jacobi inversion
problem, requires deep insight in the theory of meromorphic
functions on hyperelliptic curves and has given rise to the definition
of theta functions \cite{Dubrovin81}.
This area constituted a highlight in 19th century mathematics. 
With the advent of quantum mechanics the attention of the 
scientific community was shifted from these non-linear finite dimensional
problems to linear infinite dimensional problems.
Recently classical mechanics has
experienced a revival with two main directions. 
On  the one hand computers have induced a boom in the study of
non-integrable systems, essentially by allowing for the visualization
of chaotic phenomena like the break up of Kolmogorov-Arnold-Moser tori. 
The quantum mechanics of non-integrable systems today is a main
topic in physics. 
On the other hand the investigation of soliton equations has given deep 
insights into the theory of integrable systems and a lot of 
the knowledge about integrable systems of the 19th century has
been revived.

In this paper on ellipsoidal quantum billiards we explain
the quantum mechanics of an integrable system in terms of the
corresponding classical system via a semiclassical approach. The main
object will be the hyperelliptic curve of Jacobi's classical
theory. As usual the curve comes into play in order to give
a defintion of the action differential.
The action differential corresponding to the ellipsoidal billiard
defines a hyperelliptic curve of genus 
2 on which it is an Abelian differential of second kind. The
real and purely imaginary periods of this differential enter the
semiclassical quantization scheme in a very natural way.
The presentation of this unified view of classical and semiclassical
treatment is the main theme of this paper. 

According to the Liouville-Arnold theorem the phase space
of an integrable system with $f$ degrees of freedom is 
foliated by invariant manifolds which (almost everywhere) have the 
topology of $f$-tori. The most elegant phase space
coordinates are  action-angle
variables $({\bm I},{\bm \varphi})$, where the action variables ${\bm
  I}$ label the tori and the angles ${\bm \varphi}$
parametrize the torus for fixed ${\bm I}$.
With the original phase space variables $({\bm p},{\bm q})$ 
the action variables ${\bm I}$ are obtained from integrating the Liouville
1-form ${\bm p}\,d{\bm q}$ along $f$ independent cycles $\gamma_i$ on
the torus according to
\bege
\label{eq:actiongeneral}
I_i = \frac{1}{2\pi} \oint_{\gamma_i} {\bm p} \, d{\bm q}\,,\quad i=1,...,f\,.
\ende
Hamilton's equation reduce to
\bega
 \dot{I}_i &=& -\frac{\partial H({\bm I})}{\partial \varphi_i}=0\,,\\
 \dot{\varphi}_i &=& \frac{\partial H({\bm I})}{\partial I_i} = \omega_i \,,\quad i=1,...,f\,
\enda
with $\omega_i$ the constant frequencies. The time evolution
becomes trivial.
Although the importance of action-angle variables is
stressed in any text book on classical mechanics, especially as the
starting point for the study of non-integrable perturbations of
integrable systems \cite{Berry81}, there  can be found only few
non-trivial examples in the literature for which the action 
variables are explicitely calculated. 
P.~H.~Richter \cite{Richter90} started to fill this gap
for integrable tops
and recently this presentation has been given for various systems,
e.g.\ for the Kovalevskaya top \cite{DJR94,Dullin94b},
integrable billiards with and without potential
\cite{RW95,WR96,Schwebler97,Wiersig98}, 
the integrable motion of a particle with respect to the Kerr 
metric \cite{Heudecker95}, and the  motion of a particle  
in the presence of two Newton potentials - the so-called
two-center-problem. 
It turns out that the presentation of energy surfaces $H(\BIf )=E$ in
action space may be considered as the most compact description of an 
integrable system \cite{DHJPSWWW97}.

The importance of action variables extends to quantum
mechanics in the following way.
In a semiclassical sense the  Liouville-Arnold tori carry the
quantum mechanical wave functions. Stationary solutions of
Schr\"odinger's equation result from single valuedness conditions
imposed on the wave functions carried on the tori. These give
the semiclassical quantization conditions
\bege
\label{eq:ebkquantization}
I_i = (n_i+\alpha_i/4)\hbar\,,\quad i=1,...,f\,,
\ende
with quantum numbers $n_i$ and Maslov indices $\alpha_i$. The
$\alpha_i$  are purely classical indices of the corresponding 
Liouville-Arnold torus which is a Lagrangian manifold. They  
come into play because quantum mechanics is considered with
respect to only half the phase space variables $({\bm p},{\bm q})$ -
usually in configuration space representation, i.e. with respect to
the $q_i$ \cite{Keller58}.  The Maslov indices characterize the
singularities of the projection of the Lagrangian manifold to
configuration space which lead to phase shifts of semiclassical wave
functions supported on the tori \cite{Arnold78,Keller58}. The Maslov
indices depend on the choice of the cycles $\gamma_i$ in
Eq.~(\ref{eq:actiongeneral}). In the case of a separable system and a
canonical choice of the cycles on the torus according to
\bege
\gamma_i  : dq_j \equiv 0 \,,\quad j\neq i\,,
\ende 
we simply have $\alpha_i=0$ if the 
$i$th degree of freedom is of rotational type and $\alpha_i=2$ if the
$i$th degree of freedom is of oscillatory type.
The \EBK\ quantization (\ref{eq:ebkquantization}) was the center of
the old quantum mechanics of Bohr and Sommerfeld before 1926. 
The fact that this quantization assumes that
phase space is foliated by invariant tori was realized by Einstein,
but his 1917 paper on this matter \cite{Einstein17} 
was hardly recognized at that time.

The phase space of the ellipsoidal billiard is foliated
by four types of tori which have different Maslov indices. Two crossing
separatrix surfaces divide the action space into four regions - one
four each type of tori.
This means that the simple \EBK\ quantization condition
(\ref{eq:ebkquantization}) is not uniformly 
applicable to the ellipsoidal billiard: 
the quantum mechanical 
tunneling between the different types of tori has to be taken into
account. Both effects can semiclassically be incorporated  by a  \WKB\
ansatz for the wave function. The tunneling between tori
is then described by tunnel matrices which connect the amplitudes of
\WKB\ wave functions in different classically allowed configuration
space areas. The main ingredient for the tunnel matrices is a
penetration integral. For the ellipsoidal
billiard there exist two such penetration integrals - one for each
separatrix. 

The differentials for both penetration integrals are
identical. They are even identical to the differential for the action
integrals of the three degrees of freedom, the only difference is 
the intergration path. The action and penetration
integrals therefore appear as the real and purely imaginary periods
of a single Abelian differential of second kind. This is how
semiclassical quantum mechanics 
extends the meaning of the originally classical hyperelliptic curve
and how quantum mechanics appears as a ``complexification'' of
classical mechanics. 

Within the last few years the study of billiards has become
very popular in connection with the investigation 
of the quantum mechanics of classically chaotic
systems. The quantum mechanics of two-dimensional billiards can easily 
be investigated experimentally by
flat microwave cavities for which one component of the electric
field vector mimics the scalar quantum mechanical wave function
\cite{SS90,SS92,AGHHLRRS94}. The relation between Schr{\"o}dinger's
equation for a quantum 
billiard and Maxwell's equations for the electromagnetic field in a
three-dimensional cavity is complicated by the vector character of the
electromagnetic field \cite{DSBK98}.
Three-dimensional billiards have a direct physical
interpretation as models for atomic nuclei \cite{Strut77} and  metal
clusters \cite{Brack93}. Recently their importance has been
rediscovered in connection with lasing droplets \cite{GCNNSFSC98}.
The semiclassical analysis of rotationally symmetric ellipsoids  can
be found e.g. in \cite{AyantArvieu86,AyantArvieu86a,ABCT87}. 

This paper is organized as follows. In
Section~\ref{sec:classic} we summarize the classical aspects of the
ellipsoidal billiard. We introduce constants of the motion, discuss the
different types of tori and give a regularization of the ellipsoidal
coordinates. 
In Section~\ref{sec:actionintegrals} the hyperelliptic curve associated with
the ellipsoidal billiard is investigated. The separated Schr\"odinger
equation is solved in Section~\ref{sec:qm}. 
In Section~\ref{sec:semiqm} a uniform semiclassical quantization scheme
in terms of a {\sl WKB} 
ansatz is performed and a representation of the quantum eigenvalues
in classical action space is given. In Section~\ref{sec:deg} we
comment on how the degenerate versions of the ellipsoidal billiard,
i.e. the prolate and the oblate ellipsoidal billiard and the billiard
in the sphere, appear as special cases of the general triaxial
ellipsoidal billiard.
We conclude with some brief remarks and an outlook in 
Section~\ref{sec:conclu}. 

%% file: classic.tex
\section{The Classical System}
\label{sec:classic}
We consider the free motion of a particle of unit mass inside the
general triaxial ellipsoid in $\R^3$ defined by
\bege
x^2+\frac{y^2}{1-b^2}+\frac{z^2}{1-a^2}=1
\ende
with $0<b<a<1$. The particle is elastically reflected when it hits the
boundary ellipsoid.  Throughout this paper we take $(a,b)=(0.7,0.3)$
in our numerical calculations.

Hamilton's equations of motion and the reflection condition are
separable in
ellipsoidal coordinates $(\xi,\eta,\zeta)$. 
\rem{
If we keep one of them
fixed the remaining two coordinates parametrize  
 a confocal quadric defined by}
Each of them parametrizes a family of confocal quadrics
\bege
\label{eq:confocalsurfaces}
\frac{x^2}{s^2}+\frac{y^2}{s^2-b^2}+\frac{z^2}{s^2-a^2} = 1,
\ende
where $s\in \{\xi,\eta,\zeta\}$. For $1\ge s=\xi \ge a$  all terms
in \equ~(\ref{eq:confocalsurfaces}) are positive and the equation
defines  
a family of confocal 
ellipsoids.  Their
intersections with the $(x,y)$-plane, the $(x,z)$-plane and the
$(y,z)$-plane   are planar ellipses with  foci at
$(x,y)=(\pm b,0)$, $(x,z)=(\pm a,0)$ and $(y,z)=(\pm (a^2-b^2)^{1/2},0)$, 
respectively. For $a\ge s=\eta \ge b$ the third term in  
\equ~(\ref{eq:confocalsurfaces}) becomes
negative. \equ~(\ref{eq:confocalsurfaces}) thus gives confocal one
sheeted hyperboloids.  
Their intersections with the $(x,y)$-plane are planar ellipses with foci
$(x,y)=(\pm b,0)$; 
the intersections with the $(x,z)$-plane and the
$(y,z)$-plane   are planar hyperbolas with  foci at
$(x,z)=(\pm a,0)$ and $(y,z)=(\pm (a^2-b^2)^{1/2},0)$, respectively.
For $b\ge s=\zeta \ge 0$ 
the second and third
terms in \equ~(\ref{eq:confocalsurfaces}) are negative giving
confocal two sheeted hyperboloids. 
Their intersections with the $(x,y)$-plane and  the $(x,z)$-plane  are
planar  hyperbolas 
with  foci at 
$(x,y)=(\pm b,0)$ and $(x,z)=(\pm a,0)$, respectively; they do not
intersect the $(y,z)$-plane.  

Inverting \equ~(\ref{eq:confocalsurfaces}) within the positive 
$(x,y,z)$-octant gives 
\begin{eqnarray}
\label{eq:transxeztoxyz}
(x,y,z)=\left(\frac{\xi\eta\zeta}{ab},\frac{\sqrt{(\xi^2-b^2)(\eta^2-b^2)(b^2-\zeta^2)}}{b\sqrt{a^2-b^2}},\frac{\sqrt{(\xi^2-a^2)(a^2-\eta^2)(a^2-\zeta^2)}}{a\sqrt{a^2-b^2}} \right)
\end{eqnarray}
with
\begin{equation}
\label{eq:xezranges}
0\le \zeta \le b \le \eta \le a \le \xi \le 1.
\end{equation}
The remaining octants are obtained by appropriate reflections.
Note that the transformation $(x,y,z) \leftrightarrow (\xi,\eta,\zeta)$ is
singular on the Cartesian planes  $(x,y)$, $(x,z)$ and $(y,z)$, i.e.\
at the branch points of \equ~(\ref{eq:transxeztoxyz}), see
Fig.~\ref{fig:singularplane}. 

With $(p_\xi,p_\eta,p_\zeta)$, the momenta conjugate to
$(\xi,\eta,\zeta)$, Hamilton's function for a freely moving particle 
in ellipsoidal coordinates reads 
\bega
H= 
\frac{(\xi^2-a^2)(\xi^2-b^2)}{(\xi^2-\eta^2)(\xi^2-\zeta^2)}\frac{p_\xi^2}{2}
  +
  \frac{(a^2-\eta^2)(\eta^2-b^2)}{(\xi^2-\eta^2)(\eta^2-\zeta^2)}
  \frac{p_\eta^2}{2}    
  + \frac{(a^2-\zeta^2)(b^2-\zeta^2)}{(\xi^2-\zeta^2)(\eta^2-\zeta^2)}
  \frac{p_\zeta^2}{2}\,.
\enda
The reflection at the billiard boundary $\xi =1$ is simply described by
\bege
(\xi,\eta,\zeta,p_\xi,p_\eta,p_\zeta) \rightarrow 
        (\xi,\eta,\zeta,-p_\xi,p_\eta,p_\zeta).
\ende
Especially for the quantization it is  useful to  consider also the
symmetry reduced billiard. The billiard is then confined to one
$(x,y,z)$-octant, e.g. the positive one, with the particle
being elastically reflected when it hits the boundary or 
one of the planes $(x,y)$, $(x,z)$, or $(y,z)$.

\def\figsingularplane{%
On the planes $(x,y)$ and $(x,z)$  the coordinate surfaces of
$\xi$, $\eta$ and $\zeta$ reduce to planar conic sections, on which
the transformation $(x,y,z)\leftrightarrow (\xi,\eta,\zeta)$ is singular. 
The figure shows them for the positive $(x,y,z)$-octant.}
\def\FIGsingularplane{\centerline{\psfig{figure=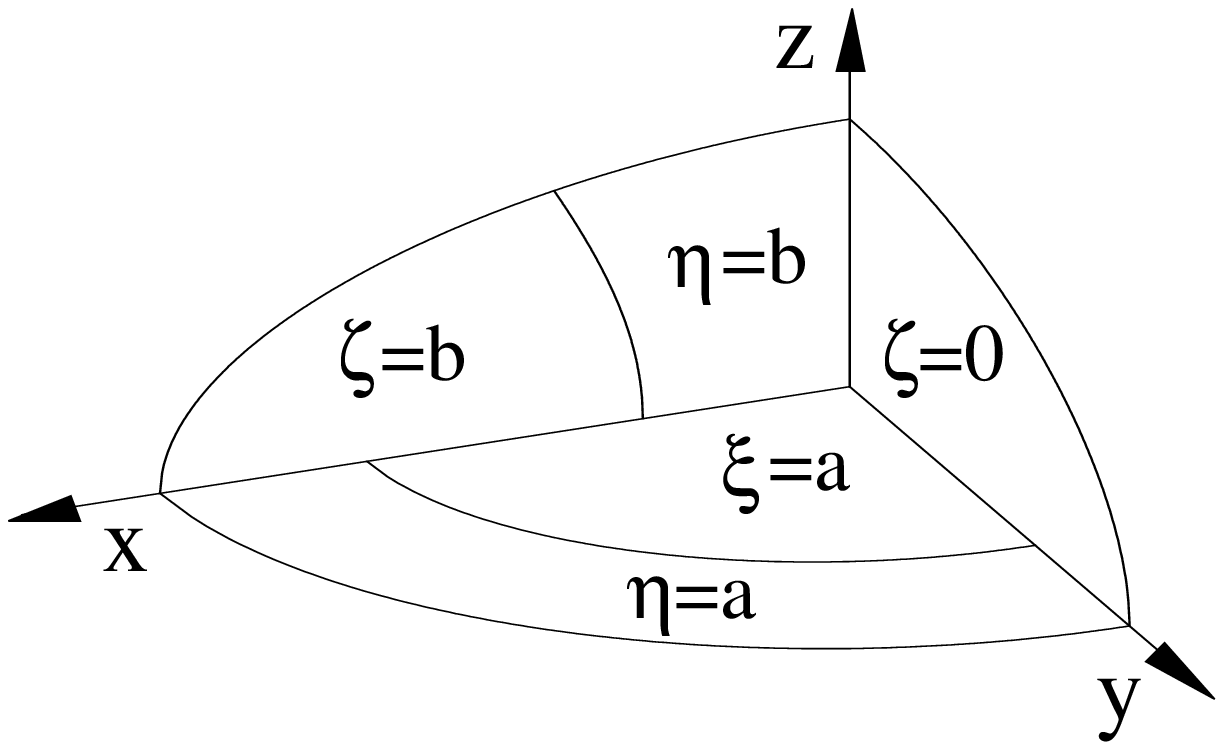,width=5cm,angle=-90}}}
\FIGo{fig:singularplane}{\figsingularplane}{\FIGsingularplane}
The separation of Hamilton's equations in these variables can, e.g., be 
found in \cite{WR96}. 
Because the Hamiltonian and the reflection condition can be separated
the system is completely integrable. Besides the energy there are
two independent conserved quantities
\bega\label{eq:xtok1EB}
K^2 & \equiv & 4\h k = |{\bm L}|^2+(a^2+b^2)p_x^2+a^2p_y^2+b^2p_z^2\,, \\
\label{eq:xtok2EB}
L^2 & \equiv & \frac{2\h l}{a^2} = \frac{b^2}{a^2}L_y^2+L_z^2+b^2p_x^2 \, ,
\enda
where $L_x$, $L_y$, $L_z$ denote the components of the total angular
momentum ${\bm L} = {\bm r} \times {\bm p}$,
but $L \not = |{\bm L}|$.
In the spherical limit $a=b=0$ we have $K=|{\bm L}|$. Thus $K$ is a  
generalization of the absolute value of the total
angular momentum. The meaning of $L$ becomes clear
in the limiting cases of rotationally symmetric ellipsoids. In the oblate
case  ($b=0$) $L$ is  the angular momentum about the shorter
semiaxis, $L=L_z$. In the prolate case ($a=b$) $L$ is related to
the angular momentum about the longer semiaxis, $L^2=K^2-2\h a^2-L_x^2$. 

After separation the squared momentum can  be 
written as
\bege
\label{eq:xietazetamomenta}
p_s^2 = 2E \frac{s^4 -2ks^2 + l}{(s^2-a^2)(s^2-b^2)}
\ende
with $s\in \{ \xi ,\eta ,\zeta \}$. 
It is convenient to take the turning points $s_1$ and $s_2$ of
$(\xi,\eta,\zeta)$ to parametrize the possible values
of $K$ and $L$, such that
\bege
\label{eq:sihk}
        s^4 - 2ks^2 + l = (s^2-s_2^2)(s^2-s_1^2).
\ende
Note that the transformation from $k$, $l$ to $s_i$ is singular
for $s_1 = s_2$.

In order to ensure real valued momenta for some configuration
$(\xi,\eta,\zeta)$, Equations~(\ref{eq:xietazetamomenta}) and 
(\ref{eq:sihk}) give the conditions
\bege
\label{eq:s1s2orders}
s_1\le s_2,\quad 0\le  s_1\le a,\quad b\le s_2 \le 1.
\ende
\def\figbifurcationdiagram{%
Bifurcation diagram (thick lines). The four regions correspond to
smooth two parameter families of 3-tori, see Fig.~\ref{fig:caustics}.
The meaning of the dashed lines $c_\lambda$ and $c_\nu$ and
the points $P_\lambda$ and $P_\nu$ is explained in the text.
The angle $\varphi$ is considered in
Section~\ref{sec:actionintegrals}. 
}
\def\FIGbifurcationdiagram{\centerline{\psfig{figure=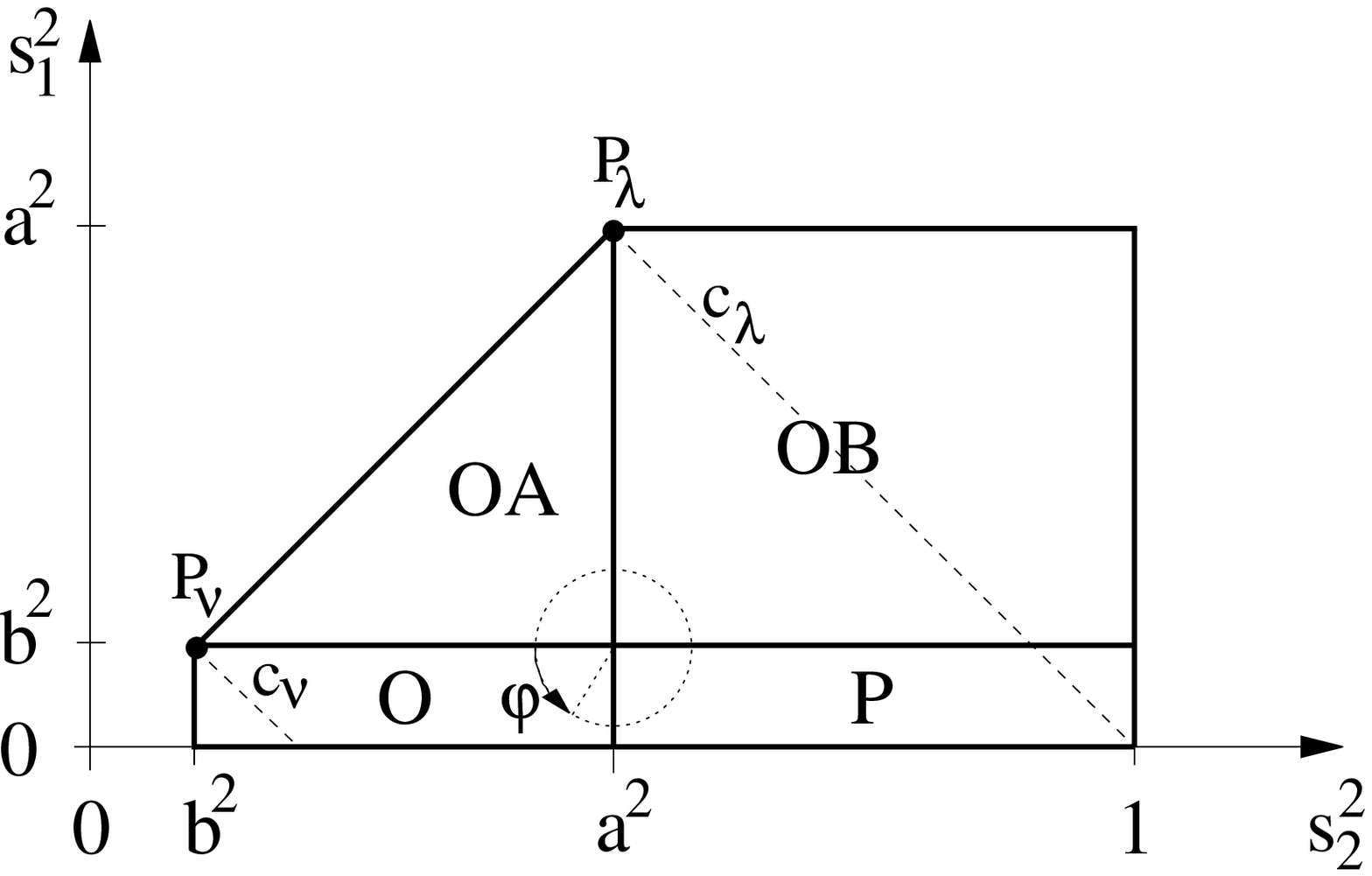,angle=-90,width=9cm}}}
\FIGo{fig:bifurcationdiagram}{\figbifurcationdiagram}{\FIGbifurcationdiagram}
As for billiards in general, the energy dependence can be removed by a simple
scaling, see \equ~(\ref{eq:xietazetamomenta}). 

The bifurcation diagram of an integrable system shows the critical
values of the energy momentum mapping from phase space to the 
constants of motion. Typically the critical values correspond to
the double roots of a certain polynomial, and the different types
of motion correspond to the ranges of regular values of the energy
momentum mapping. 

In the ellipsoidal billiard the type of motion is determined by 
the ordering of the numbers $b$, $s_1$, $a$ and $s_2$. 
Equality in \equ~(\ref{eq:s1s2orders}) gives the five
outer lines of the bifuraction diagram, while
the lines $s_1=b$ and $s_2=a$ give the inner lines,
see Fig.~\ref{fig:bifurcationdiagram}.
The bifurcation diagram divides the parameter plane into four
patches. In Fig.~\ref{fig:caustics} the corresponding types of 3-tori
are represented by their 
caustics, i.e.\ by their envelopes in configuration space.  The
ellipsoidal boundary itself is usually not considered as a caustic. 
The caustics are pieces of the
quadric surfaces in \equ~(\ref{eq:confocalsurfaces}). 
Motion of type \tO\ is purely oscillatory in
all variables $(\xi,\eta,\zeta)$. The
oscillations in the ellipsoidal direction $\xi$ is given by
reflections at the boundary ellipsoid $\xi=1$. $\eta$ and $\zeta$
oscillate between their caustics.
The remaining types of motion are best understood by considering
the two limiting cases of rotationally symmetric ellipsoids.
Type \tP\ involves a rotation about the $x$-axis described by the
coordinate~$\eta$. $\xi$~now oscillates
between the caustic and the boundary ellipsoid. $\zeta$~oscillates
between its caustics.
This is the only generic type of  motion in
prolate ellipsoids. Motion types \tOA\ and \tOB\ both involve
rotations about the $z$-axis, described by the coordinate~$\zeta$. They
are the two generic types of  
motion in oblate ellipsoids. For \tOA\, $\xi$ oscillates between the
boundary ellipsoid, for \tOB\, $\xi$ oscillates between the caustic and the
boundary ellipsoid. The way $\eta$ oscillates between its caustics
is different in the two cases.
Motion type \tO\ can only occur in the general
triaxial ellipsoid without any rotational symmetry. A given 
value of the constants of motion $(E,K^2,L^2)$ or $(E,s_1^2,s_2^2)$ in
region \tO\  
corresponds to a single 3-torus in phase space. In all the other regions
there exist two disjoint tori in phase space which have the same
constants of motion. They just differ by a sense of rotation.
The non-generic
motions on lower dimensional tori corresponding to the critical lines 
in Fig.~\ref{fig:bifurcationdiagram} are discussed in detail in
\cite{WR96,Wiersig98}. \\
\def\FIGcaustics{%
  \centerline{\psfig{figure=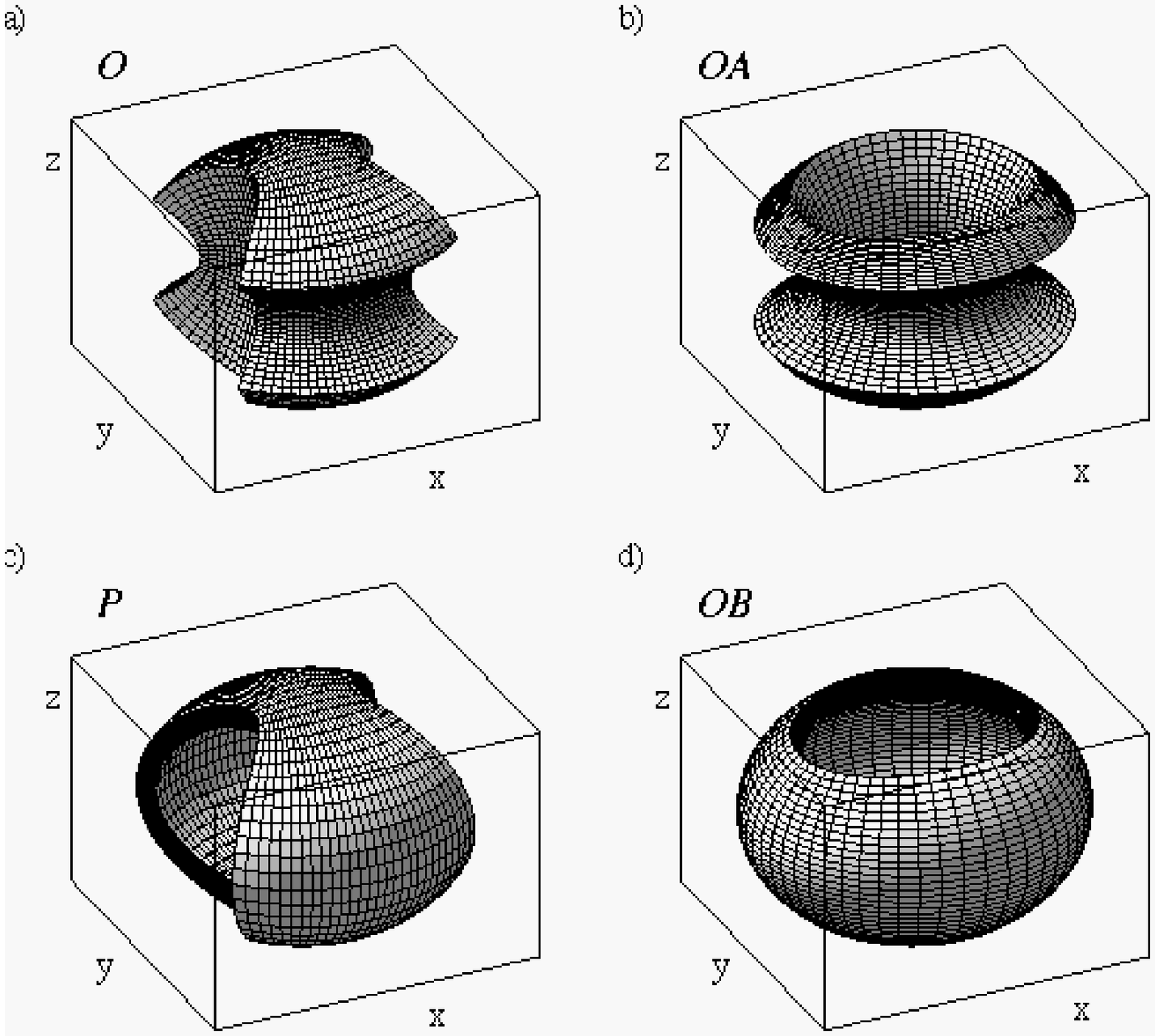,angle=-90,width=12.5cm}}
  }
\def\figcaustics{%
Caustics and boundary ellipsoid of the four types of invariant 3-tori for
the ellipsoid with 
constants of the  motion
a) $(s_1^2,s_2^2) = (0.05,0.4)$,
b) $(s_1^2,s_2^2) = (0.25,0.4)$,
c) $(s_1^2,s_2^2) = (0.05,0.8)$,
d) $(s_1^2,s_2^2) = (0.25,0.8)$.
}
\FIGo{fig:caustics}{\figcaustics}{\FIGcaustics}

%% file: regularization.tex
The description of the free motion inside the ellipsoid in terms of
the phase space variables $(\xi,\eta,\zeta,p_\xi,p_\eta,p_\zeta)$ is
rather complicated because of the change of coordinate sheets each
time a boundary of the intervals in \equ~(\ref{eq:xezranges}) is reached. Upon
crossing one of the Cartesian coordinate planes $(x,y)$ or $(x,z)$
one of the momenta $p_\xi$,
$p_\eta$ or $p_\zeta$ changes from $\pm \infty $ to $\mp \infty$, see
\equ~(\ref{eq:xietazetamomenta}) 
and Fig.~\ref{fig:singularplane}.
The singularities in \equ~(\ref{eq:xietazetamomenta}) can be removed
by a canonical transformation. The new coordinates are 
better suited for the semiclassical considerations in Section~\ref{sec:semiqm}.

For the generating function of this canonical transformation we choose
the ansatz
\bege
F_2 = \lambda (\xi ) p_\lambda + \mu (\eta ) p_\mu + \nu (\zeta) p_\nu.
\ende
The index 2 indicates that this is a generating function of type
2 in the notation of H. Goldstein, see \cite{Gold80}.
Then
\bege
\lambda = \frac{\partial F_2}{\partial p_\lambda },\,\, \mu = \frac{\partial
  F_2}{\partial p_\mu} ,\,\, \nu = \frac{\partial F_2}{\partial p_\nu}
\ende
are the new coordinates with $(p_\lambda ,p_\mu ,p_\nu )$ the
conjugate momentum variables. The transformation is completed by
relating the old and new momentum variables:
\bege
p_\xi = \frac{d \lambda }{d \xi} p_\lambda,\,\,p_\eta =
\frac{d\mu}{d\eta}p_\mu ,\,\, p_\zeta = \frac{d\nu}{d\zeta}p_\nu.
\ende
To remove the singularities in Eq.~(\ref{eq:xietazetamomenta}) we
require the above derivatives to be
\bega
\frac{d\lambda}{d\xi} &=& \quad\frac{a}{\sqrt{(\xi^2-a^2)(\xi^2-b^2)}},\\
\label{eq:dmudeta}
\frac{d\mu}{d\eta} &=& -\frac{a}{\sqrt{(a^2-\eta^2)(\eta^2-b^2)}},\\
\frac{d\nu}{d\zeta} &=& \quad\frac{a}{\sqrt{(a^2-\zeta^2)(b^2-\zeta^2)}}.
\enda 
Note the negative sign of the derivative  $d\mu/d\eta$.
These equations involve square roots of fourth order polynomials, 
i.e.\ they lead to elliptic integrals. 
Their inversion leads  to elliptic functions. One finds 
\begin{eqnarray}
\label{eq:xietazetaoflambdamunufirst}
\xi(\lambda) &=& a\frac{\dn(\lambda,q)}{\cn(\lambda,q)},\\
\eta(\mu) &=& a\dn(\mu,q'),\\
\label{eq:xietazetaoflambdamunulast}
\zeta(\nu) &=& b\sn(\nu,q),
\end{eqnarray}
where $\sn(\phi,q)$, $\cn(\phi,q)$ and $\dn(\phi,q)$ are Jacobi's
elliptic functions with 'angle' $\phi$ and modulus $q$
\cite{GradRyzh65}. Here the modulus 
is given by $q=b/a$. $q'=(1-q^2)^{1/2}$ denotes the conjugate
modulus. This is the standard parameterization of the elliptic coordinates
by elliptic functions, see e.g.~\cite{Morse53}.
For the momenta one finds
\bege \label{eq:lmnmomenta}
p_{\hat{s}}^2=\sigma_{\hat{s}} \frac{2E}{a^2}\left( s^4(\hat{s}) -2ks^2(\hat{s}) + l\right),
\ende
with $\hat{s}\in\{\lambda,\mu,\nu\}$ and 
$s(\hat{s})\in\{\xi(\lambda),\eta(\mu),\zeta(\nu)\}$ from
Equations~(\ref{eq:xietazetaoflambdamunufirst})-(\ref{eq:xietazetaoflambdamunulast}).
The coefficients $\sigma_{\hat{s}}$ are the signs
$\sigma_\lambda=\sigma_\nu=+$ and $\sigma_\mu=-$.

Transforming
the coordinate ranges in \equ~(\ref{eq:xezranges}) for the old coordinates
$(\xi ,\eta ,\zeta)$ to the new coordinates gives
\begin{eqnarray}
\label{eq:lmnrangesreduced1}
0 \le & \lambda & \le \Fell \left(\chi,q\right),\\
\label{eq:lmnrangesreduced2}
0 \le & \mu & \le \Kell(q')=\Kell'(q),\\
\label{eq:lmnrangesreduced3}
0 \le & \nu & \le \Kell(q),
\end{eqnarray}
for the motion in one octant.
Here $\Fell(\chi,q)$ denotes Legendre's incomplete elliptic integral of
first kind with amplitude $\chi$ and modulus $q$
\cite{GradRyzh65,Byrd71}. The amplitude is 
given by
\bege
 \sin^2 \chi = \frac{1-a^2}{1-b^2} \, .
\ende
$\Kell(q)$ is the complete elliptic integral of first
kind with modulus $q$ and $\Kell'(q) = \Kell(q')$ its complement. 
In the following we will  omit the modulus in the notation for
elliptic integrals because the modulus will not change in the course
of this paper. 
The appearance  of the incomplete integral is due to the
fact that we cut off the coordinate range in the ellipsoidal direction,
i.e. to the billiard character of the underlying motion.
In terms of the Cartesian coordinates the coordinate ranges in
Equations (\ref{eq:lmnrangesreduced1})-(\ref{eq:lmnrangesreduced3}) yield the octant $x,y,z\ge 0$ within the ellipsoid.
Inserting $(\lambda ,\mu ,\nu )$ into the expressions for the Cartesian
coordinates in  Eq.~(\ref{eq:transxeztoxyz}) gives 
\bega
\label{eq:translmntoxyz1}
x &=& a   \frac{\dn(\lambda ,q)\dn(\mu,q')\sn(\nu,q)}{\cn(\lambda,q)},\\
\label{eq:translmntoxyz2}
y &=& q'a \frac{\cn(\mu,q')\cn(\nu,q)}{\cn(\lambda,q)},\\
\label{eq:translmntoxyz3}
z &=& q'a \frac{\sn(\lambda,q)\sn(\mu,q')\dn(\nu,q)}{\cn(\lambda,q)}.
\enda 
The functions $\sn(\phi)$ and $\cn(\phi)$ both have period
$4\Kell$ on the real axis, $\dn(\phi)$ has period $2\Kell$.  
Extending the ranges in Equations
(\ref{eq:lmnrangesreduced2}) and (\ref{eq:lmnrangesreduced3})  to the full
real axis for $\mu$ and $\nu$ thus gives  $x$, $y$ and $z$ as
periodic functions of $\mu$ and $\nu$. If in addition to
that we let $\lambda$ vary in the interval $[- \Fell(\chi),\Fell(\chi)]$ 
the billiard dynamics becomes smooth across
the planes  $(x,y)$, $(x,z)$ and $(y,z)$. We thus have a coordinate system
that both separates Hamilton's equations and the reflection condition and
yields smooth dynamics inside the ellipsoid. 
The motion is thus best described as a geodesic flow on the
product of an interval and a $2$-torus,
\bege
(\lambda,\mu,\nu) \in [- \Fell(\chi),\Fell(\chi)] \times T^2,
\ende
i.e.\ on a solid 2-torus as depicted in
Fig.~\ref{fig:massivetorus}. The flow 
is smooth except for the reflections at the boundaries $\lambda = \pm
\Fell(\chi)$ which are still desribed by the sign change
\bege
(\lambda,\mu,\nu,p_\lambda,p_\mu,p_\nu) \rightarrow (\lambda,\mu,\nu,-p_\lambda,p_\mu,p_\nu).
\ende
The whole torus 
\begin{eqnarray}
- \Fell \left(\chi\right) \le & \lambda & \le \Fell\left(\chi \right)\,,\\
0 \le & \mu & \le 4\Kell'\,,\\
0 \le & \nu & \le 4\Kell
\end{eqnarray}
gives a fourfold cover of the interior of the ellipsoid. 
\def\figmassivetorus{%
Solid $2$-torus as the fourfold cover of the configuration space 
of the billiard inside the ellipsoid.}%
\def\FIGmassivetorus{\centerline{\psfig{figure=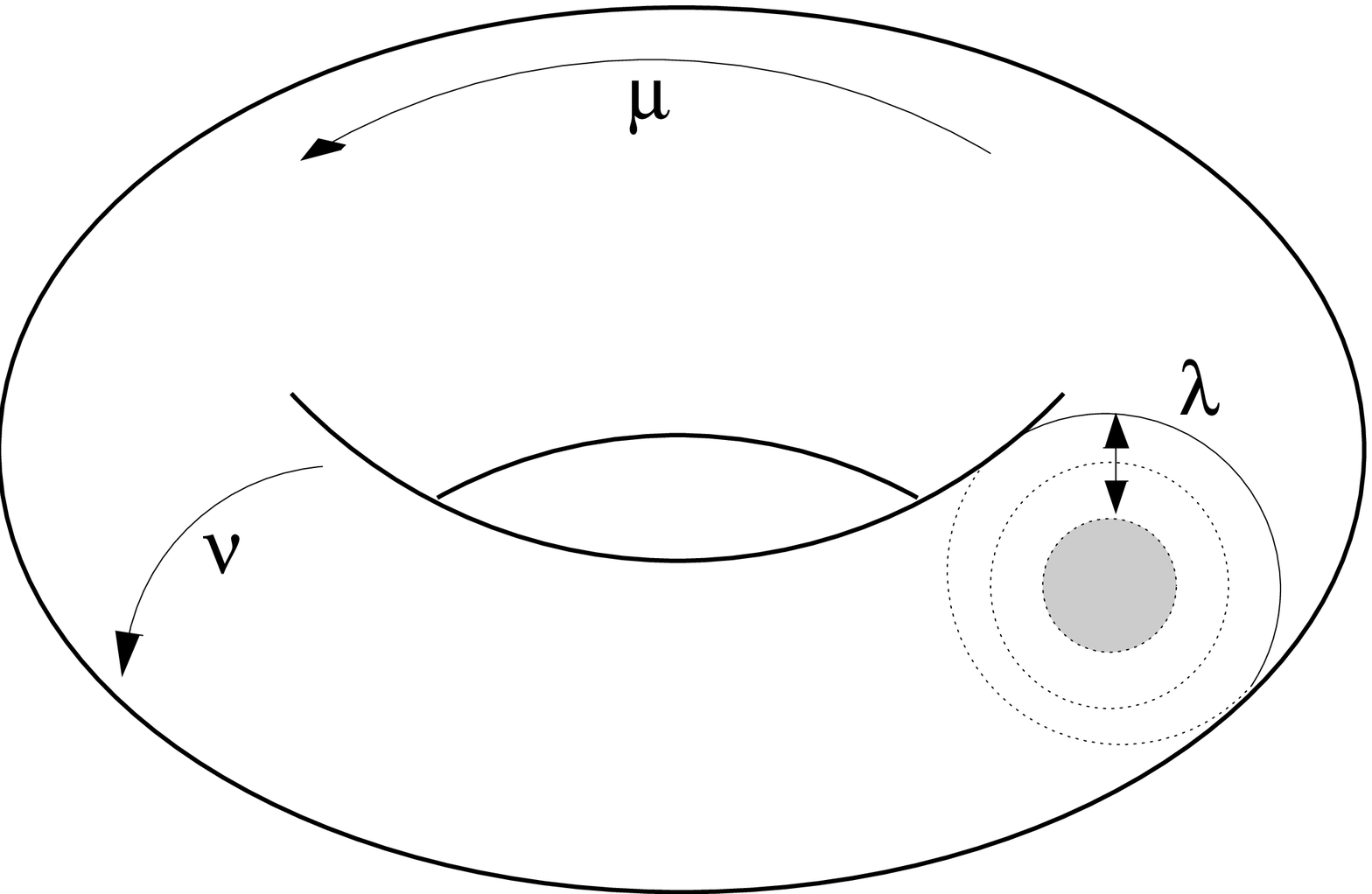,width=6cm}}}%
\FIGo{fig:massivetorus}{\figmassivetorus}{\FIGmassivetorus}%
\def\figconfigtorus{%
Representation of the solid $2$-torus (Fig.~\ref{fig:massivetorus})
of the configuration space as
a cube with periodic boundaries in the directions of $\mu$ and
$\nu$. Each small cube represents one $(x,y,z)$-octant. 
They are labeled in a 'binary' way with
respect to the signs of $x$,$y$ and $z$, i.e. $(-,-,-)$ corresponds to
$0$, $(-,-,+)$ corresponds to 1, ..., $(+,+,+)$ corresponds to
$7$. The labels are put on the right side of each cube.}
\def\FIGconfigtorus{\centerline{\psfig{figure=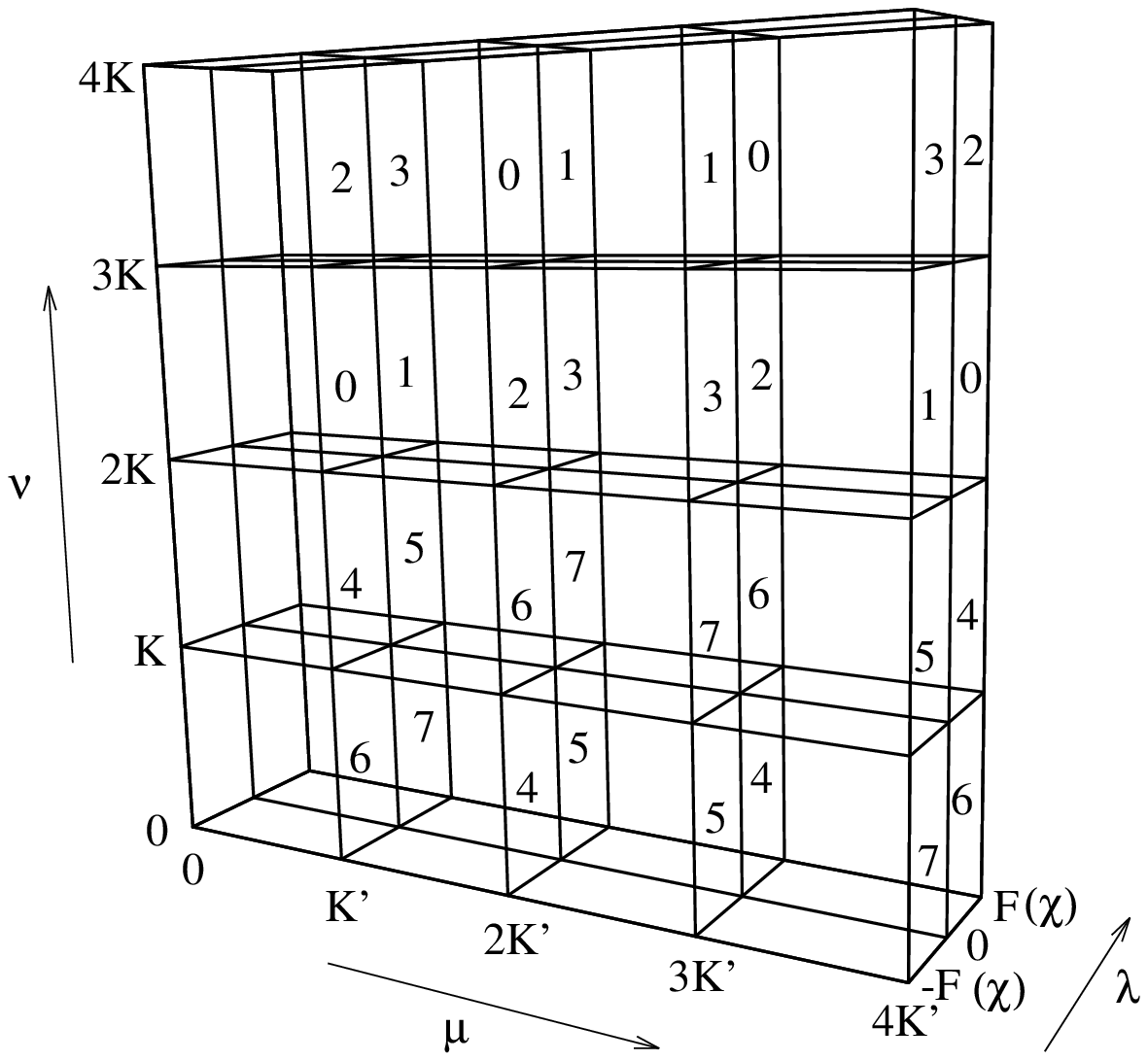,width=9cm,angle=-90}}}%
\FIGo{fig:configtorus}{\figconfigtorus}{\FIGconfigtorus}%
In Fig.~\ref{fig:configtorus} we represent the solid torus as a cube
and mark the boundaries between the preimages of the different
$(x,y,z)$-octants. Each $(x,y,z)$-octant gives a small cube
\bege
[0,\pm \Fell(\chi)] \times [n_\mu \Kell',(n_\mu+1)\Kell'] \times [n_\nu \Kell,(n_\nu+1)\Kell]
\ende
with $(n_\mu,n_\nu) \in \Z^2$. The fact that each  of the
small cubes has to be bounded by 5 neighbouring small cubes 
to make the dynamics smooth can be
understood in terms of the old variables $\xi$, $\eta$ and
$\zeta$. Each $(x,y,z)$-octant is bounded by five singular sheets of the
coordinates $(\xi,\eta,\zeta)$, see \fig~\ref{fig:singularplane}. 
Note that instead of considering the three real
Equations~(\ref{eq:xietazetaoflambdamunufirst})-(\ref{eq:xietazetaoflambdamunulast}) it is equivalent to consider only the $\zeta$ equation but for complex
$\nu$ in the fundamental domain and use the idendities 
$\sn(u+\Kell+i\Kell',q) = q^{-1}\dn(u,q)/\cn(u,q)$ and 
$\sn(-iu+\Kell+i\Kell',q) = q^{-1}\dn(u,q')$.

The four covers of the ellipsoid are related by the group of
involutions which leave the Cartesian coordinates in
Equations~(\ref{eq:translmntoxyz1})-(\ref{eq:translmntoxyz3}) fixed.   
This group has three non-trivial elements
\bega
\label{eq:invol1}
        S_1(\lambda,\mu,\nu) &=& (-\lambda,-\mu,\nu)\,, \\
\label{eq:invol2}
        S_2(\lambda,\mu,\nu) &=& (\lambda,-\mu-2\Kell',2\Kell-\nu)\,, \\
\label{eq:invol3}
        S_3(\lambda,\mu,\nu) &=& (-\lambda,\mu-2\Kell',2\Kell-\nu)\,.
\enda
Any two of them generate the group which is isomorphic to the 
dihedral group $D_2$ (also called ``Kleinsche Vierergruppe'' \cite{LudFal88}).

Inspection of Equations
(\ref{eq:translmntoxyz1})-(\ref{eq:translmntoxyz3})
shows that it is 
justifiable to think of $\mu$ as a kind of rotational angle about the
$x$-axis. In the $y$-component and $z$-component $\mu$ appears as the
argument of the 
elliptic functions sn and cn which are similar to the
trigonometric functions sine and cosine. Similarly $\nu$ 
can be considered as a rotation angle about the $z$-axis. The types of
motion in Fig.~\ref{fig:caustics} therefore have the interpretations of
$\mu$-rotations for type \tP\ and $\nu$-rotations with different
$\mu$-oscillations for types \tOA\ and \tOB.

In \fig~\ref{fig:configtorus} each column 
\bege
[-\Fell(\chi),\Fell(\chi)] \times [n_\mu \Kell',(n_\mu+1)\Kell'] \times [0,4\Kell]
\ende 
with $n_\mu \in \Z$ fixed 
gives a single cover of the interior of the ellipsoid. 
 This does not hold analogously for $\nu$. This is familiar from
 the polar coordinates of the sphere where $\nu$ should be compared to the
 azimutal angle and $\mu$ is similar to the polar angle.

For the 
semiclassical quantization in Section \ref{sec:semiqm} it is helpful to
deal with a simple kinetic-plus-potential-energy Hamiltionian. We
therefore write \equ~(\ref{eq:lmnmomenta}) in the form
\bege
\label{eq:effham}
E_{\hat{s}}=\frac{p_{\hat{s}}^2}{2} + V_{\hat{s}}(\hat{s})
\ende
with 
\bege
\label{eq:effen}
E_{\hat{s}} = \sigma_{\hat{s}}\frac{E}{a^2} l
\ende
and
\bege
\label{eq:effpot}
V_{\hat{s}}(\hat{s}) =
-\sigma_{\hat{s}}\frac{E}{a^2}(s^4(\hat{s})-2ks^2(\hat{s}))\,. 
\ende
The effective potentials
$V_{\mu}$ and $V_{\nu}$ are periodic
functions with periods $2\Kell'$ and $2\Kell$, respectively.
$V_{\lambda}$ is symmetric about 0. The number of
potential wells per period changes 
across the lines
$2k=s_1^2+s_2^2=2a^2$ and $2k=s_1^2+s_2^2=2b^2$ 
indicated as the dashed lines $c_\lambda$ and $c_\nu$ in
Fig.~\ref{fig:bifurcationdiagram}. Between $c_\lambda$ and 
$c_\nu$, $V_{\mu}$ has two maxima per period at integer
multiples of $\Kell'$, $V_{\nu}$ has one maximum per period
at odd integer multiples of $\Kell$ and $V_{\lambda}$ has
a single maximum at $\lambda=0$. 
The effective potentials and energies for this
region in Fig.~\ref{fig:bifurcationdiagram} are shown
in Fig.~\ref{fig:effpotentials}.
Above  $c_\lambda$, $V_{\lambda}$ has a minimum at $\lambda
=0$ and two symmetric maxima, and $V_{\mu}$ has only one
maximum per period at odd integer multiples of $\Kell'$. It is easy to
check that the effective
energy $E_{\lambda}$ is always less than the potential
energy at the minimum at $\lambda =0$. This minimum thus has no
consequences for the classical dynamics. At $c_\lambda$ in
Fig.~\ref{fig:bifurcationdiagram}, $V_{\lambda}$ changes
from one to two maxima, and at $P_\lambda$ we additionally have
$V_{\lambda}(0)=E_{\lambda}$.
$E_{\mu}$ reaches its minimum value relative to
$V_{\mu}$  here.
Below the line $c_\nu$ the maxima of $V_{\nu}$ at odd
integer multiples of $\Kell$ change into local minima and
$V_{\nu}$ has two maxima per period. The maxima of
$V_{\mu}$ at odd integer multiples of $\Kell'$ have
vanished here and $V_{\mu}$ has one maximum per
period. Again it is easy to check that the effective energy
$E_{\nu}$ is always less then local minima of 
$V_{\nu}$ at odd integer multiples of $\Kell$. The local
minima thus do not influence the classical dynamics. At $c_\nu$
in Fig.~\ref{fig:bifurcationdiagram}, $V_{\nu}$ changes
from one to two minima and at $P_\nu$ we additionally have
$V_{\nu}((2n+1)\Kell)=E_{\nu}$ for $n\in
\Z$. $E_{\mu}$ reaches its minimum value relative to
$V_{\mu}$  here. These cases are summarized in
Fig.~\ref{fig:effpotentialsspecialsall}. 
\def\figeffpotentials{%
Effective potentials $V_{\hat{s}}$  (solid) together with
the effective energies $E_{\hat{s}}$ (dotted) for the parts
of the regions \tO, \tOA, \tP\  and \tOB\  with $2b^2 \le
s_1^2+s_2^2 \le 2a^2$. At the top the 
reflections are indicated which determine the symmetry of the
wave functions according to the parity at the reflection point, see Section~\ref{sec:qm}.} 
\def\FIGeffpotentials{\centerline{\psfig{figure=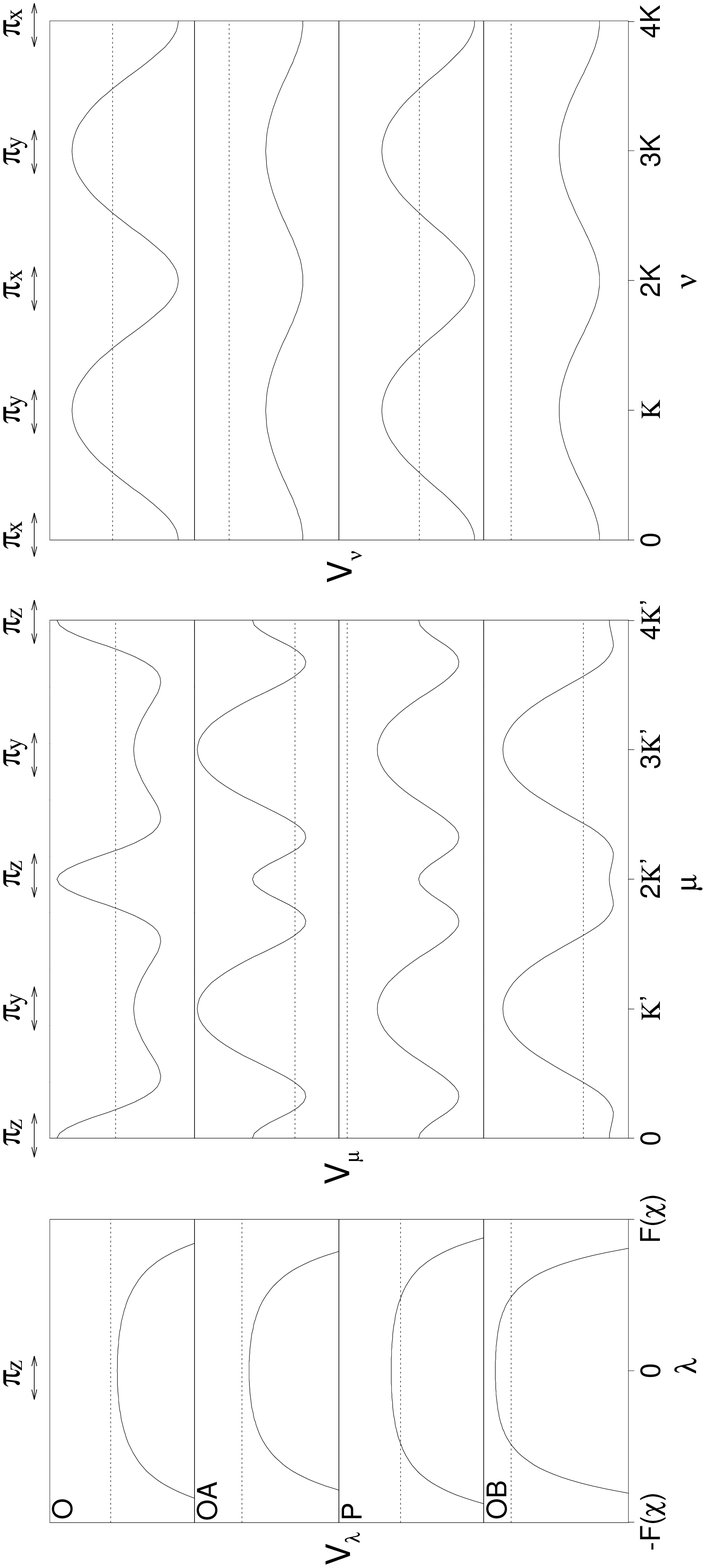,angle=-90,width=16cm}}}
\FIGo{fig:effpotentials}{\figeffpotentials}{\FIGeffpotentials}
\def\figeffpotentialsspecialsall{%
Analogue of Fig.~\ref{fig:effpotentials} for parameter combinations in
Fig.~\ref{fig:bifurcationdiagram} 
above the line $c_\lambda$ (first row),
at $P_\lambda$ (second row), below the line $c_\nu$ (third row) and at
$P_\nu$ (last row).} 
\def\FIGeffpotentialsspecialsall{\centerline{\psfig{figure=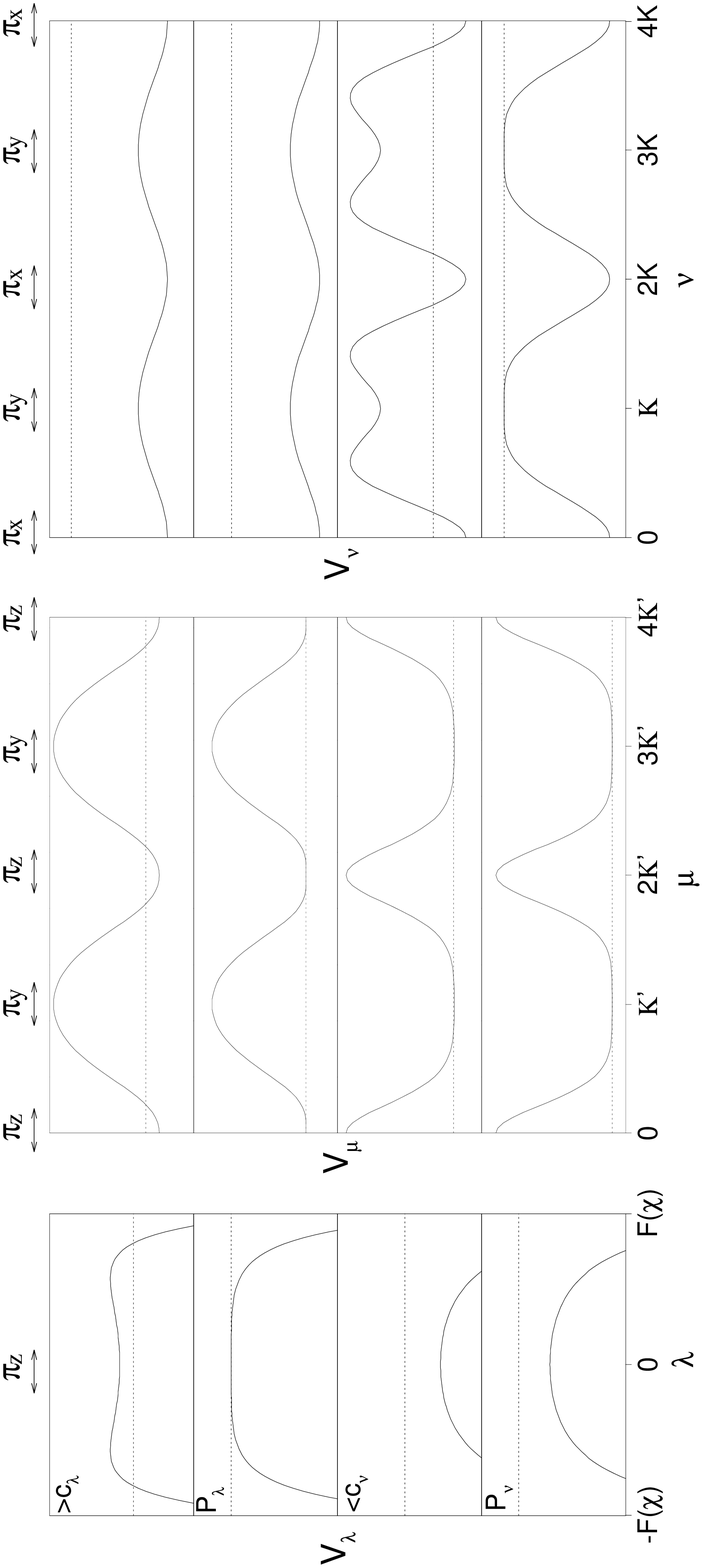,angle=-90,width=16cm}}}
\FIGo{fig:effpotentialsspecialsall}{\figeffpotentialsspecialsall}{\FIGeffpotentialsspecialsall}

%% file: actionintegrals.tex
\section{The Action Integrals}
\label{sec:actionintegrals}

For the calculation of actions it is useful to inspect the
caustics in Fig.~\ref{fig:caustics}. The action integrals are written
in the form
\bege
\label{eq:actions}
I_{\hat{s}}\equiv I_s =\frac{1}{2\pi} \oint p_s \, ds =\frac{m_s}{2\pi} \int_{s_-}^{s_+} p_s \,ds
\ende 
with $(\hat{s},s) \in \{(\lambda,\xi),(\mu,\eta),(\nu,\zeta) \}$. 
The integers $m_s$ and the integration boundaries~$s_-$ and~$s_+$ 
can be found in Tab.~\ref{tab:actions}, see \cite{WR96}, also the 
final comments in Section~\ref{sec:classic}.  
\begin{table}[!h]
\begin{center}
{ 
\tabstart  \small
\begin{tabular}{|c|c|c|c|c|c|c|c|c|c|}\hline
type & $m_\xi$ & $m_\eta$ & $m_\zeta$ & $\xi_-$ & $\xi_+$ & $\eta_-$ &
$\eta_+$ & $\zeta_-$ & $\zeta_+$ \\ \hline 
\tO  & $4$    & $4$   & $4$ & $a$ & $1$ & $b$ & $s_2$ & $0$ & $s_1$\\ 
\tOA  & $4$    & $2$ & $\pm 4$ & $a$ & $1$ & $s_1$ & $s_2$ & $0$ & $b$\\ 
\tP &$2$  & $\pm 4$   & $4$ & $s_2$ & $1$ & $b$ & $a$ & $0$ & $s_1$\\ 
\tOB & $2$ & $4$ & $\pm 4$ & $s_2$ & $1$ & $s_1$ & $a$ & $0$ & $b$\\ \hline
  \end{tabular}
\tabend
}
\caption[]{\label{tab:actions} \capsty Integration boundaries $s_-$ and
  $s_+$ and
  multipliers $m_s$ in \equ~(\ref{eq:actions}) for the four types of
  motion \tO , \tOA , \tP\ and \tOB.}
\end{center}
\end{table}
For the symmetry reduced ellipsoidal billiard any motion is of
oscillatory type always giving
$m_s=2$. To distinguish the symmetry reduced actions from the actions of
the full ellipsoid we write the former with tildes, i.e.
\bege
\tilde{I}_{\hat{s}}\equiv \Ids = \frac{1}{\pi} \int_{s_-}^{s_+} p_s\,ds.
\ende
\def\figsep{%
Energy surface and separatrix surfaces in the space of the actions $\BJf$.} 
\def\FIGsep{\centerline{\psfig{figure=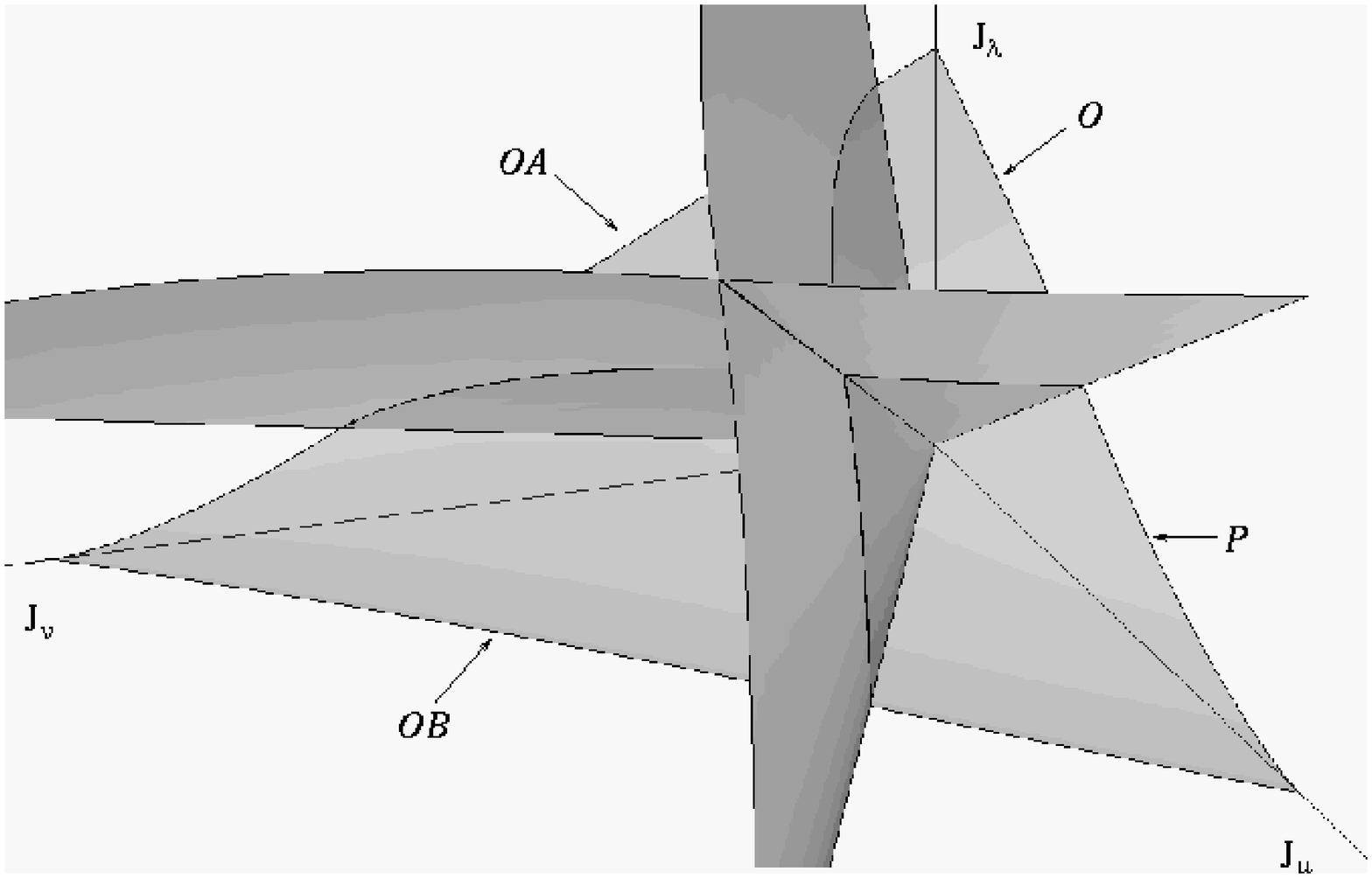,width=10cm,angle=-90}}}
\FIGo{fig:sep}{\figsep}{\FIGsep}
\noindent The presentation of the energy surface $H(\BIf )=E$ in the space of
the actions $\BIf$ is not smooth because an action variable can change
discontinuously upon traversing a separatrix.
In contrast to that the  symmetry reduced system $\tilde{H}(\BId)$ is
continuous.
For the quantum mechanical considerations it is advantagous to have a
continuous energy surface even for the full system. We therefore
introduce the actions 
\bege
\label{eq:Jdefinition}
\BJf = 2\BId,
\ende
which have the property that the phase space volume below the energy
surface $\tilde{H}(\BJf /2)=E$ in the space of the actions $\BJf$ is
equal 
to the phase space volume below 
the energy surface $H(\BIf )=E$ in the space of the actions $\BIf$ for
the same energy  $E$. In  
Fig.~\ref{fig:sep} $\tilde{H}(\BJf /2)=E$ is shown together with the
separatrix surfaces $s_1^2=b^2$ and $s_2^2=a^2$. Because the 
action variables scale with the energy the separatrix surfaces 
are foliated by rays through the origin. They divide the action space 
into the four regions corresponding to the different types of motion
\tO, \tP, \tOA\ and \tOB.
\def\figslitriemannsphere{%
Riemann sphere $\overline{\C}$ with the critical points
$z_1,...,z_6$. The rectangular boundary with the point $z_6$ should be
considered as the point $\infty$. The dashed line marks the boundary ellipsoid extented
to the complex plane. The points $-i$ and $i$ are marked for reasons
of orientation.}
\def\FIGslitriemannsphere{\centerline{\psfig{figure=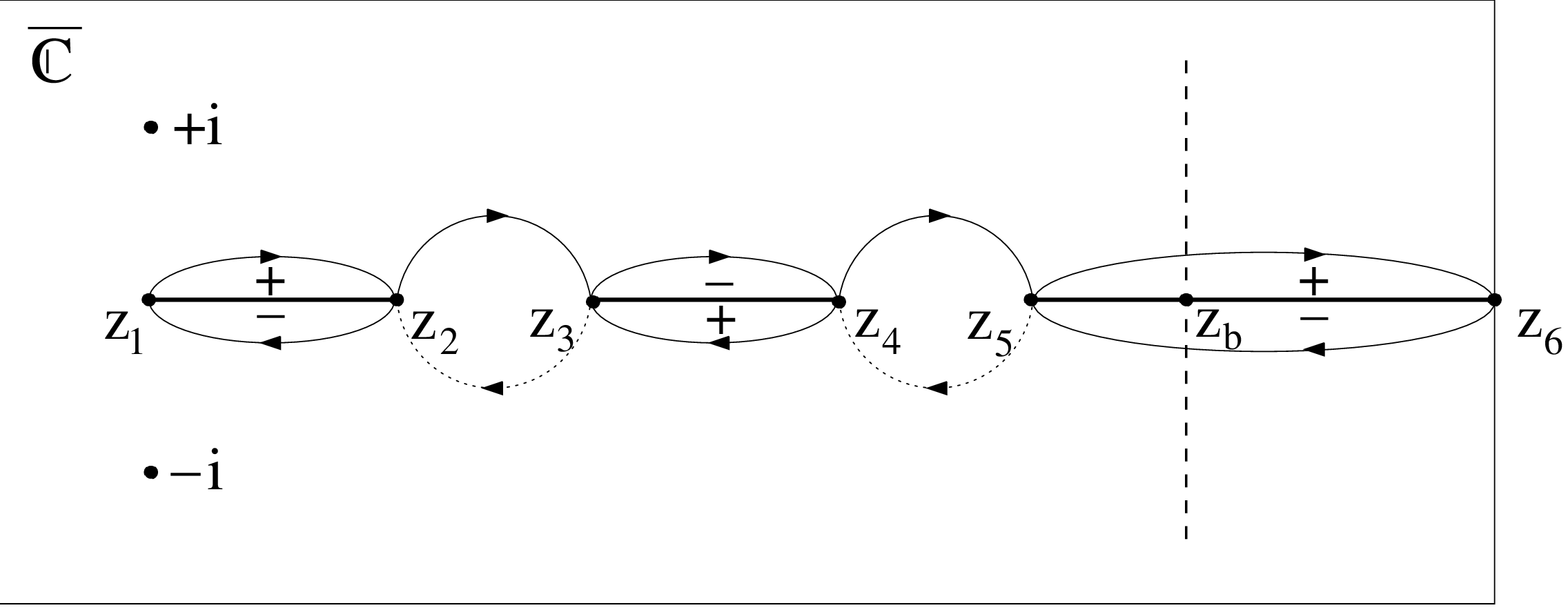,width=4.5cm,angle=-90}}}
\FIGo{fig:slitriemannsphere}{\figslitriemannsphere}{\FIGslitriemannsphere}

\def\fighyperellipticcurve{%
The glueing of 2 slit Riemann spheres to give the hyperelliptic curve
${\cal R}_w$. The horizontal plane marks the billiard boundary
extended to $\overline{C}^2$. The points $-i$ and $i$ are marked for reasons
of orientation.}
\def\FIGhyperellipticcurve{\centerline{\psfig{figure=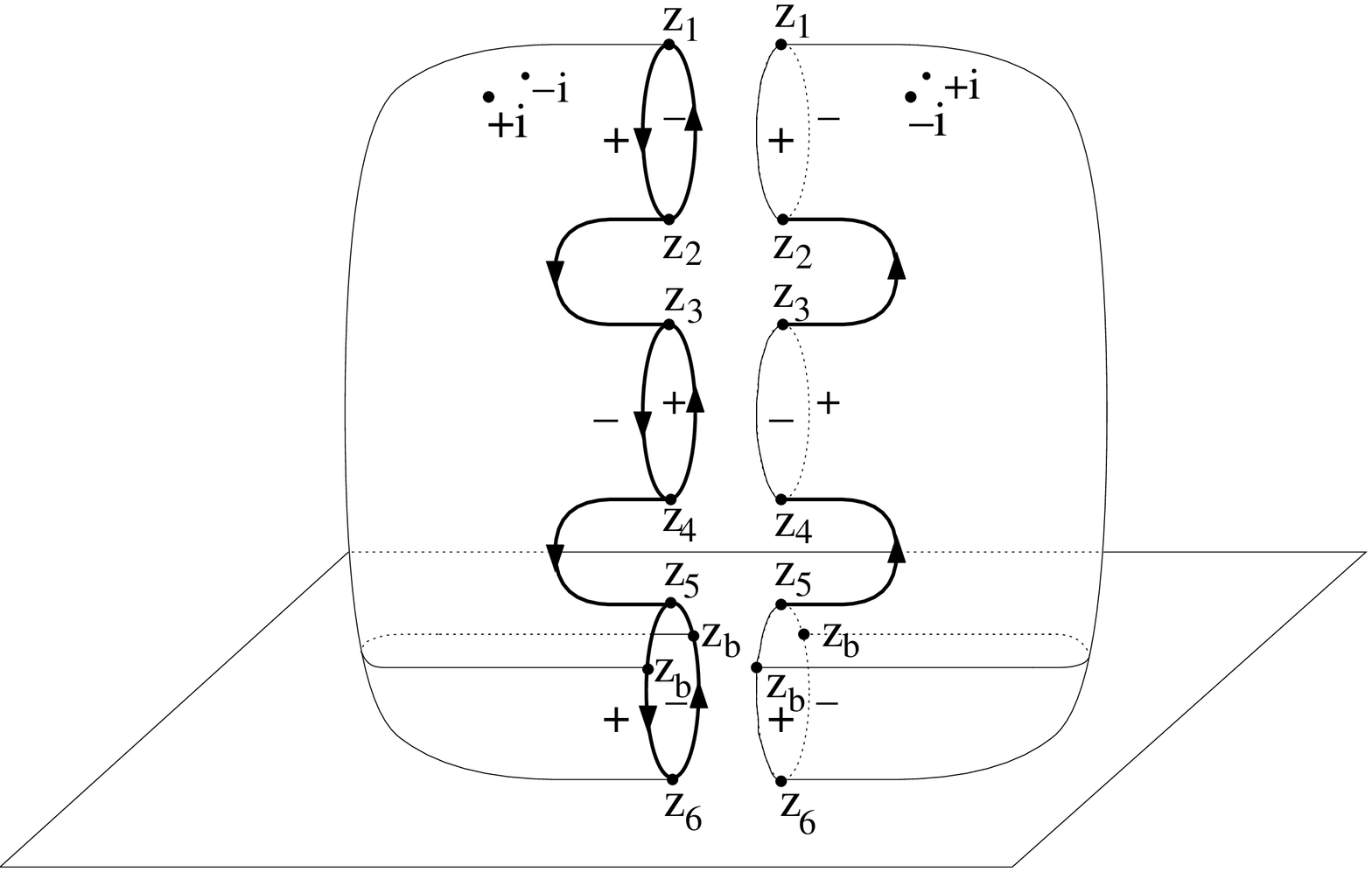,width=7.5cm,angle=-90}}}
\FIGo{fig:hyperellipticcurve}{\fighyperellipticcurve}{\FIGhyperellipticcurve}

\noindent Inserting the momenta from \equ~(\ref{eq:lmnmomenta}) and substituting $z=s^2$
in \equ~(\ref{eq:actions}) shows that the action integrals are of the form 
\bege
\label{eq:actionintegrals}
\int_{z_-}^{z_+} (z-s_2^2)(z-s_1^2)\,\frac{dz}{w}
\ende
with 
\bege
w^2= \prod_{i=1}^5(z-z_i)
\ende
where $z_-$, $z_+$ are sucessive members of 
$\{z_1,z_2,z_3,z_4,z_5,z_b\}=\{0,s_1^2,b^2,s_2^2,a^2,1\}$.
$z_b=1$ corresponds to the boundary of the billiard. 
The differential $dz/w$ has the six critical points $z_1,...,z_5$ and
$z_6=\infty$ which implies that the integrals in
\equ~(\ref{eq:actionintegrals}) are hyperelliptic. There do
not exist tabulated standard forms for these integrals but 
there is the well developed theory of so called Abelian integrals.
The main object of this theory is a Riemann surface, in our case
the hyperelliptic curve
\bege
{\cal R}_w = \{ (z,w)\in \overline{\C}^2 : w^2 = \prod_{i=1}^5 (z-z_i) \}.
\ende
Here $\overline{\C}$ denotes the compactified complex plane, i.e.\ the
Riemann sphere.
To construct a picture of ${\cal R}_w$ we proceed in the following 
manner. We order the critical points $z_i$, $i=1,...,5$, according to their
magnitudes. This gives $4$ orderings, one for each type of motion,
e.g. $z_1=0$, $z_2=s_1^2$, $z_3=b^2$, $z_4=s_2^2$, and $z_5=a^2$ 
for type \tO, see Tab.~\ref{tab:actions}. 
The points $z_i$ are marked on the Riemann
sphere, see Fig.~\ref{fig:slitriemannsphere}. We then slit the
Riemann sphere along the real axis between the points $z_i$
and $z_{i+1}$ for $i=1,3,5$. Excluding the three slits from $\overline{\C}$
the sign of $w$ is everywhere well defined on this manifold when
it is fixed at one arbitrary point. In
Fig.~\ref{fig:slitriemannsphere} we choose the sign of $w$ to be
positive right above the slit $[z_1,z_2]$. Then the sign is negative
right above the slit $[z_3,z_4]$ and again positive right above the
slit $[z_5,z_6]$. Right below the slits the sign of $w$ is opposite
to the sign right above. 
Around the slits we have the  closed paths $z_1 \rightleftharpoons z_2$,
$z_3 \rightleftharpoons z_4$ and $z_5 \rightleftharpoons z_6$.
\rem{
Along the paths $z_1 \rightleftharpoons z_2$ and
$z_3 \rightleftharpoons z_4$ marked in
Fig.~\ref{fig:slitriemannsphere} the actions $I_\mu$ and
$I_\nu$ can be integrated. The path of integration for the 
action integral $I_\lambda$ is not connecting two critical points, i.e.\ it
is not a cloed path on the Riemann surface. It is taken along the slit
$[z_5,z_6]$, but only between $z_5$ and $z_b$.
The action integral $I_\lambda$ is said to be
incomplete, $I_\mu$ and $I_\nu$ are complete.
}
On another copy of
$\overline{\C}$ we introduce the same slits but choose the sign of $w$
opposite to the choice on the former copy. 
The path from $z_2$ to $z_3$ on the former copy in
Fig.~\ref{fig:slitriemannsphere} is assumed to be
the first half of a closed path $z_2 \rightleftharpoons z_3$ of which
the second half from $z_3$ back to $z_2$ lies on the latter copy. The
same is assumed to hold for the closed path $z_4 \rightleftharpoons
z_5$. 
To unify the view glue the two
copies at the corresponding slits such that the corresponding
critical points coincide and such that $w$ changes smoothly across the
seams, see Fig.~\ref{fig:hyperellipticcurve}. The result is a compact
Riemann surface, i.e. a
manifold which carries a complex structure and to
which the full machinery of Cauchy integration theory is applicable. 
The surface has genus $g=2$ and there are 4 non-contractable paths
on the manifold which cannot be transformed smoothly into each
other. They form a basis of the four-dimensional homology group
corresponding to this surface. The homology basis may be specified by
the choice of the closed paths 
 $z_1 \rightleftharpoons z_2$,
$z_2 \rightleftharpoons z_3$, $z_3 \rightleftharpoons z_4$ and $z_4
\rightleftharpoons z_5$. 
The path $z_5 \rightleftharpoons z_6$ is homologous to the sum of
$z_1 \rightleftharpoons z_2$ and
$z_3 \rightleftharpoons z_4$ then.
From the non-trivial topology of the Riemann surface ${\cal R}_w$
it follows that there may exist non-vanishing closed integrals 
(even for vanishing residues). The action integral
\equ~(\ref{eq:actionintegrals}) is of this type. It is an integral
with singularities but vanishing residues - a so called Abelian
integral of second kind. 
The actions integrals $I_\nu$ and $I_\mu$ 
of \equ~(\ref{eq:actionintegrals}) are taken along the closed paths $z_1
\rightleftharpoons z_2$ and $z_3 \rightleftharpoons z_4$. Due 
 to the reflection at the boundary ellipsoid the action integral
 $I_\lambda$ is  not taken along a closed path. It is taken along the
 slit $[z_5,z_6]$, but only between $z_5$ and $z_b$. It is therefore
 called incomplete. 
The integrals $I_\mu$
 and $I_\nu$ are called complete. 
These three integrals give real numbers. In contrast to this the
integration of \equ~(\ref{eq:actionintegrals}) along the closed paths
$z_2 \rightleftharpoons z_3$ and $z_4 \rightleftharpoons z_5$ yields
purely imaginary numbers.  
These integrals have an  important  physical
meaning for the semiclassical quantization scheme in
Section~\ref{sec:semiqm}.  They give the penetration integrals which
will be needed for 
the discussion of quantum mechanical tunneling. At this stage we already
mention that there are only two such penetration integrals and we
define them as follows:
\bega
\label{eq:thetanu}
\Theta_\nu   &\equiv& \Theta_\zeta = -2i \int_{s_1}^b p_\zeta \,d\zeta
, \\
\label{eq:thetalambda}
\Theta_\lambda &\equiv& \Theta_\xi = -2i \int_a^{s_2} p_\xi \,d\xi.
\enda
The factor $i$ in the definition turns both integrals into real
numbers. 
It is useful to take these definitions independent from the type of
motion \tO, \tOA, \tP\  and \tOB, i.e. for any ordering of 
$b$, $s_1$, $a$ and $s_2$. This will become clear in
Section~\ref{sec:semiqm}.\\
\def\figriemannsurffamily{%
The parameter ranges for $s^2 \in \{\zeta^2,\eta^2,\xi^2 \}$ define a
family of hyperelliptic curves parametrized by the angle $\varphi$ 
in Fig~\ref{fig:bifurcationdiagram}.
}
\def\FIGriemannsurffamily{\centerline{\psfig{figure=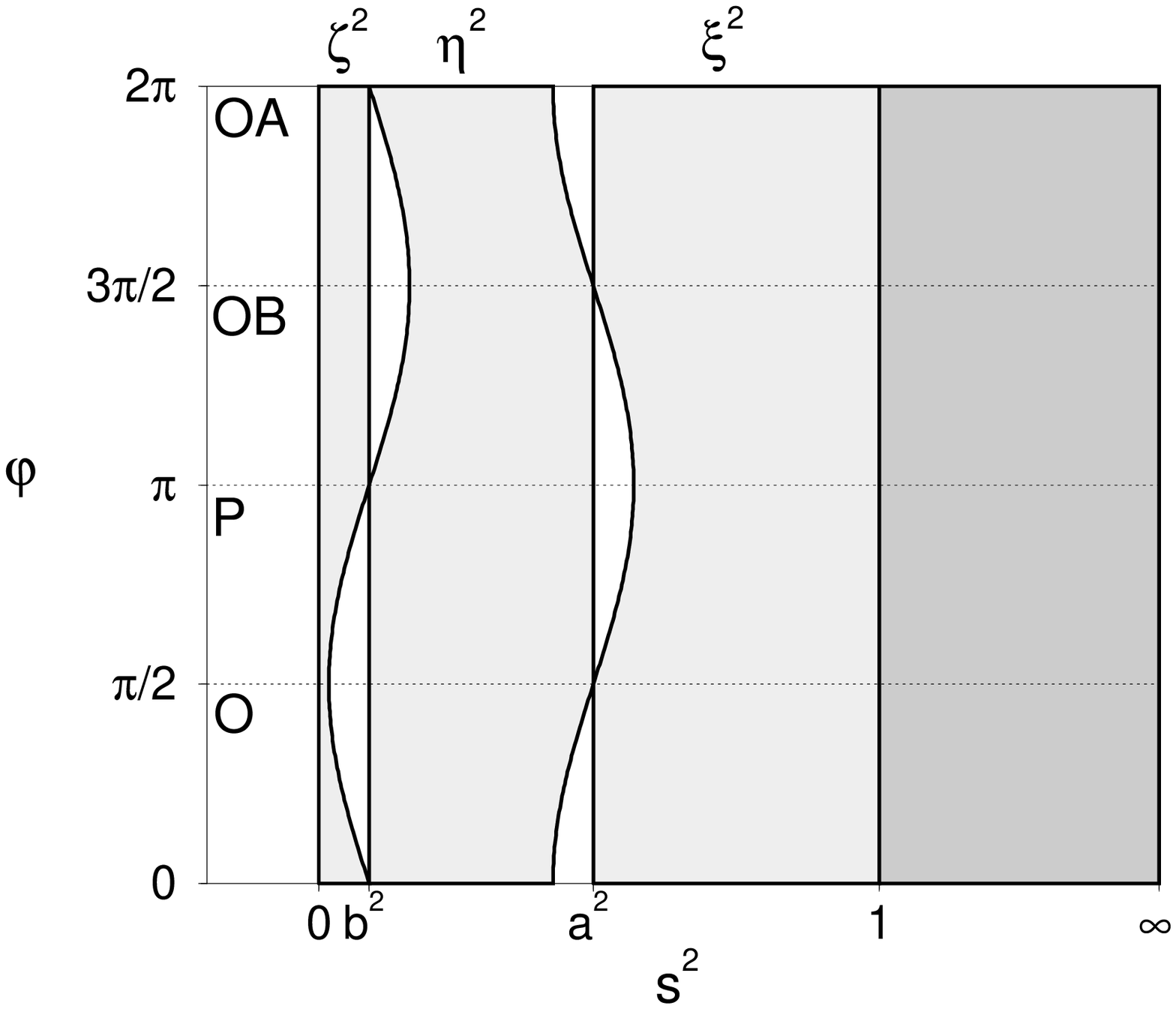,width=10cm}}}
\FIGo{fig:riemannsurffamily}{\figriemannsurffamily}{\FIGriemannsurffamily}

In Fig.~\ref{fig:riemannsurffamily} we show the ranges for the
coordinates $s^2 \in \{\zeta^2,\eta^2,\xi^2\}$ on the circle in the
parameter plane in
Fig.~\ref{fig:bifurcationdiagram}. Generically the ranges for $\zeta^2$
and $\eta^2$ and the ranges for $\eta^2$ and $\xi^2$ are separated
by finite gaps. This gives the Riemann surfaces of genus 2 as described above. 
Now consider the circle in Fig.~\ref{fig:bifurcationdiagram}.
On the bifurcation lines which are reached for 
angle $\varphi$ equal to integer multiples of $\pi/2$ one of the two gaps
vanishes. One of the penetration integrals in \equ~(\ref{eq:thetanu})
then vanishes too.
This means that two of the three slits in 
Fig.~\ref{fig:slitriemannsphere} merge and the genus
of the Riemann surface is diminished by one. On the bifurcation lines
we thus find elliptic curves (genus 1) with an additional pole in
the differential for the actions. They can also be considered as
singular hyperelliptic curves.
The three action
integrals and the one remaining penetration integral are of elliptic
type then. In \cite{WR96} analytic expression in terms of Legendre's
standard integrals are calculated for these cases. 

%% file: qm.tex

\newcommand{\tsi}{\tilde\psi}

\section{The Quantum System}
\label{sec:qm}

The quantum mechanical billiard problem is the problem of
determining the spectrum of the Laplacian in the billiard domain with 
Dirichlet boundary conditions on the boundary.
Equivalently, this is given by Schr\"odinger's equation for a free particle
in the ellipsoid which in turn is Helmholtz's equation in three
dimensions,
\bege
-\frac{\hbar^2}{2}{\bm \nabla}^2 \psi = E \psi .
\ende
As in the classical case the potential vanishes
inside the ellipsoid and is infinite outside the ellipsoid. 
This potential  classically leads to elastic reflections and
quantum mechanically imposes Dirichlet boundary conditions
on the ellipsoid. The three discrete symmetries of the ellipsoid are
the reflections at  
the three Cartesian coordinate planes. The wave function can have
even or odd 
parity with respect to each discrete symmetry, $\psi(x,y,z) = 
\pi_x \psi(-x,y,z)$ etc. Combining the two parities for each
dimension we obtain a total of eight parity combinations denoted by
${\bm \pi}=(\pi_x,\pi_y,\pi_z)$ where each parity is from $\{+,-\}$.

Corresponding to the two sets of classical coordinates we get two sets of
quantum mechanical equations. In both cases the separation is the
same as in the classical case and the wave function $\psi$ is a product
of three separated wave functions. The ellipsoidal
coordinates $(\xi,\eta,\zeta)$
lead to the analogue of \equ~(\ref{eq:xietazetamomenta}), which is
\bege
\label{eq:sephelmholtz}
-\hbar^2\left(\sqrt{(s^2-a^2)(s^2-b^2)}\frac{d}{ds}\right)^2 \psi_s(s) = 
2E \left(s^4 - 2 k s^2 + l\right) \, \psi_s(s)
\ende
with  $s\in\{\xi,\eta,\zeta\}$. If we set $E=0$ but keep $K$ and $L$ finite
we obtain one of the many forms of the Lam{\'e} equation 
\cite{Morse53,WhitWats65}.
Since we are not only interested in 
the solution of the Laplace equation in the ellipsoid but in the
spectrum of the Laplacian we have to consider this generalized Lam{\'e}
equation, known as the ellipsoidal wave equation.

Transforming the equation into the regularized coordinates 
leads to the analogue of \equ~(\ref{eq:lmnmomenta}),
\bege
\label{eq:helmreg}
-\hbar^2 
 a^2\sigma_{\hat{s}} 
\frac{d^2}{d\hat{s}^2}\psi_{\hat{s}}(\hat{s})
=
2E\left(s(\hat{s})^4 - 2k s(\hat{s})^2 + l\right) \, \psi_{\hat{s}}(\hat{s})
\ende
where $\hat{s}\in \{\lambda,\mu,\nu\}$ and
$s(\hat{s}) \in \{\xi(\lambda),\eta(\mu),\zeta(\nu)\}$ from 
Equations~(\ref{eq:xietazetaoflambdamunufirst})-(\ref{eq:xietazetaoflambdamunulast}).
Comparing to the equations for the billiard in the ellipse \cite{WWD97}
\equ~(\ref{eq:helmreg})
is analogous to the Mathieu equation(s) in its 
standard form, while \equ~(\ref{eq:sephelmholtz}) is analogous to its
algebraic form. 
Note that similar to the classical case it would be sufficient 
to only consider the equation for $\nu$ in the complex domain instead of 
the three equations for real arguments.

The Dirichlet boundary conditions require that the wave function
$\psi(\lambda,\mu,\nu) = \psi_\lambda(\lambda) \psi_\mu(\mu) \psi_\nu(\nu)$
is zero on the ellipsoid, which gives $\psi_\lambda(\pm \Fell(\chi)) = 0$.
The solutions in the two angular variables $\mu$ and $\nu$ must be  periodic
with periods $4\Kell'$ and $4\Kell$, respectively, in order to give a smooth
function on the solid 2-torus described in Section~\ref{sec:classic}.

For $\mu$ and $\nu$, \equ~(\ref{eq:helmreg}) is a linear differential
equation with periodic coefficients. Floquet theory guarantees the
existence of solutions $\psi_\mu$ with period an integer multiple of
$2\Kell'$ and solutions $\psi_\nu$ with period an integer multiple of
$2\Kell$, respectively. The involutions $S_1$, $S_2$ and $S_3$ in
Equations (\ref{eq:invol1})-(\ref{eq:invol3}) relate the symmetries of
the separated wave functions to the parities $\pi_x$, $\pi_y$ and
$\pi_z$. Starting with $S_1$ the separation of the invariance
condition $\psi(\lambda,\mu,\nu)=\psi(S_1(\lambda,\mu,\nu))$ gives
\bege
\label{eq:inv1cond}
\frac{\psi_\lambda(\lambda)}{\psi_\lambda(-\lambda)} =
\frac{\psi_\mu(-\mu)}{\psi_\mu(\mu)}. 
\ende
Since the left hand side and the right hand side are functions of
$\lambda$ and 
$\mu$ alone they have to be equal to some common constant. Because we
may change the sign of $\lambda$ and $\mu$ independently giving the
reciprocals of both sides of \equ~(\ref{eq:inv1cond}) the
separation constant must have unit modulus. From
\equ~(\ref{eq:translmntoxyz3}) we see that the sign is the parity
$\pi_z$. Similarly, from the invariance of $\psi$ under $S_2$ we get
\bege
\label{eq:inv2cond}
\frac{\psi_\mu(-\mu-2\Kell')}{\psi_\mu(\mu)} =
\frac{\psi_\nu(\nu)}{\psi_\nu(2\Kell-\nu)}\,. 
\ende
From replacing $\mu$ by $-\mu-2\Kell'$ and/or $\nu$ by $2\Kell - \nu$
we see that both sides of \equ~(\ref{eq:inv2cond}) again have to
be equal to a separation constant of unit modulus. With the aid of 
\equ~(\ref{eq:translmntoxyz2}) we may identify the sign with the
parity $\pi_y$. $\pi_y$ thus gives the parity of the wave function
$\psi_\mu$ for reflections about $\Kell'$ and of $\psi_\nu$ for
reflections about $\Kell$. From the invariance of $\psi$ with respect
to $S_3$ we get 
\bege
\label{eq:inv3cond}
\frac{\psi_\lambda(-\lambda)}{\psi_\lambda(\lambda)}
\frac{\psi_\mu(\mu-2\Kell')}{\psi_\mu(\mu)} =
\frac{\psi_\nu(\nu)}{\psi_\nu(2\Kell-\nu)}  
\ende
or with the results from above 
\bege
\label{eq:inv4cond}
\psi_\mu(\mu-2\Kell') = \pi_y \pi_z \psi_\mu(\mu)\,.
\ende
This relates the product of the parities $\pi_y$ and $\pi_z$ to the
period of $\psi_\mu$. $\psi_\mu$ is $2\Kell'$-periodic 
for $\pi_y\pi_z=+$ and  $4\Kell'$-periodic
(i.e.\ not $2\Kell'$-periodic)  for $\pi_y\pi_z=-$. 
Similar arguments hold for the wave function $\psi_\nu$. Here $\pi_x$
gives the symmetry of $\psi_\nu$ for reflections about 0. The
product of the parities $\pi_x$ and $\pi_y$ determines its
period: $\psi_\nu$ is $2\Kell$-periodic for
$\pi_x\pi_y=+$ and $4\Kell$-periodic (i.e.\ not $2\Kell$-periodic)
for $\pi_x\pi_y=-$. 
The parities for the separated wave functions are shown at the
top of Fig.~\ref{fig:effpotentials}.

Even though the ellipsoidal coordinates $(\xi,\eta,\zeta)$ are not regular, 
the parities are most simply expressed by properties of the
wave functions in these singular coordinates.
Let us first rewrite 
\equ~(\ref{eq:sephelmholtz}) in the form
\bege
\label{eq:helmholtzpolyn}
        f \psi'' + g \psi' + h \psi / \hbar^2 = 0
\ende
with the polynomials
\bega
      f(s) &=& (s^2-a^2)(s^2-b^2),\\
      g(s) &=& f'(s)/2= s(2s^2-a^2-b^2), \\
      h(s) &=& 2E(s^4 - 2ks^2 + l) \,.
\enda
The singularities of ~\equ~(\ref{eq:helmholtzpolyn}) and 
equivalently of \equ~(\ref{eq:sephelmholtz}) are given by the
 zeroes $\pm a$ and $\pm b$ of $f$.
We postpone the question of additional singularities at infinity to
Section \ref{sec:deg} because they are not important for our
numerical calculations.
In order to look at the asymptotics of the solutions of
\equ~(\ref{eq:helmholtzpolyn}) at the singular points 
we calculate the corresponding indicial equations. Denoting the
position of the singularity under consideration by $c$, the exponents
$\alpha$ of the solutions are the solutions of the indicial equation
(see e.g.~\cite{Morse53})
\bege
\alpha^2 + (p_c-1)\alpha + q_c = 0, 
\ende
where
\bege
p_c = \lim_{s\rightarrow c} \frac{g(s)}{f(s)}(s-c), \quad
   q_c = \lim_{s\rightarrow c} \frac{h(s)}{f(s)}(s-c)^2.
\ende
To calculate $p_c$ it is best to perform the partial fraction decomposition 
of $g/f$,
\bege \label{eq:gfparfrac}
        \frac{g(s)}{f(s)} = \frac12 \left( 
        \frac1{s-b} + \frac1{s+b} + \frac1{s-a} + \frac1{s+a}
                \right).
\ende
From this it is obvious that $p_c = 1/2$ for all singular points.
Since $h/f$ only has simple poles $q_c = 0$, the exponents
are $0$ and $1/2$. 
Because the singularities of the wave equations do not produce
essential singularities in its solutions these singularities are
called regular.
In our case the two exponents refer to the two parities
possible at a regular singular point. We require $\psi_\eta(b) =\psi_\zeta(b) =
0$ for $\pi_y = -$ and  $\psi_\eta(b) = \psi_\zeta(b) = 1$  (up to
normalization) for $\pi_y = +$.  Similarly the value at the regular
singular 
point $a$  of the wave functions $\psi_\xi$ and $\psi_\eta$ determines
the parity $\pi_z$. For $\pi_x$ it is a little simpler, 
because it is determined by the value of $\psi_\zeta$ at the 
ordinary point $\zeta = 0$.
The boundary condition at $\xi = 1$ always is $\psi_\xi(1) = 0$.
The need for the solution to be invariant under
an additional symmetry group (arising e.g.\ if we work on a covering space) 
does not appear, because we only solve the wave function
in one octant.  
The boundary conditions are
summarized in Tab.~\ref{tab:parities}. Note that in line with the
above considerations the table 
shows a very simple structure: the sign $-$ or $+$ in the first
three parity columns successively determine the value $0$ or $1$ of the
wave functions at $0$, $b$ and $a$.
\begin{table}[!h]
\begin{center}
{ 
\tabstart  \small
\begin{tabular}{|c|c|c|c|c|c|c|c|c|c|c|c|}\hline
   $\pi_x$& $\pi_y$ & $\pi_z$ & $\psizeta (0)$ & $\psizeta (b)$ &
  $\psieta (b)$ & $\psieta (a)$ &$\psixi (a)$ &
  $\psixi (1)$ & $\psi_\nu$ period & $\psi_\mu$ period &
  nodal planes \\ \hline
  $-$ & $-$ & $-$ & 0 & 0 & 0 & 0 & 0 & 0 &$2\Kell$ &$2\Kell'$  & $(x,y)$, $(x,z)$,
  $(y,z)$ \\ 
  $-$ & $-$ & $+$ & 0 & 0 & 0 & 1 & 1 & 0 &$2\Kell$ &$4\Kell'$ & $(x,z)$, $(y,z)$\\  
  $-$ & $+$ & $-$ & 0 & 1 & 1 & 0 & 0 & 0 &$4\Kell$ &$4\Kell'$ & $(x,y)$, $(y,z)$\\
  $-$ & $+$ & $+$ & 0 & 1 & 1 & 1 & 1 & 0 &$4\Kell$ &$2\Kell'$ & $(y,z)$\\
  $+$ & $-$ & $-$ & 1 & 0 & 0 & 0 & 0 & 0 &$4\Kell$ &$2\Kell'$ & $(x,y)$, $(x,z)$\\ 
  $+$ & $-$ & $+$ & 1 & 0 & 0 & 1 & 1 & 0 &$4\Kell$ &$4\Kell'$ & $(x,z)$\\
  $+$ & $+$ & $-$ & 1 & 1 & 1 & 0 & 0 & 0 &$2\Kell$ &$4\Kell'$ & $(x,y)$ \\
  $+$ & $+$ & $+$ & 1 & 1 & 1 & 1 & 1 & 0 &$2\Kell$ &$2\Kell'$ & $-$ \\   \hline
\end{tabular}
\tabend
}
\caption[]{\label{tab:parities} \capsty Parities, boundary
  conditions of the separated  wave functions $\psi_s$, periods of the
  separated wave functions $\psi_\nu$ and $\psi_\mu$, and Cartesian
  nodal planes.}
\end{center}
\end{table}

In our numerical procedure we are going to start integrating at the
regular singular points. Since this is impossible for initial conditions
belonging to the solution with exponent $\alpha =1/2$ we have to
factor out this behaviour analytically. To find solutions with the
parities $\pi_z=-$ and/or $\pi_y=-$ 
we employ the transformations
\bega
\pi_y = +, \pi_z = - & : & \psi = \sqrt{s^2-a^2} \tsi_{+-}, \\
\pi_y = -, \pi_z = + & : & \psi = \sqrt{s^2-b^2} \tsi_{-+}, \\
\pi_y = -, \pi_z = - & : & \psi = \sqrt{s^2-a^2} \sqrt{s^2-b^2} \tsi_{--},
\enda
and leave  $\psi = \tsi_{++}$ unchanged  for $\pi_y = \pi_z = +$. The
polynomials $h$ 
and $g$ in \equ~(\ref{eq:helmholtzpolyn}) change according to
\bega
\begin{array}{ll}
\tilde{h}_{+-}(s) =  h(s)+2s^2-b^2,     & \tilde{g}_{+-}(s) = g(s)+2s(s^2-b^2),\\
\tilde{h}_{-+}(s) =  h(s)+2s^2-a^2,     & \tilde{g}_{-+}(s) = g(s)+2s(s^2-a^2),\\
\tilde{h}_{--}(s) =  h(s)+6s^2-b^2-a^2, & \tilde{g}_{--}(s) = g(s)+2s(2s^2-b^2-a^2).
\end{array}
\enda
The functions $\tilde{h}_{++} = h $ and $\tilde{g}_{++} = g$ remain unchanged. 
The resulting transformed equations change the prefactor $1/2$
in \equ~(\ref{eq:gfparfrac}) to $3/2$ for the terms involving $\pm a$, $\pm
b$, or  both, respectively.
Hence $p_c = 3/2$ at the correponding regular singular point and
$\alpha = 0, -1/2$.
We are now able to start integrating at the singular points $c=a$ or
$c=b$, or to be more precise, a distance $\Delta s$ away from them, always with
the special velocity that corresponds to the regular 
solution with exponent $\alpha = 0$. The initial conditions  are
\bege
        \tsi'(c \pm \Delta s) = -\frac{\tilde{h}(c)}{\tilde{g}(c)}, \quad
        \tsi(c \pm \Delta s) = 1 \pm \Delta s \, \tsi'(c\pm\Delta s).
\ende
In order to find $\tsi$, three conditions on the three separated
wave functions $\tsi_s$ have to be fulfilled simultaneously. This 
is possible because there are three parameters $E$, $k$ and $l$  in
the three equations.
However, each equation depends on all the three separation constants; 
the equations are separated but the constants are not.
We use a numerical procedure similar to that described in
\cite{WWD97}, the essential 
difference being that for the ellipsoid the wave function $\tsi_\eta$ 
has a regular singular point on both ends of the interval. Since it is not
possible to integrate a regular solution into a singular point, but
only away from it, we divide the interval into two equal parts
$[b,(a+b)/2]$ and $[(a+b)/2,a]$ and require the solution to match
smoothly at $s = (a+b)/2$. This is called shooting to a fitting
point \cite{Press88}. These two and the two remaining intervals $[0,b]$
and $[a,1]$ are all transformed to $[0,1]$, and the resulting
system of four equations is simultaneously solved. With  Newton's method
\cite{Press88} the three free parameters are adjusted  to satisfy 
the three remaining conditions 
$\tilde{\psi}_\xi(1) = 0$, the smoothness condition at the 
fitting point $(a+b)/2$ for $\tilde{\psi}_\eta$ and $\tilde{\psi}_\zeta'(0)=0$ for $\pi_x=+$ or
$\tilde{\psi}_\zeta(0)=0$ for $\pi_x=-$, respectively.
Taking the semiclassical values for $E$, $k$ and $l$ from
Section~\ref{sec:semiqm} as an 
initial guess, the method always converges to the exact 
eigenvalues. 
\def\figtsi{%
The eight transformed eigenfunctions $\tsi$ with all quantum numbers
1 and all possible parities, i.e. $|1,1,1;\pm \pm \pm\rangle$.
The parity $\pi_x$ is $-$ for wave functions starting at 0
and $+$ otherwise.
The remaining two parities $\pi_y, \pi_z$ are distinguished by
the dashing.}
\def\FIGtsi{\centerline{\psfig{figure=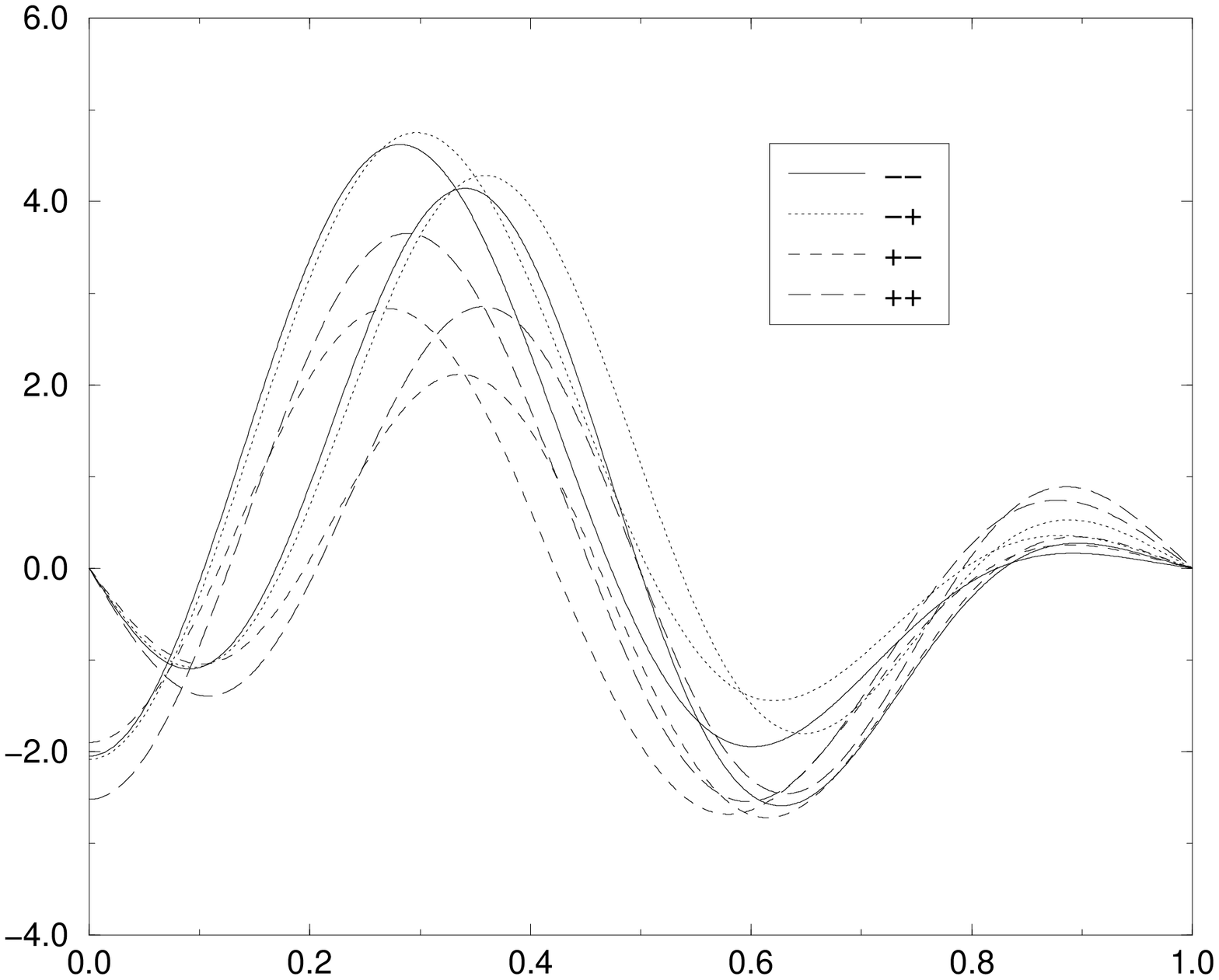,width=7cm,angle=-90}}}
\FIGo{fig:tsi}{\figtsi}{\FIGtsi}

Because we are free in the normalization of the three separated wave
functions $\tsi_s$ they can be multiplied by constant factors to give
one smooth function on the interval $[0,1]$, see Fig.~\ref{fig:tsi}.
$\tsi$ has
$n$ zeroes $\in (0,b)$, $m$ zeroes $\in (b,a)$ and $r$
zeroes $\in (a,1)$. 
The quantum numbers $(r,m,n)$ together 
with the parities ${\bm \pi}=(\pi_x,\pi_y,\pi_z)$ completely determine
the state which we denote by $|r,m,n;\pi_x\pi_y\pi_z\rangle$. 
The quantum numbers $(r,m,n)$ belong to the reduced system 
in one octant; the parities determine the wave function on the boundaries
of the octant.
It is not so simple to count the corresponding number of nodal 
surfaces in the full system. It is complicated by the fact that
the $(x,y)$-plane  and $(x,z)$-plane are composed of two different types
of quadrics, see Fig.~\ref{fig:singularplane}. 
Away from the three Cartesian coordinate planes  
the number of ellipsoidal nodal surfaces
is counted by $r$,
the number of one sheeted hyperboloidal
nodal surfaces (``rotating'' about the shortest semiaxis $z$) is given 
by $m$ and
the number of two sheeted hyperboloidal nodal surfaces (``rotating''
about the longest semiaxis $x$) is given by $n$, but because each surface
has two sheets this makes $2n$ nodal surfaces. Depending on the parity
combinations ${\bm \pi}$ the Cartesian coordinate planes give additional
nodal planes according to the last
column of Tab.~\ref{tab:parities}.

%% file: semiqm.tex
\section{Semiclassical Quantization}
\label{sec:semiqm}

The semiclassical quantization of the ellipsoidal billiard is obtained
from single valuedness conditions that are imposed on \WKB\ wave
functions on the 
fourfold cover discussed in Section~\ref{sec:classic}. Let us
consider Schr{\"o}dinger's equation for a general one dimensional
Hamiltonian of the form 
\bege
\label{eq:generalhamilt}
\hat{H}=-\frac{\hbar^2}{2}\frac{\partial^2}{\partial q^2}+V(q)\,.
\ende 
For a fixed energy $E$ in each region $j$ between two
successive classical turning points a \WKB\ wave function of the form
\bege \label{eq:wkbwavef}
\psi^{(j)}(q) =
\left(A_+^{(j)}\exp(iS_j(q)/\hbar)+A_-^{(j)}\exp(-iS_j(q)/\hbar)\right)/\sqrt{p(q)}
\ende
is reasonable. Its phase is given by 
\bege
S_j(q)=\int_{q_j}^q p(q')\,dq'
\ende
with the classical momentum $p(q)=\sqrt{2(E-V(q))}$. $A_+^{(j)}$ and
$A_-^{(j)}$ are 
constants. The reference point $q_j$ for the phase integral is an 
    arbitrary point in the region under consideration but it is
    convenient to take it as the left or right classical turning point
    although the \WKB\ wave function is a good approximation only away
    from the classical turning points. 

In the following we will consider \WKB\ wave functions only in
classically allowed regions although they are valid
even in regions where $E<V(q)$ giving real
exponentials in \equ~(\ref{eq:wkbwavef}). The amplitude vectors ${\bm
  A}^{(1)}=(A_+^{(1)},A_-^{(1)})^t$ and ${\bm
  A}^{(2)}=(A_+^{(2)},A_-^{(2)})^t$ of \WKB\ wave functions in two
classically allowed regions 1 and 2 separated by a classically
forbidden region are related by the matrix equation ${\bm
  A}^{(2)} = M(\Theta) {\bm
  A}^{(1)}$ with the tunnel matrix (see
\cite{BM72} and 
the references therein)
\bege
\label{eq:tunnelconnection}
M(\Theta) = e^{\Theta/\hbar}\left(\begin{array}{cc}
  \sqrt{1+e^{-2\Theta/\hbar}} & -i \\
  i & \sqrt{1+e^{-2\Theta/\hbar}}\\
  \end{array}\right) \,,
\ende
where 
\bege
\label{eq:penint}
\Theta=-i\int_{q_1}^{q_2}p(q)\,dq
\ende
is the penetration integral of the potential barrier.
Here $q_1$ and $q_2$ are the turning points to the left and 
right of the barrier, i.e. $V(q)<E$ for $q<q_1$ and $q>q_2$
and $V(q)>E$ for $q_1<q<q_2$. The matrix (\ref{eq:tunnelconnection})
remains valid if we increase the energy $E$ above the barrier's
maximum. Then the
classical turning points become complex ($q_1$ complex conjugate to
$q_2$) giving a negative penetration integral in
\equ~(\ref{eq:penint}). For $-\Theta \gg \hbar$ the matrix $M(\Theta)$
becomes the  identity matrix.

The amplitude vectors ${\bm A}^{(1)}$ and ${\bm A}^{(2)}$ of
two \WKB\ wave 
functions defined in the same classically allowed region but with
different reference points for the phase integral are related by the
phase shift ${\bm A}^{(2)}=P(\phi){\bm A}^{(1)}$ with the matrix
\bege
\label{eq:phaseshift}
P(\phi)=
\left(
\begin{array}{cc}
\exp(i \phi/\hbar) & 0 \\
0 & \exp(-i \phi /\hbar)
\end{array}
\right)\,,
\ende
where $\phi = \int_{q_1}^{q_2} p(q)\,dq$.
\def\FIGeffpotentialmutunneling{%
\centerline{\psfig{figure=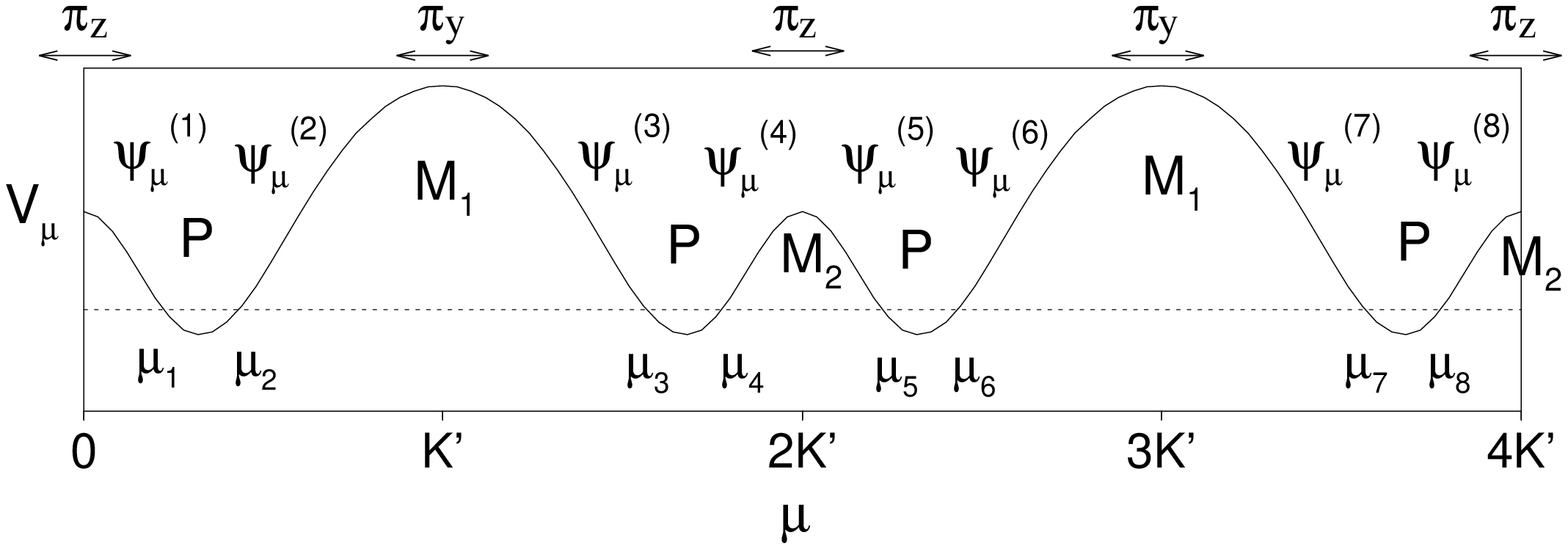,angle=-90,width=14cm}}
}
\def\figeffpotentialmutunneling{%
Effective potential $V_{\mu}$ (solid line) and effective energy
$E_{\mu}$ (dashed line) for motion type 
\tOA. The turning points $\mu_i$ ($i=1,...,8$) define {\sl WKB} wave functions
$\psi_\mu^{(i)}$ whose amplitudes are connected by the matrices $P$, $M_1$
and $M_2$. 
}
\FIGo{fig:effpotentialmutunneling}{\figeffpotentialmutunneling}{\FIGeffpotentialmutunneling}

Let us now specify the Hamiltonian (\ref{eq:generalhamilt}) for the
ellipsoidal billiard  
by the consideration of the effective potentials and energies defined
in Equations~(\ref{eq:effham})-(\ref{eq:effpot}). 
To illustrate the 
semiclassical quantization scheme we concentrate on the $\mu$ degree
of freedom and again present the effective
potential $V_{\mu}$ and energy $E_{\mu}$ for
motion type \tOA\ in 
Fig.~\ref{fig:effpotentialmutunneling}.  In the range
$[0,4\Kell']$ we have the eight turning points $\mu_i$ marked in the
figure. Taking them as the reference points for the definition of \WKB\
wave functions we get two wave functions in each of the four
classically allowed regions. The amplitude vectors ${\bm A}^{(1)}$
and ${\bm A}^{(2)}$ are connected by the phase shift
matrix (\ref{eq:phaseshift}) with $\phi =
\int_{\mu_1}^{\mu_2}p_\mu\,d\mu$. From Tab.~\ref{tab:actions} and the
negative sign in \equ~(\ref{eq:dmudeta}) it becomes clear that
$\phi=-(\pi/2)\Jmu$ with $\Jmu$ the action defined in
\equ~(\ref{eq:Jdefinition}). The matrix
\bege
P =
\left(
\begin{array}{cc}
\exp(-i (\pi/2)\Jmu/\hbar) & 0 \\
0 & \exp(i (\pi/2)\Jmu/\hbar)
\end{array}
\right)
\ende
then also relates the pairs of amplitude vectors ${\bm A}^{(3)}$ and
${\bm A}^{(4)}$, ${\bm A}^{(5)}$ and ${\bm A}^{(6)}$, ${\bm A}^{(7)}$
and ${\bm A}^{(8)}$, see Fig.~\ref{fig:effpotentialmutunneling}. 

The amplitude vectors ${\bm A}^{(2)}$ and ${\bm A}^{(3)}$  are related
by the tunnel matrix (\ref{eq:tunnelconnection}) where the penetration
integral 
$\Theta=-i\int_{\mu_2}^{\mu_3} p_\mu\,d\mu$ is 
the penetration integral $\Theta_\nu$ defined in
\equ~(\ref{eq:thetanu}) (again see 
Tab.~\ref{tab:actions} and keep in mind the negative sign in
\equ~(\ref{eq:dmudeta})). 
The integration boundaries in the definition of $\Theta_\nu$ were
independent of the classical 
type of motion. Therefore the connection relation remains valid if the
effective energy and potential
change such that we classically have a different type of motion,
especially for motion types \tO\ and \tP\ where the turning points
$\mu_2$ and $\mu_3$ become complex, see Fig~\ref{fig:effpotentials}.
We set 
\bege
M_1 = M(-\Theta_\nu)\,.
\ende
The matrix $M_1$ also connects the amplitude vectors ${\bm
  A}^{(6)}$ and ${\bm A}^{(7)}$.

Similarly one finds that the pairs of amplitude vectors ${\bm
  A}^{(4)}$, ${\bm  A}^{(5)}$ and ${\bm
  A}^{(8)}$, ${\bm  A}^{(1)}$
are related by the tunnel matrix
\bege
M_2 = M(-\Theta_\lambda)
\ende
with $\Theta_\lambda$ from \equ~(\ref{eq:thetalambda}) where we
have taken into account the $4\Kell'$-periodicity of the wave function
$\psi_\mu$.

Starting at $\mu_1$ the quantization of the $\mu$ 
degree of freedom now reduces  to finding an effective
energy $E_{\mu}$ and an effective potential $V_\mu$ for which there
exists a non-zero amplitude ${\bm   A}^{(1)}$  which is mapped onto
itself upon one traversal through 
the interval $[0,4\Kell']$, see Fig.~\ref{fig:effpotentialmutunneling}. 
This is equivalent to the quantization condition 
\bege
\label{eq:millerquant1}
\det ((M_2 P M_1 P)^2 -{\bm 1})=0
\ende
with ${\bm 1}$ the identity matrix.
Similar quantization conditions can be found in
\cite{Miller68,Connor68,Child74,RDWW96}. 
Because of $\det M(\Theta) = \det P(\phi) = 1$,
\equ~(\ref{eq:millerquant1}) may be rewritten as
\bege
\label{eq:millerquant2}
\trace (M_2 P M_1 P)^2  = 2\,.
\ende
The eigenvalues of \equ~(\ref{eq:millerquant1}) include all parity
combinations $\pi_y$ and $\pi_z$. To distinguish between the different
parities more information is needed.
The parities give the additional conditions
\bega
\label{eq:tunnelrel1}
{\bm A}^{(3)}=
\pi_y
\left(
\begin{array}{cc}
0 & 1 \\
1 & 0
\end{array}
\right)
{\bm A}^{(2)}
&,&
{\bm A}^{(7)}=
\pi_y
\left(
\begin{array}{cc}
0 & 1 \\
1 & 0
\end{array}
\right)
{\bm A}^{(6)}
\,,\\
\label{eq:tunnelrel2}
{\bm A}^{(5)}=
\pi_z
\left(
\begin{array}{cc}
0 & 1 \\
1 & 0
\end{array}
\right)
{\bm A}^{(4)}
&,&
{\bm A}^{(1)}=
\pi_z
\left(
\begin{array}{cc}
0 & 1 \\
1 & 0
\end{array}
\right)
{\bm A}^{(8)}
\,.
\enda
These conditions have to be solved consistently with the above tunnel 
relations. From the various possibilities to do this we choose 
the following. 
We map the amplitude vector ${\bm A}^{(1)}$ from a point
$\mu\in(\mu_1,\mu_2)$ to the point $\mu+2\Kell'\in(\mu_5,\mu_6)$. From
\equ~(\ref{eq:inv4cond}) we know that this produces the sign $\pi_y \pi_z$.
There are two possibilities to replace one of the tunnel matrices in
this map by the corresponding condition in Equations
(\ref{eq:tunnelrel1}) and (\ref{eq:tunnelrel2}). We thus get the equations
\bege
\label{eq:matrixequat}
B{\bm A}^{(1)}=0\,, \quad C{\bm A}^{(1)}=0 
\ende
with the matrices
\bega
B &=& 
\pi_z
\left(
\begin{array}{cc}
0 & 1 \\
1 & 0
\end{array}
\right)
P M_1 P
-
\pi_y \pi_z 
{\bm 1}\,,
\\
C &=& 
M_2 P
\pi_y
\left(
\begin{array}{cc}
0 & 1 \\
1 & 0
\end{array}
\right)
P
-
\pi_y \pi_z 
{\bm 1}\,.
\enda
\equ~(\ref{eq:matrixequat}) 
is the analogue of \equ~(\ref{eq:millerquant1}) for half the
interval $[0,4\Kell']$. 
We are free in the normalization of the \WKB\ wave function. Setting
$A_+^{(1)}=1$ we find $A_-^{(1)}=-B_{11}/B_{12}$. If we insert this into
the equation involving the matrix $C$ and decompose the resulting
equations into their real and imaginary parts the remaining independent
conditions are
\bege
\label{eq:mcondcos}
  \cos (\pi \Jmu/\hbar) =
\frac{\pi_z \pi_y
  e^{(\Theta_{\lambda}+\Theta_{\nu})/\hbar
    }-1}{\sqrt{\left(1+e^{2\Theta_{\lambda}/\hbar
      }\right)\left( 
    1+e^{2\Theta_{\nu}/\hbar} \right)}}
\ende
and
\bege
\label{eq:mcondsin}
\sin (\pi \Jmu/\hbar) = \frac{\pi_z e^{\Theta_{ \lambda}/\hbar } + \pi_y
  e^{\Theta_{\nu}/\hbar}}{\sqrt{\left(
1+e^{2\Theta_{\lambda}/\hbar }\right)\left(1+e^{2\Theta_{\nu}/\hbar} \right)}}\,.
\ende
These equations have to be fulfilled simultaneously. They
are not independent of each another, but the relation is simple: 
the second equation is fulfilled on every second  solution of the
first equation.

For the $\lambda$ and $\nu$ degree of freedom we have to comment on
the additional potential barriers in
Fig.~\ref{fig:effpotentialsspecialsall} appearing below the line 
$c_\nu$ and above the line $c_\lambda$ of the bifurcation diagram in 
Fig.~\ref{fig:bifurcationdiagram}.
As we have mentioned in Section~\ref{sec:classic} the effective
energies always lie below the additional local potential minima. They
only  reach the local minima at the points $P_\nu$ and $P_\lambda$,
respectively.
The corresponding classical turning points are always complex and
therefore these barriers would enter the quantization scheme almost
always with
large negative penetration integrals, i.e. with tunnel matrices close
to the identity matrix. The only exceptions occur in the regions 
close to the points $P_\nu$ and $P_\lambda$ which lie at the border of
the bifurcation diagram. Here the action $J_\mu$ goes to zero, i.e. the
semiclassical approximation is expected to give poor results anyway.
The additional barriers will therefore not be taken into
account. 
The quantization conditions for $\lambda$ and $\nu$ are then exactly the
same as in the case of the planar elliptic billiard  discussed in
\cite{WWD97}. We only state the results. 
For the $\lambda$ degree of freedom one finds the two conditions  
\bege
\label{eq:lcondcos}
\cos  (\pi \Jlambda/\hbar) = \frac{-\pi_z}{\sqrt{1+e^{2\Theta_\lambda/\hbar}}} 
\ende
and
\bege
\label{eq:lcondsin}
\sin  (\pi \Jlambda/\hbar) = \frac{-1}{\sqrt{1+e^{-2\Theta_\lambda/\hbar}}} 
\ende
and for the $\nu$ degree of freedom the conditions
\bege
\label{eq:ncondcos}
  \cos (\pi \Jnu/\hbar ) = \frac{\pi_x \pi_y}{\sqrt{1+e^{2\Theta_\nu/\hbar}}} 
\ende
and
\bege
\label{eq:ncondsin}
  \sin (\pi \Jnu/\hbar ) = \frac{\pi_x }{\sqrt{1+e^{-2\Theta_\nu/\hbar}}}.
\ende
The actions $J_\lambda$ and $J_\nu$ are again taken from
\equ~(\ref{eq:Jdefinition}) and the only penetration integrals that
appear are those defined in Equations
(\ref{eq:thetanu}) and (\ref{eq:thetalambda}). \\
\begin{table}[!h]
\begin{center}
{ 
\tabstart  \small
\begin{tabular}{|c|lll|lll|}\hline
  \equ  & \multicolumn{3}{|c|}{type \tO : $-\Theta_\lambda,\Theta_\nu \gg \hbar
   $} & 
\multicolumn{3}{|c|}{type \tOA : $-\Theta_\lambda,-\Theta_\nu \gg \hbar$} \\  \hline
  (\ref{eq:lcondcos}) &$-\pi_z$& &$\Iflambda =\Jlambda =
  (n_\lambda+\frac{4}{4})\hbar$&$-\pi_z$&&$\Iflambda = \Jlambda =
  (n_\lambda+\frac{4}{4})\hbar$\\
  (\ref{eq:lcondsin}) &$0$&\raisebox{1.5ex}[-1.5ex]{$\bigg\}$}&$n_\lambda =
  2r+(1-\pi_z)/2$&$0$&\raisebox{1.5ex}[-1.5ex]{$\bigg\}$}&$n_\lambda = 2r+(1-\pi_z)/2$\\[1.3ex]
  (\ref{eq:mcondcos}) &$0$& &$\Ifmu =\Jmu=(n_\mu +\frac{1}{2})\hbar$ & $-1$ & &  $\Ifmu
   =\Jmu /2=(n_\mu +\frac{1}{2})\hbar$  \\
  (\ref{eq:mcondsin}) &$\pi_y$&\raisebox{1.5ex}[-1.5ex]{$\bigg\}$}
  &$n_\mu = 2m+(1-\pi_y)/2$&$0$&\raisebox{1.5ex}[-1.5ex]{$\bigg\}$}&$n_\mu=m$  \\[1.3ex]
  (\ref{eq:ncondcos}) &$0$& &$\Ifnu = \Jnu
  =(n_\nu+\frac{1}{2})\hbar$&$\pi_x\pi_y$& &$\Ifnu= \pm \Jnu = \pm
  n_\nu \hbar$ \\
  (\ref{eq:ncondsin}) &$\pi_x$&\raisebox{1.5ex}[-1.5ex]{$\bigg\}$}&$n_\nu = 2n
  +(1-\pi_x)/2$&$0$&\raisebox{1.5ex}[-1.5ex]{$\bigg\}$}&$n_\nu = 2n + (2-\pi_x-\pi_y)/2$ \\ \hline
\equ &
 \multicolumn{3}{|c|}{type $\tP$: $\Theta_\lambda,\Theta_\nu \gg
    \hbar$} & 
\multicolumn{3}{|c|}{type
  \tOB : $\Theta_\lambda,- \Theta_\nu \gg \hbar   $} \\  \hline
  (\ref{eq:lcondcos}) &$0$& &$\Iflambda = \Jlambda /2 =(n_\lambda
  +\frac{3}{4})\hbar$&$0$& &$\Iflambda = \Jlambda /2 = (n_\lambda+\frac{3}{4})\hbar$\\
  (\ref{eq:lcondsin}) &$-1$&\raisebox{1.5ex}[-1.5ex]{$\bigg\}$}&$n_\lambda =r$&$-1$&\raisebox{1.5ex}[-1.5ex]{$\bigg\}$}&$n_\lambda
  =r$\\[1.3ex]
  (\ref{eq:mcondcos}) & $\pi_y \pi_z$  & &
   $\Ifmu =\pm \Jmu =\pm  n_\mu \hbar$
   &$0$& &$\Ifmu =\Jmu =(n_\mu +\frac{1}{2})\hbar$ \\
  (\ref{eq:mcondsin}) & $0$ & 
 \raisebox{1.5ex}[-1.5ex]{$\bigg\}$} &$n_\mu=2m+(1-\pi_y \pi_z)/2$&$\pi_z$&
\raisebox{1.5ex}[-1.5ex]{$\bigg\}$}
&$n_\mu=2m+(1-\pi_z)/2$ \\[1.3ex]
  (\ref{eq:ncondcos}) &$0$& &$\Ifnu = \Jnu
  =(n_\nu+\frac{1}{2})\hbar$&$\pi_x \pi_y$& &$\Ifnu = \pm \Jnu = \pm
  n_\nu \hbar$ \\
  (\ref{eq:ncondsin}) &$\pi_x$&\raisebox{1.5ex}[-1.5ex]{$\bigg\}$}&$n_\nu = 2n +
  (1-\pi_x)/2$&$0$&\raisebox{1.5ex}[-1.5ex]{$\bigg\}$}&$n_\nu = 2n +(2-\pi_x -\pi_y )/2$ \\ \hline
  \end{tabular}
\tabend
}
\caption[]{\label{tab:limitcases} \capsty 
Limiting quantization conditions for the 4 types of classical
motion. The left hand sides of the braces in each box gives the
limiting value of the 
right hand side of the equation cited in the very first column. 
The non-negative integers $n_\lambda$, $n_\mu$ and $n_\nu$ are related
to the quantum numbers $r$, $m$ and
$n$ introduced in Section~\ref{sec:qm}, see Tab.~\ref{tab:actions}.
}
\end{center}
\end{table}

We first inspect the quantization conditions in
Equations~(\ref{eq:mcondcos})-(\ref{eq:ncondsin}) for the limiting
cases $|\Theta_\lambda|,|\Theta_\nu| \gg\hbar$. The signs
of the penetration integrals $\Theta_\lambda$ and $\Theta_\nu$
determine the type of classical motion. The limiting cases
$|\Theta_\lambda|,|\Theta_\nu| \gg\hbar$ thus correspond to the four regions in
classical action space far away from the separatrix surfaces in
Fig.~\ref{fig:sep}. From the limiting quantization conditions for the
actions ${\bm J}$ the quantization of the original action variables
${\bm I}$ can be deduced from Tab.~\ref{tab:actions}. We summarize the
results in Tab.~\ref{tab:limitcases}. The limiting quantization
conditions for ${\bm I}$ may be compared with the {\sl EBK}
quantization in \equ~(\ref{eq:ebkquantization}). From the
identification of the Maslov phases ${\bm
  \alpha}=(\alpha_\lambda,\alpha_\mu,\alpha_\nu)$ in the equations in
Tab.~\ref{tab:limitcases} we find
\bega
\label{eq:maslovphases1}
{\bm \alpha} =
(4 , 2 , 2)
\mbox{ for type \tO},\quad
{\bm \alpha} =
(4 , 2 , 0)
\mbox{ for type \tOA},\\
\label{eq:maslovphases2}
{\bm \alpha} =
(3 , 0 , 2)
\mbox{ for type \tP},\quad
{\bm \alpha} =
(3 , 2 , 0)
\mbox{ for type \tOB}.
\enda
The Maslov indices $\alpha_\mu$ and $\alpha_\nu$ are in agreement with
the simple {\sl EBK} rule stated in the introduction: For motion of type
\tO\, $\mu$ and $\nu$ oscillate, for motion of type \tP\ the motion is
rotational in $\mu$ and oscillatory in $\nu$, \tOA\ and \tOB\ are
oscillatory in the $\mu$ degree of freedom and rotational in $\nu$,
see Section~\ref{sec:classic}. For the $\lambda$ degree of
freedom we have to take into account the reflection at the boundary
ellipsoid which wave mechanically leads to Dirichlet boundary
condition. For motion types \tO\ and \tOA\ $\lambda$ oscillates with
two reflections giving $\alpha_\lambda=4$. For motion types \tP\ and
\tOB\ $\lambda$ oscillates between the boundary ellipsoid and the
caustic giving $\alpha_\lambda = 3$.
\def\FIGquantumboxes{%
\centerline{ 
        \raisebox{0.2cm}{\mbox{\tO}}
        \psfig{figure=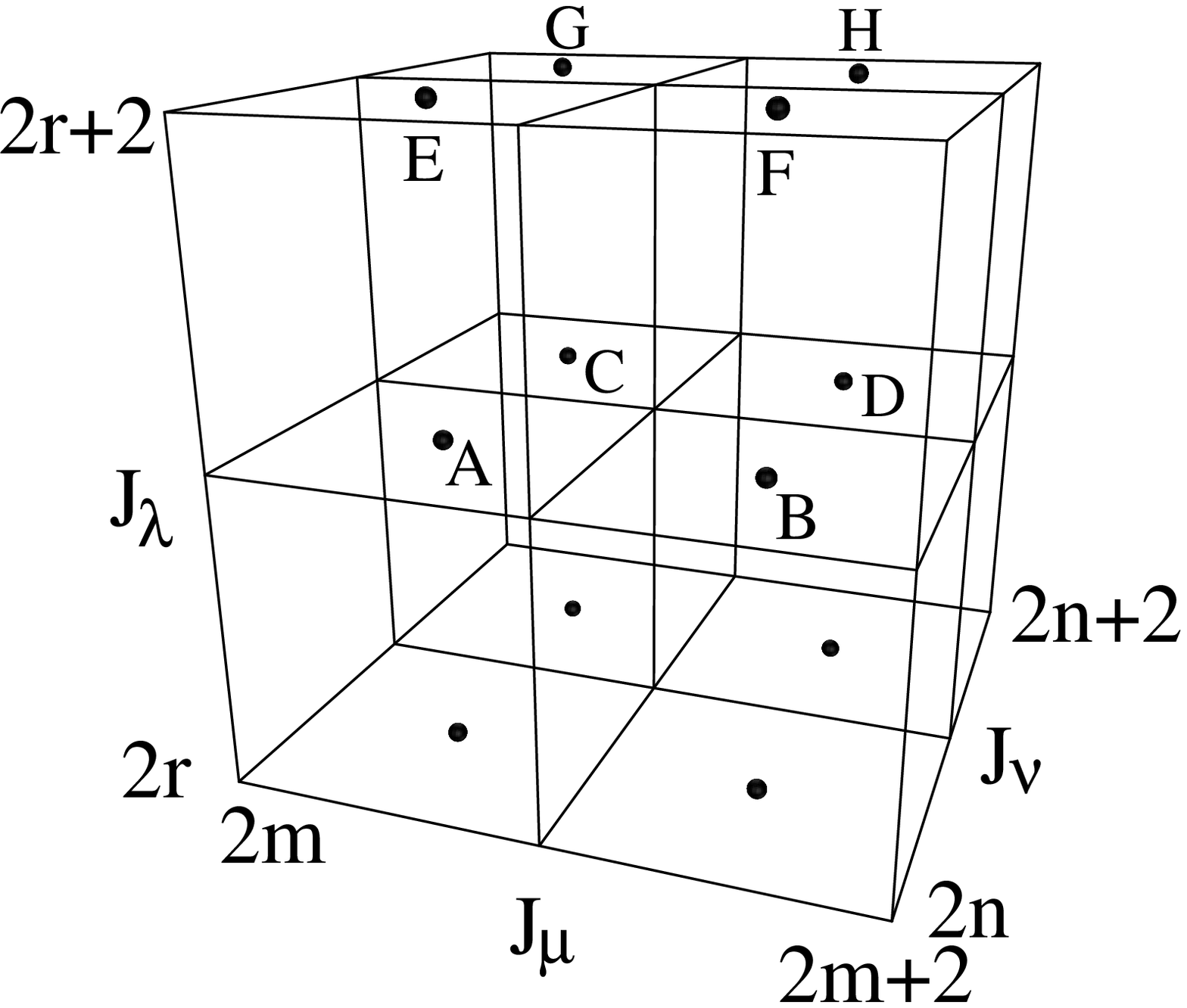,angle=-90,width=5.5cm}
        \hspace*{0.5cm}
        \raisebox{0.2cm}{\mbox{\tOA}}
        \psfig{figure=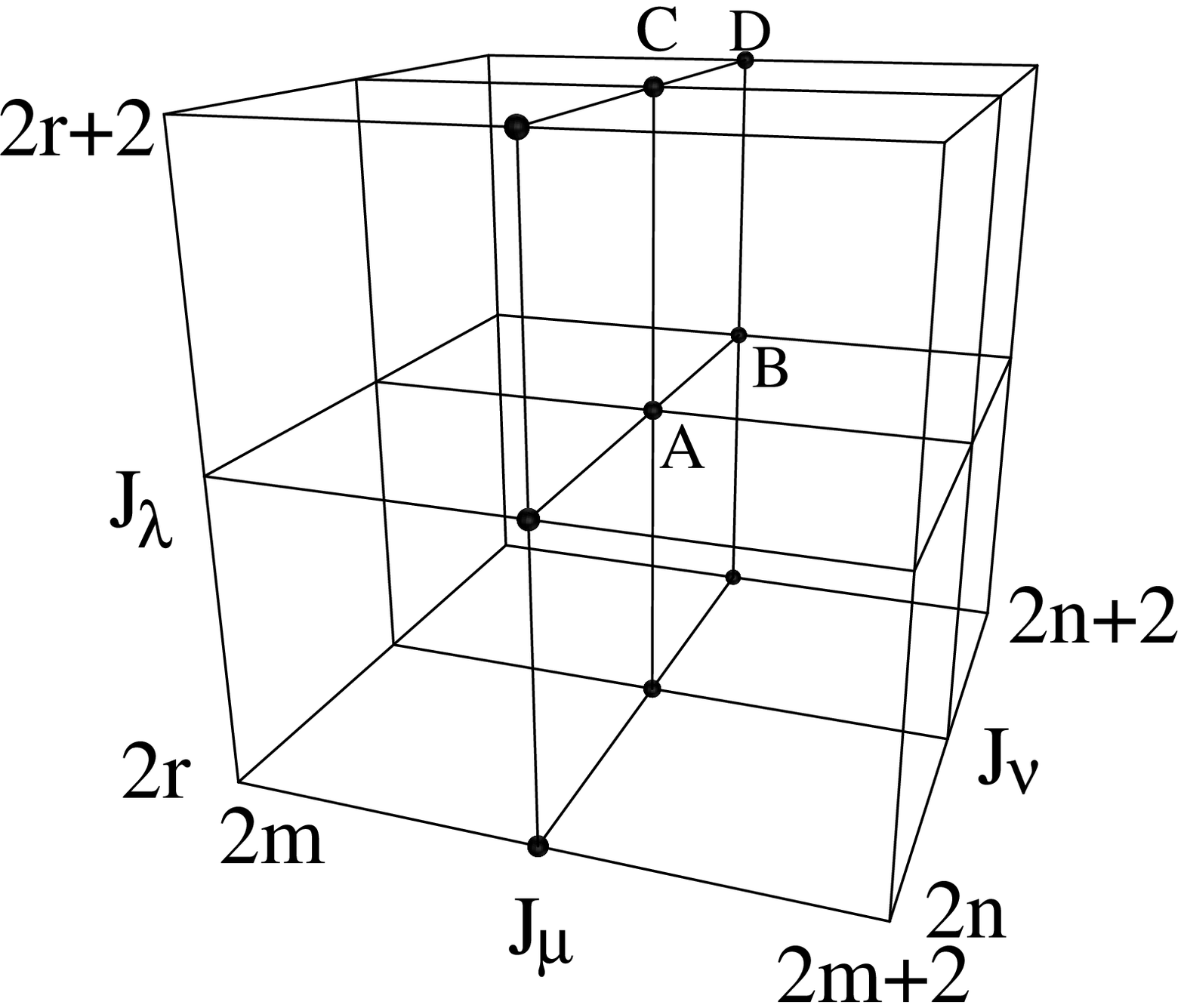,angle=-90,width=5.5cm}
   } 
\centerline{
        \raisebox{0.2cm}{\mbox{\tP}}
        \psfig{figure=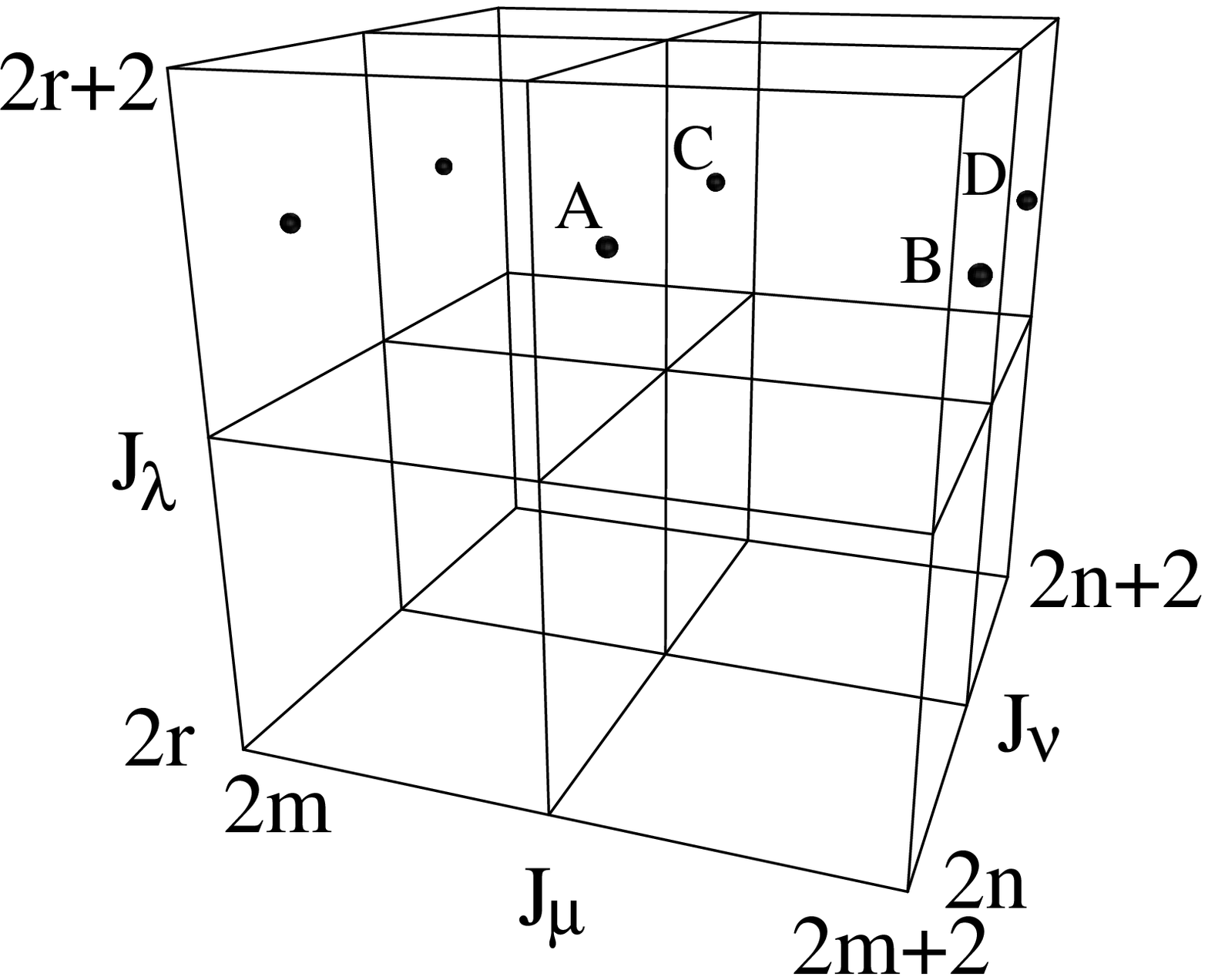,angle=-90,width=5.5cm} 
        \hspace*{0.5cm}
        \raisebox{0.2cm}{\mbox{\tOB}}
        \psfig{figure=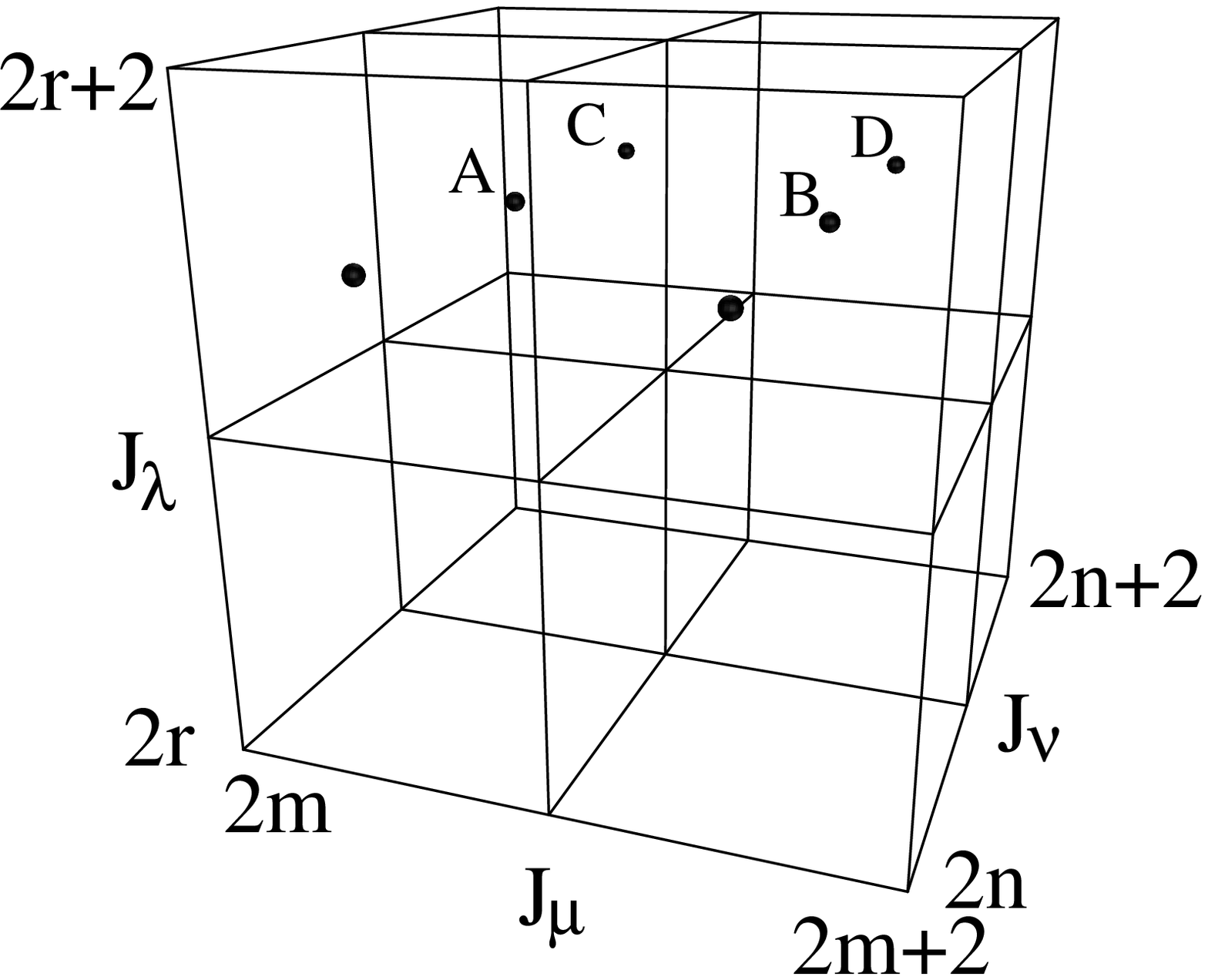,angle=-90,width=5.5cm}
} }
\def\figquantumboxes{%
Quantum cells $\Delta J_\lambda =
\Delta J_\mu = \Delta J_\nu = 2$ in classical action space for the 4
limiting cases of 
classical types of motion. $\BJf$ is measured in units of $\hbar$. 
}
\FIGo{fig:quantumboxes}{\figquantumboxes}{\FIGquantumboxes}

The {\sl EBK} quantization condition in \equ~(\ref{eq:ebkquantization})
defines a lattice in classical action space. The Maslov indices
determine how this lattice is shifted relative to the simple lattice
$({\bm n}\hbar)$. Since we have four different vectors of Maslov
indices for the ellipsoidal billiard we have four different lattice
types away from the separatrix surfaces in Fig.~\ref{fig:sep}. We
present the different lattices in Fig.~\ref{fig:quantumboxes} for
quantum cells of width $\Delta J_\lambda = \Delta J_\mu = \Delta J_\nu
= 2\hbar$. Each cell contains eight quantum states. For motion type \tO\  all
states are non-degenerate. For motion types \tOA\ and \tOB\ each states
is twofold quasidegenerate according to the two senses of rotation in
$\nu$. Analogously for motion type \tP\ each state is twofold
quasidegenerate according to the two senses of rotation in the
variable~$\mu$.
From the quantum mechanical point of view the quasidegeneracy can be
understood in terms of the effective energies and potentials in
Fig.~\ref{fig:effpotentials}. For eigenvalues which classically correspond to
rotational motions far away from the classical separatrices the
effective energy is much larger then the effective potential. The
energy is then dominated by the kinetic energy, the specific shape of
the potential becomes irrelevant. The effective energy then only
depends on
the net number of nodes of the wave function and not on the
location of the nodes. Therefore wave functions with different
symmetries but the same net number of nodes give the same effective
energy.  With the aid of Tab.~\ref{tab:limitcases} we can identify the
states corresponding to the capital letters in
Fig.~\ref{fig:quantumboxes}, see Tab.~\ref{tab:limitquntumnumbers}. 
\begin{table}[!h]
\begin{center}
{ 
\tabstart  \small
\begin{tabular}{|lll|lll|}\hline
\tO\ & & & \tOA\ & & \\ \hline
 & A: & $|r,m,n;+++\rangle$   & & A: & $|r,m,n;+-+\rangle$,\\
 & B: & $|r,m,n;+-+\rangle$   & &    & $|r,m,n;-++\rangle$ \\
 & C: & $|r,m,n;-++\rangle$   & & B: & $|r,m,n;--+\rangle$, \\
 & D: & $|r,m,n;--+\rangle$   & &    & $|r,m,n+1;+++\rangle$ \\
 & E: & $|r,m,n;++-\rangle$   & & C: & $|r,m,n;+--\rangle$, \\
 & F: & $|r,m,n;+--\rangle$   & &    & $|r,m,n;-+-\rangle$ \\
 & G: & $|r,m,n;-+-\rangle$   & & D: & $|r,m,n;---\rangle$, \\
 & H: & $|r,m,n;---\rangle$   & &    & $|r,m,n+1;++-\rangle$ \\ \hline
\tP\ & & & \tOB\ & &  \\ \hline
 & A: & $|r,m,n;++-\rangle$,  & & A: & $|r,m,n;+-+\rangle$, \\
 &    & $|r,m,n;+-+\rangle$   & &     & $|r,m,n;-++\rangle$ \\
 & B: & $|r,m,n;+--\rangle$,  & & B: & $|r,m,n;+--\rangle$, \\
 &    & $|r,m,n+1;+++\rangle$ & &     & $|r,m,n;-+-\rangle$ \\
 & C: & $|r,m,n;-+-\rangle$,  & & C: & $|r,m,n;--+\rangle$, \\
 &    & $|r,m,n;--+\rangle$   & &     & $|r,m,n+1;+++\rangle$ \\
 & D: & $|r,m,n;---\rangle$,  & & D: & $|r,m,n;---\rangle$, \\
 &    & $|r,m,n+1;-++\rangle$ & &     & $|r,m,n+1;++-\rangle$ \\
  \hline
  \end{tabular}
\tabend
}
\caption[]{\label{tab:limitquntumnumbers} \capsty 
Quantum states in Fig.~\ref{fig:quantumboxes}. 
}
\end{center}
\end{table}

The quantization conditions in Equations
(\ref{eq:lcondcos})-(\ref{eq:ncondsin}) are uniform, i.e.\ they do not
only give the limiting {\sl EBK} lattices in
Fig.~\ref{fig:quantumboxes} but also specify how these lattices join
smoothly across the separatrix surfaces of Fig.~\ref{fig:sep}. 
In the following we will refer to the uniform lattice in action space
as {\sl WKB} lattice.
The transitions may be described in terms of effective Maslov
phases. In order to 
see this we insert the {\sl EBK}-like quantization conditions for the
symmetry reduced ellipsoid
\bege\label{eq:wkbsymred}
\BId = \BJf /2 =  (\tilde{{\bm n}} +\tilde{{\bm \alpha}}/4)\hbar
\ende
with $\tilde{{\bm n}}=(r,m,n)$ into the left hand sides of Equations
(\ref{eq:lcondcos})-(\ref{eq:ncondsin}). The parities $\pi_x$, $\pi_y$
and $\pi_z$ on the right hand sides determine whether we have Dirichlet
or Neumann boundary conditions on the corresponding piece of the
Cartesian coordinate plane bounding the symmetry reduced
ellipsoid. The different parity combinations altogether give the
quantum states of the full ellipsoidal billiard. 
We now solve Equations~(\ref{eq:lcondcos})-(\ref{eq:ncondsin}) for the
effective Maslov phases $\tilde{{\bm \alpha}}$. The quantum numbers
$\tilde{{\bm n}} = (r,m,n)$ drop out because of the $2\pi$-periodicity of the
trigonometric functions and
 it remains to invert the tangent
on the correct branch which is determined by the parity
combination. A little combinatorics gives
\bega
\label{eq:Maslovlambda}
 \tilde{\alpha}_\lambda &=& \pi_z \frac{2}{\pi} \arctan e^{\Theta_\lambda
   /\hbar} +3 - \pi_z,\\ 
\label{eq:Maslovnu}
 \tilde{\alpha}_\nu &=& \pi_y \frac{2}{\pi} \arctan e^{\Theta_\nu /\hbar} + 2
 - \pi_x - \pi_y.
\enda
For $\tilde{\alpha}_\mu$ this simple form cannot be achieved. Instead we write
\bege
\label{eq:Maslovmu}
\tilde{\alpha}_\mu = \frac{2}{\pi} \mbox{arg} \left(\pi_z\pi_y e^{(\Theta_{\lambda}+\Theta_{\nu})/\hbar} -1 + i \left( \pi_z
e^{\Theta_{\lambda}/\hbar} + \pi_y 
e^{\Theta_{\nu}/\hbar}\right)  \right),
\ende
where arg maps the polar angle of a complex number to the
interval $[0,2\pi)$.
Essentially the effective Maslov phases  consist
of the simple switching function $(2/\pi)\arctan e^x$ which changes from 0 to
1 when $x$ changes from $-\infty$ to $+\infty$.
From the simple form of the effective Maslov phases it follows that the ranges
in which the semiclassically quantized action variables ${\bm J}$ may
vary are restricted according to
${\bm J} = (J_\lambda,J_\mu,J_\nu)\mbox{ mod } 2\hbar \in P_{\bm \pi}$,
where the parity boxes $P_{\bm \pi}$ have side length $\hbar/2$ in the
directions of $J_\lambda$ and $J_\nu$ and side length $\hbar$ in the
direction $J_\mu$. For the different parity combinations~${\bm \pi}$
we find 
\bege
\begin{array}{lcccccc}
P_{---} &=& [(3/2)\hbar,2\hbar]& \times &[\hbar,2\hbar] &\times &[(3/2)\hbar,2\hbar]\,, \\
P_{--+} &=& [\hbar,(3/2)\hbar] &\times &[(1/2)\hbar,(3/2)\hbar]& \times & [(3/2)\hbar,2\hbar]\,, \\
P_{-+-} &=& [(3/2)\hbar,2\hbar]& \times & [(1/2)\hbar,(3/2)\hbar]& \times &[\hbar,(3/2)\hbar]\,, \\
P_{-++} &=& [\hbar,(3/2)\hbar] &\times& [0,\hbar]& \times& [\hbar,(3/2)\hbar]\,, \\
P_{+--} &=& [(3/2)\hbar,2\hbar]& \times& [(\hbar,2\hbar]& \times &[(1/2)\hbar,\hbar]\,, \\
P_{+-+} &=& [\hbar,(3/2)\hbar] &\times& [(1/2)\hbar,(3/2)\hbar] &\times& [(1/2)\hbar,\hbar]\,, \\
P_{++-} &=& [(3/2)\hbar,2\hbar]& \times& [(1/2)\hbar,(3/2)\hbar] &\times &[0,(1/2)\hbar]\,, \\
P_{+++} &=& [\hbar,(3/2)\hbar]& \times& [0,\hbar] &\times& [0,(1/2) \hbar]\,,
\end{array}
\ende
see Fig.~\ref{fig:parityboxes}.\\
\def\FIGparityboxes{%
\centerline{\psfig{figure=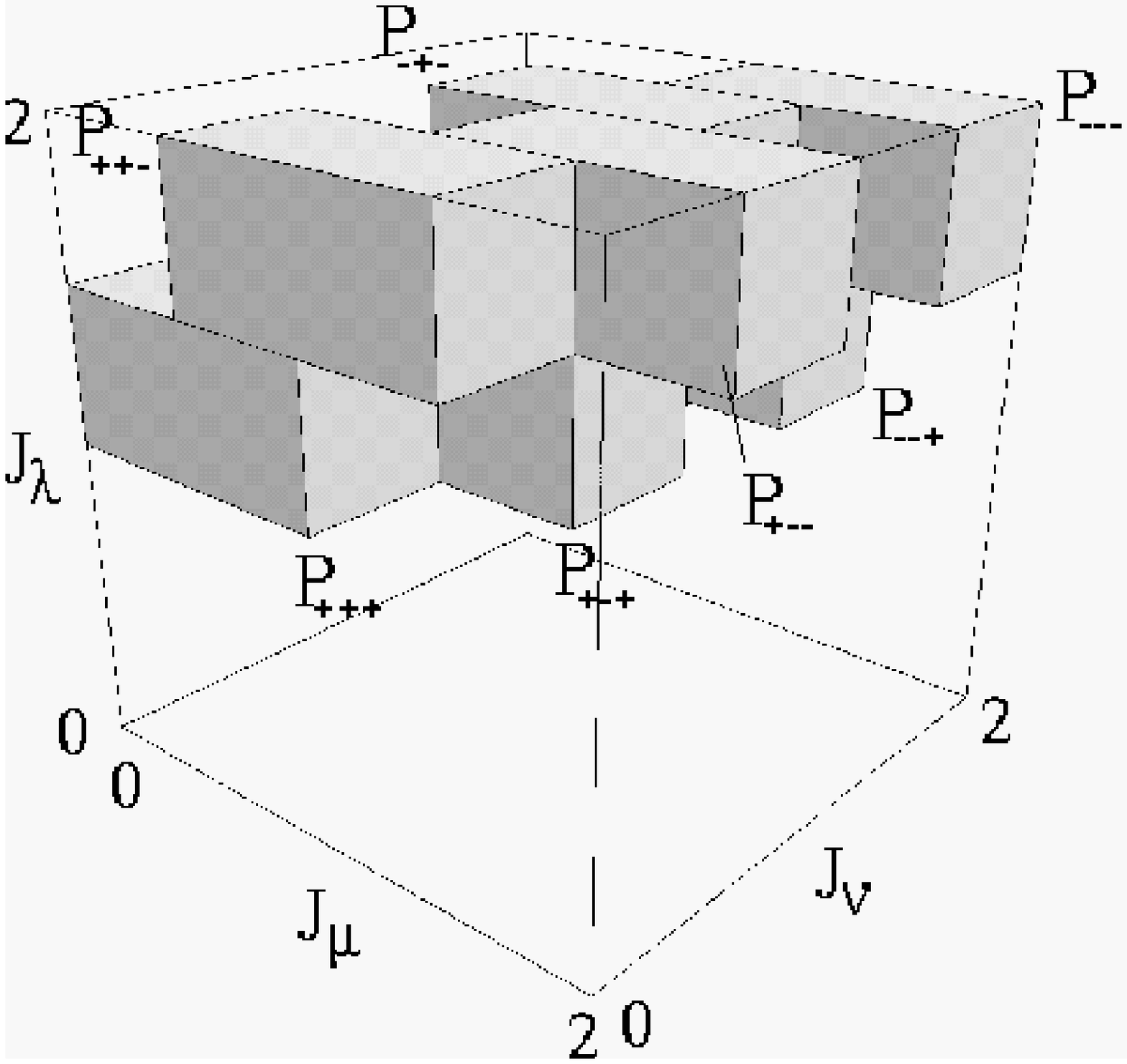,angle=-90,width=5.5cm}}
}
\def\figparityboxes{%
The parity boxes $P_{{\bm \pi}}$ that contain the semiclassical states. 
$P_{-++}$ is obscured by the other parity boxes. 
$\BJf$ is measured in units of $\hbar$.
}
\FIGo{fig:parityboxes}{\figparityboxes}{\FIGparityboxes}

Let us comment on the numerical procedure to solve the 
 quantization condition~(\ref{eq:wkbsymred}).
For given  quantum numbers $\tilde{\bm n}$ and parity combination
${\bm \pi}$ we have to find the corresponding zero of the function
\bege\label{eq:Gwkb}
{\bm G}(E,s_1^2,s_2^2;{\bm \pi},\tilde{{\bm n}}) = \BId(E,s_1^2,s_2^2)
-(\tilde{{\bm 
  n}}+\Bmaslovd(E,s_1^2,s_2^2,{\bm \pi})/4)\hbar \,.
\ende
As in the exact quantum mechanical problem this problem is not
separable for the separation constants and we have to apply Newton's
method in three dimensions.  
In order to find good starting values for Newton's method we
introduce an approximate function $\Gapp$
for ${\bm G}$. The functional dependence of $\Gapp$, i.e. of the
approximate actions $\Iapp$ and Maslov phases $\maslovapp$,  on the
parameters $(E,s_1^2,s_2^2)$ should be very simple such that the 
zeroes of  $\Gapp$ can be found analytically.
For $\maslovapp$ we simply take the mean value of $\tilde{\bm \alpha}$
for a given parity combination ${\bm \pi}$.
In order to get an expression for $\Iapp$
we take advantage of two properties of the energy surface in action space. 
Firstly the shape of the energy surface  is very similar to a triangle
and secondly up to  a simple
scaling the shape does not change with the energy. 
If we denote the intersections of the energy surface $\h=1/2$ with the
coordinate axes in action space by ${\bar I}_\lambda$, ${\bar
  I}_\mu$, and  ${\bar I}_\nu$ the action variables can be
approximated by
\bege\label{eq:Iapp}
\Iapp(E,s_1^2,s_2^2) = \sqrt{2\h}\left[\left(
\begin{array}{c}
{\bar I}_\lambda \\
0 \\
0 
\end{array}\right)   
+ \gamma_1 \left(
\begin{array}{c}
-{\bar I}_\lambda \\
0 \\
{\bar I}_\nu 
\end{array}\right)
+ \gamma_2 \left(   
\begin{array}{c}
-{\bar I}_\lambda \\
{\bar I}_\mu \\
0 
\end{array}\right)\right] \ ,
\ende
where $\gamma_1$ and $\gamma_2$ parametrize the approximate triangular
energy surface 
$\h=1/2$.
This can be considered as a crude periodic orbit quantization:
we take the actions of the three stable isoltated periodic orbits 
of the system and approximate the whole energy surface by that
of a harmonic oscillator that would have isolated stable orbits
with those actions. The analogy has to be taken with care because
our actions scale with $\sqrt{E}$, while those of the true 
harmonic oscillator are linear in the energy.
The edges of this approximate energy surface are given by
$\gamma_1 = 0$, $\gamma_2 = 0$, and $\gamma_1+\gamma_2 = 1$,
respectively. 
In order to give
$s_1$ and $s_2$ as functions of $\gamma_1$ and $\gamma_2$ it is useful
to have a look at the asymptotic behavior of the actions and their
approximations~(\ref{eq:Iapp}) upon approaching the edges of the
energy surface. 
A simple calculation shows that $\tilde{I}_\nu$ behaves quadratically
in $s_1$ for $s_1 \to 0$. We set
\bege
s_1^2 = a^2\gamma_1 \,.
\ende 
A similar consideration of the asymptotics of $\tilde{I}_\lambda$ for 
$s_2 \to 1$ gives 
\bege 
s_2^2 = 1-(1-b^2)(1-(\gamma_1+\gamma_2))^{2/3} \ .
\ende 
In our numerical calculation the starting values obtained from these
approximations always were sufficiently good  to make Newton's method
converge to the right state.
The quasidegeneracy of the states is no problem here because the
degenerate states correspond to different parity combinations ${\bm
  \pi}$.
In Tab.~\ref{tab:eigenvalues} the
semiclassical eigenvalues are compared to the exact quantum 
mechanical results. The semiclassical energy eigenvalues are always a little
too low. The same is true for $k_{\rm sc}$ while $l_{\rm sc}$ tends to
be too low. We do not have a good explanation for this. \\
\begin{table}[!h]
\begin{center}
{
\tabstart \small
\begin{tabular}{|ccc|ccc|ccc|ccc|c|}\hline
  \rule[-3mm]{0mm}{8mm}
  $E_{{\text{qm}}}$ & $k_{\text{qm}}$ & $l_{\text{qm}}$ & 
  $E_{{\text{sc}}}$ & $k_{\text{sc}}$ & $l_{\text{sc}}$ & 
  $r$ & $m$ & $n$ & $\pi_x$ & $\pi_y$ & $\pi_z$ &
  $\Delta E$ \\ \hline
  6{.}65202 & 0{.}17113 & 0{.}01217 & 6{.}14810 & 0{.}20240 & 0{.}02052 & 
  0& 0 & 0 & $+$ & $+$ & $+$ & 7.6\\
  12{.}1738 & 0{.}25651 & 0{.}04766 & 11{.}6003 & 0{.}27426 & 0{.}05007 &
  0 & 0 & 0 & $-$ & $+$ & $+$ & 4.7\\ 
  12{.}5791 & 0{.}23997 & 0{.}02728 & 11{.}9202 & 0{.}26066 & 0{.}03416 &
  0 & 0 & 0 & $+$ & $-$ & $+$ & 5.2\\ 
  16{.}0174 & 0{.}14696 & 0{.}00966 & 15{.}5690 & 0{.}15868 & 0{.}01338 &
  0 & 0 & 0 & $+$ & $+$ & $-$ & 2.8\\ 
  19{.}3555 & 0{.}30548 & 0{.}07460 & 18{.}7075 & 0{.}31769 & 0{.}07694 & 
  0 & 0 & 1 & $+$ & $+$ & $+$ & 3.3\\ 
  19{.}4998 & 0{.}30161 & 0{.}06998 & 18{.}8698 & 0{.}31328 & 0{.}07198 &
  0 & 0 & 0 & $-$ & $-$ & $+$ & 3.2\\ 
  21{.}2740 & 0{.}25695 & 0{.}01646 & 20{.}7491 & 0{.}26577 & 0{.}01936 &
  0 & 1 & 0 & $+$ & $+$ & $+$ & 2.5\\    
  23{.}4713 & 0{.}22232 & 0{.}03399 & 22{.}9153 & 0{.}23135 & 0{.}03520 &
  0 & 0 & 0 & $-$ & $+$ & $-$ & 2.4\\ 
  23{.}9671 & 0{.}21138 & 0{.}01734 & 23{.}3068 & 0{.}22222 & 0{.}02081 &
  0 & 0 & 0 & $+$ & $-$ & $-$ & 2.8\\ 
  \vdots & & & & & & & &  &  &  &  &\\ 
  1000{.}25 & 0{.}31235 & 0{.}08368 & 999{.}421 & 0{.}31266 & 0{.}08375 &
  4 & 1 & 8 & $+$ & $+$ & $-$ & 0.08\\ 
  1000{.}25 & 0{.}31235 & 0{.}08368 & 999{.}421 & 0{.}31266 & 0{.}08375 &
  4 & 1 & 7 & $-$ & $-$ & $-$ & 0.08\\ 
  1000{.}34 & 0{.}27684 & 0{.}04333 & 999{.}654 & 0{.}27705 & 0{.}04322 & 
  4 & 4 & 4 & $+$ & $-$ & $-$ & 0.07\\  
  1001{.}11 & 0{.}43872 & 0{.}07161 & 998{.}606 & 0{.}43941 & 0{.}07165 &
  0 & 10 & 5 & $-$ & $+$ & $-$ & 0.25\\ 
  1001{.}36 & 0{.}46428 & 0{.}19051 & 1000{.}49 & 0{.}46452 & 0{.}19067 &
  1 & 2 & 12 & $-$ & $-$ & $+$ & 0.09\\ 
  1001{.}36 & 0{.}46428 & 0{.}19051 & 1000{.}49 & 0{.}46452 & 0{.}19067 &
  1 & 2 & 13 & $+$ & $+$ & $+$ & 0.09\\ 
  1001{.}39 & 0{.}21559 & 0{.}01869 & 1000{.}78 & 0{.}21583 & 0{.}01876 &
  6 & 4 & 1 & $-$ & $-$ & $+$ & 0.06\\ 
  1001{.}52 & 0{.}34082 & 0{.}09365 & 1001{.}38 & 0{.}34082 & 0{.}09365 &
  3 & 2 & 8 & $-$ & $+$ & $-$ & 0.01\\ 
  1001{.}52 & 0{.}34082 & 0{.}09365 & 1001{.}38 & 0{.}34082 & 0{.}09365 &
  3 & 2 & 8 & $+$ & $-$ & $-$ & 0.01\\ 
  1001{.}63 & 0{.}35433 & 0{.}03148 & 1000{.}62 & 0{.}35461 & 0{.}03156 &
  1 & 11 & 2 & $+$ & $-$ & $+$ & 0.10\\ 
  1001{.}63 & 0{.}35433 & 0{.}03148 & 1000{.}62 & 0{.}35461 & 0{.}03156 &
  1 & 11 & 2 & $+$ & $+$ & $-$ & 0.10\\ 
  1001{.}72 & 0{.}27396 & 0{.}03156 & 1000{.}31 & 0{.}27449 & 0{.}03168 &
  4 & 6 & 2 & $-$ & $-$ & $+$ & 0.14\\ 
  1001{.}79 & 0{.}18851 & 0{.}02737 & 1001{.}45 & 0{.}18863 & 0{.}02732 &
  7 & 1 & 3 & $-$ & $+$ & $+$ & 0.03\\ 
  1002{.}44 & 0{.}39897 & 0{.}00410 & 999{.}751 & 0{.}39971 & 0{.}00415 &
  0 & 15 & 0 & $+$ & $-$ & $+$ & 0.27\\ 
  1002{.}44 & 0{.}39897 & 0{.}00410 & 999{.}751 & 0{.}39971 & 0{.}00415 &
  0 & 15 & 0 & $+$ & $+$ & $-$ & 0.27\\ 
  1002{.}51 & 0{.}18815 & 0{.}02683 & 1002{.}23 & 0{.}18823 & 0{.}02672 &
  7 & 1 & 3 & $+$ & $-$ & $+$ & 0.03\\ 
  1002{.}73 & 0{.}26921 & 0{.}00980 & 1002{.}49 & 0{.}26927 & 0{.}00986 &
  3 & 9 & 0 & $-$ & $-$ & $+$ & 0.02\\ 
  1002{.}73 & 0{.}43764 & 0{.}06971 & 999{.}919 & 0{.}43852 & 0{.}07009 &
  0 & 10 & 5 & $+$ & $-$ & $-$ & 0.28\\ 
  1002{.}95 & 0{.}39083 & 0{.}08511 & 1002{.}02 & 0{.}39107 & 0{.}08516 &
  1 & 6 & 7 & $-$ & $+$ & $-$ & 0.09\\ 
  1002{.}95 & 0{.}39083 & 0{.}08511 & 1002{.}02 & 0{.}39107 & 0{.}08516 &
  1 & 6 & 7 & $+$ & $-$ & $-$ & 0.09\\  
 \hline
\end{tabular}
\tabend
}
\caption[]{\label{tab:eigenvalues} \capsty The quantum mechanical eigenvalues 
  $(E_{{\text{qm}}},k_{\text{qm}},l_{\text{qm}})$ and the
  semiclassical eigenvalues
  $(E_{{\text{sc}}},k_{\text{sc}},l_{\text{sc}})$ 
  of the ellipsoidal billiard for the ranges $E_{{\text{qm}}} < 24$ 
  and $1000 < E_{{\text{qm}}} < 1003$.
  The relative error $\Delta \h =
  (\h_{{\text{qm}}}-\h_{{\text{sc}}})/\h_{{\text{qm}}}$ is given in percent.} 
\end{center}
\tabend
\end{table}

Let us first consider the four transitions of the {\sl WKB} lattice in
action space across the separatrix surfaces of
Fig.~\ref{fig:sep} away from the 
intersection line of the separatrix surfaces. Upon each crossing only
two effective Maslov phases change appreciably. We therefore take the
action component of ${\bm J}$ corresponding to the effective Maslov
phase that stays approximately constant as being semiclassically
quantized. To do so we have to fix the quantum number belonging to
this action component 
and the parities appearing in its effective Maslov
phase. The quantum numbers corresponding to the two other action
components and the remaining free parities then define a family of
surfaces in action space, which intersect the plane corresponding to the
action component that already fulfills the semiclassical quantization
condition. In Fig.~\ref{fig:maslovtransitions} we represent the plane
corresponding to the semiclassically quantized action component and
the intersection lines projected onto the plane of the two remaining
free action components. The semiclassical eigenvalues appear as the
intersection points of the intersection lines as far as they are
contained in a parity box. 

Let us first consider the transition from region \tO\ to region
\tOA\ in Fig.~\ref{fig:maslovtransitions}a. Upon this transition we
always have $-\Theta_\lambda\gg\hbar$, see
Tab.~\ref{tab:limitcases}.
From \equ~(\ref{eq:Maslovlambda}) we see that the effective Maslov phase
$\tilde{\alpha}_\lambda$ stays approximately
$3-\pi_z$. We semiclassically quantize $J_\lambda$ by fixing the
quantum number $r=10$ and the parity $\pi_z=-$. The transition of the
{\sl WKB} lattice takes place in the action components $J_\mu$ and
$J_\nu$. In the region corresponding to motion of type \tO\ all quantum
states are non-degenerate. Upon the transition from region \tO\ to
region \tOA\ quantum states with the same product of the parities $\pi_x$
and $\pi_y$ become quasidegenerate. For $(\pi_x,\pi_y)=(\pm,\mp)$ the
quasidegenerate states have the same quantum numbers $(r,m,n)$; for
$(\pi_x,\pi_y)=(+,+)$ and $(\pi_x,\pi_y)=(-,-)$ they differ by 1 in
the quantum number $n$, see Tables~\ref{tab:limitcases} and
~\ref{tab:limitquntumnumbers}  and Fig.~\ref{fig:quantumboxes}. The
picture for $\pi_z=+$ is similar and therefore is omitted. 

In Fig.~\ref{fig:maslovtransitions}b the transition from region \tO\
to region \tP\ is presented. Here we again have a transition from
non-degeneracy to quasidegeneracy. From Tab.~\ref{tab:limitcases} we
see that we 
always have $\Theta_\nu\gg\hbar$ 
giving $\tilde{\alpha}_\nu\approx 2-\pi_x$, see
\equ~(\ref{eq:Maslovnu}). For the semiclassical quantization of
$J_\nu$ we choose $n=5$ and $\pi_x=-$ and we represent the transition
of the {\sl WKB} lattice
in the components $J_\lambda$ and $J_\mu$. In region \tP\ the quantum
states with the same product of parities $\pi_y$ and $\pi_z$ are
quasidegenerate, again see Tables~\ref{tab:limitcases} and
~\ref{tab:limitquntumnumbers}  and Fig.~\ref{fig:quantumboxes}.
For $\pi_x=+$ the picture is similar and therefore is not shown here. 

For the transition from region \tP\ to region \tOB\ we always have
$\Theta_\lambda \gg \hbar$ giving $\tilde{\alpha}_\lambda\approx
3$. The transition of the {\sl WKB} lattice takes place in the
components $J_\mu$ and $J_\nu$, see
Fig.~\ref{fig:maslovtransitions}c. For the semiclassical quantization
of $J_\lambda$ we have chosen $r=5$. Upon the transition the
quasidegeneracy in region \tP\ explained above changes to the
quasidegeneracy in region \tOB. Here quantum states with the same
product of the parities $\pi_x$ and $\pi_y$ are
quasidegenerate. For $(\pi_x,\pi_y)=(\pm,\mp)$ they have the same
quantum numbers $(r,m,n)$; for $(\pi_x,\pi_y)=(-,-)$ and
$(\pi_x,\pi_y)=(+,+)$ they differ by 1 in the quantum number $n$.

Upon the transition from \tOA\ to \tOB\ we have $-\Theta_\nu \gg
\hbar$ giving $\tilde{\alpha}_\nu\approx 2-\pi_x-\pi_y$. For the
quantization of the action component 
$J_\nu$ we choose $n=10$ and $(\pi_x,\pi_y)=(-,-)$. The transition of
the {\sl WKB} lattice takes place in the components $J_\lambda$ and
$J_\mu$, see Fig.~\ref{fig:maslovtransitions}d. In contrast to the 
transition from \tP\ to \tOB\ the change of the quasidegeneracy is not
connected to the action components presented in the figure. For
\tOA\ and \tOB\ the same pairs of states are quasidegenerate, see
Tab.~\ref{tab:limitquntumnumbers}. Only the shift of the {\sl WKB}
lattice relative to the simple  lattice $({\bm n}\hbar)$ changes, see
Fig.~\ref{fig:quantumboxes}.  For $(\pi_x,\pi_y)=(+,-)$, $(-,+)$ and
$(+,+)$ the pictures are similar and are not shown here.

Note that we represent the parity boxes in
Fig.~\ref{fig:maslovtransitions} with side length $\hbar/2$ also in
the direction of $J_\mu$. The reason is that within each region
presented in the plots we always have $\Theta_\lambda\ge \Theta_\nu$ or
$\Theta_\lambda\le \Theta_\nu$, respectively. 
These relations restrict the
range of $\tilde{\alpha}_\mu$, see \equ~(\ref{eq:Maslovmu}), and
halve the parity boxes $P_{\bm \pi}$. 
One of these relations is
always trivially fulfilled in region \tO\ and \tOB\ because of the different
signs of the tunnel integrals there. For \tOA\ and \tP\ none of
these relations holds within the whole region, i.e. it is not possible
to define some kind of reduced parity boxes that are halve the $P_{\bm
  \pi}$ valid for regions \tOA\ and \tP\ although this is possible for
regions \tO\ and \tOB.\\
\def\FIGmaslovtransitions{%
\centerline{ 
        \raisebox{0.0cm}{\mbox{a)}}
        \psfig{figure=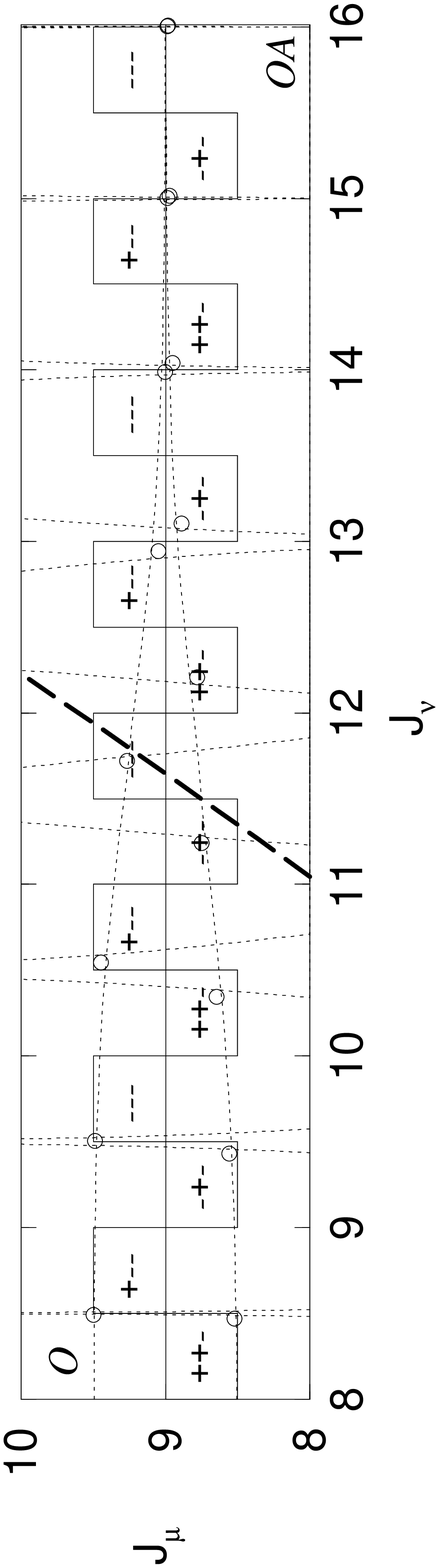,angle=-90,width=16.5cm}
       } 
\centerline{
        \raisebox{0.0cm}{\mbox{b)}}
        \psfig{figure=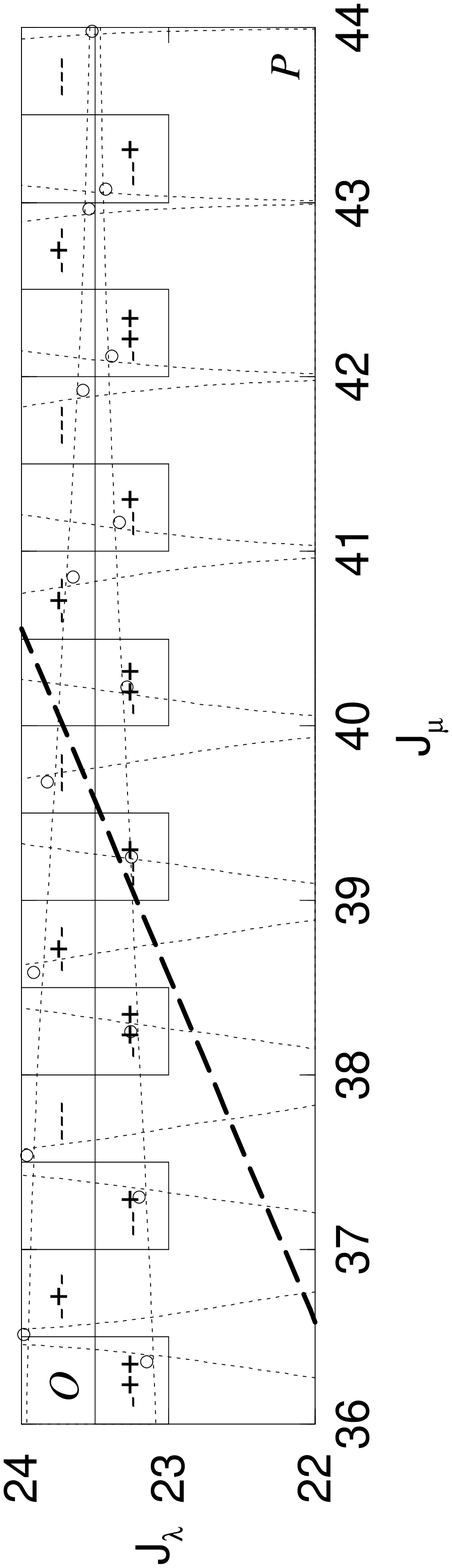,angle=-90,width=16.5cm}   
      } 
\centerline{ 
        \raisebox{0.0cm}{\mbox{c)}}
        \psfig{figure=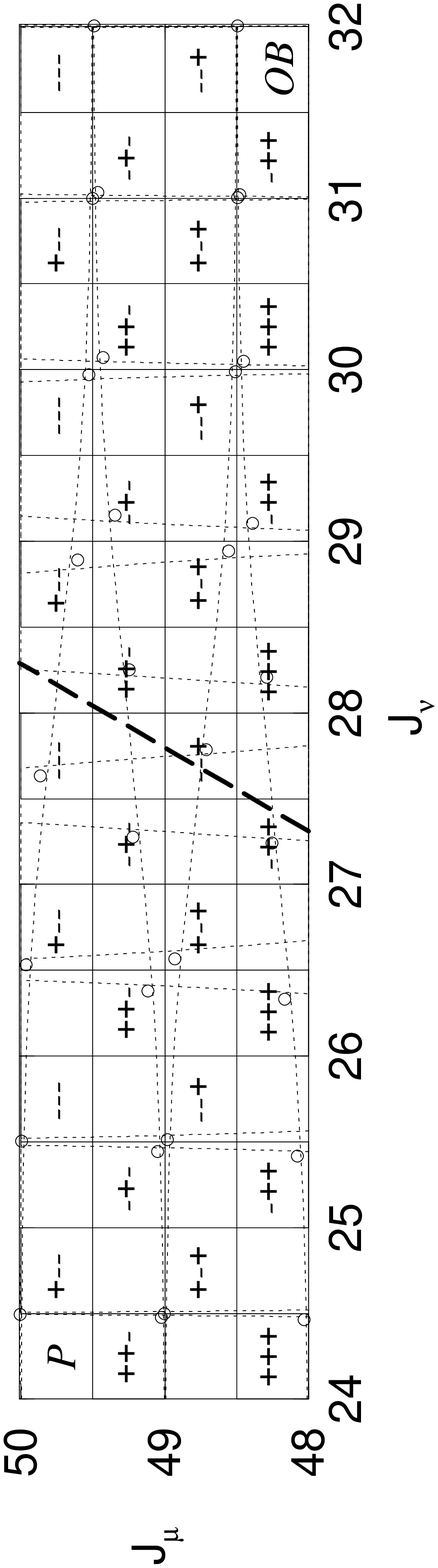,angle=-90,width=16.5cm}
       } 
\centerline{
        \raisebox{0.0cm}{\mbox{d)}}
        \psfig{figure=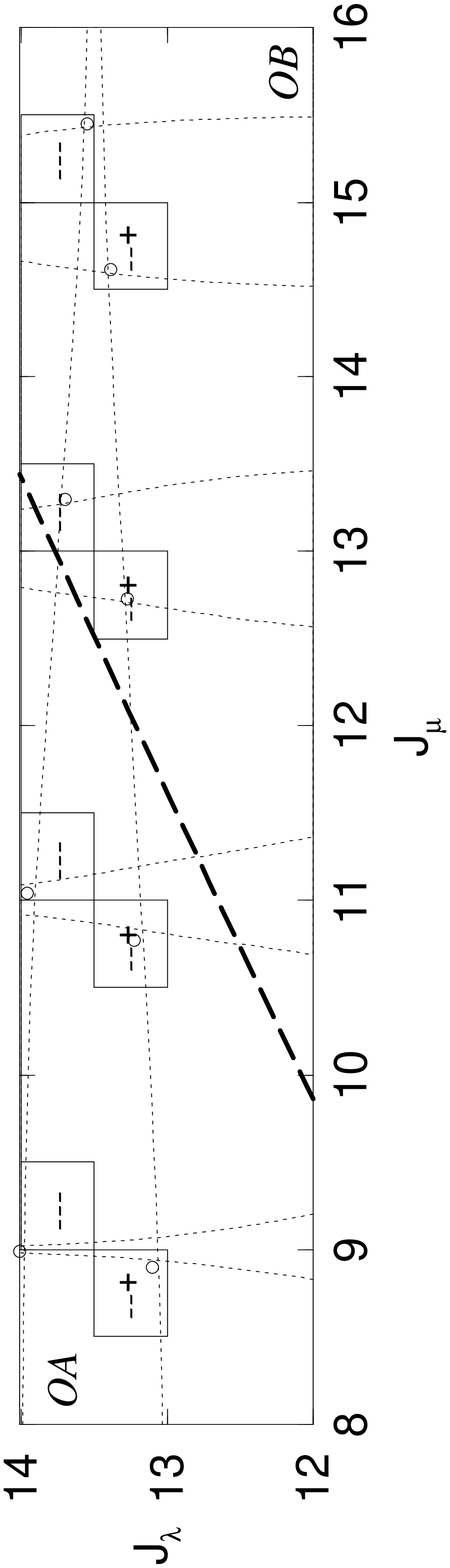,angle=-90,width=16.5cm}   
      } 
}
\def\figmaslovtransitions{%
Transition of the {\sl WKB} lattice upon crossing the classical
separatrix surfaces (bold dashed lines) in action space. 
In each figure the parity boxes for four
neighbouring quantum cells $\Delta J_\lambda =
\Delta J_\mu = \Delta J_\nu = 2$ are shown.
The exact quantum states are shown as circles. The semiclassical
eigenvalues appear as the intersection of the thin short dashed
lines within the parity boxes. $\BJf$ is measured in units of
$\hbar$. 
\rem{
a) From motion type \tO\ to motion type \tOA\ in the section $r=10$ with
$\pi_z=-1$. The phase $\tilde{\alpha}_\lambda$ is constant. The section for
$\pi_z = +1$ looks similar. b) From motion type  \tP\ to motion type
\tOB\ in the  
section $r = 5$. The phase $\tilde{\alpha}_\lambda$ is
constant. c) From motion type  \tO\ to motion type \tP\ in the
section $n = 5$ and $\pi_x = -1$. The phase $\tilde{\alpha}_\nu$ is
constant. The section for $\pi_x = +1$ looks similar. d) From motion
type 
\tOA\ to motion type \tOB\ in the 
section $n = 10$ and $(\pi_x,\pi_y) = (-,-)$. The phase
$\tilde{\alpha}_\nu$ is 
constant. The sections for $(\pi_x,\pi_y) = (+,-), (-,+)$ and $(+,+)$
look similar.
}
}
\FIGo{fig:maslovtransitions}{\figmaslovtransitions}{\FIGmaslovtransitions}

The situation is much more complicated in the neighbourhood of the
intersection line of the separatrix surfaces in Fig.~\ref{fig:sep}. Here
the transition of the {\sl WKB} lattice cannot be shown in
two dimensional sections. In Fig.~\ref{fig:maslovsurfaces} we
present the surfaces $\lambda_{\pi_y \pi_z}$,
$\mu_{\pi_y \pi_z}$ and $\nu_{\pi_y \pi_z}$ with the
action component $J_\lambda$, $J_\mu$ or $J_\nu$, respectively,
being semiclassically quantized for the parity combination  ${\bm
  \pi}=(-,\pi_y,\pi_z)$ and the corresponding  quantum number from
$(r,m,n)$. The surfaces carry the intersection lines as explained above.
We restrict the representation to $\pi_x=-$ to keep the
picture clear. For $\pi_x=+$ the picture is similar. The index $*$
indicates two surfaces with different parities located almost
on top of each other.
With $\pi_x$ being fixed the effective
Maslov phases $\tilde{\alpha}_\lambda$ and $\tilde{\alpha}_\nu$ depend
only on one further parity $\pi_y$ or $\pi_z$,
respectively. $\tilde{\alpha}_\mu$ depends on $\pi_y$ 
and $\pi_z$. We therefore show two surfaces for $J_\lambda$ and $J_\nu$
and four surfaces for $J_\mu$. The parity boxes are omitted in this
figure. In Fig.~\ref{fig:maslovsurfaces} thus not every intersection
point corresponds to a semiclassical eigenvalue. The semiclassical
eigenvalues may be identified with those intersection points lying
closest to the exact quantum eigenvalues represented by spheres in the figure.
\rem{\footnote{REM: I think we should display the values of $K$ and $L$
instead of $s_{1,2}$.}}
\def\FIGmaslovsurfaces{%
\centerline{
 \raisebox{9.0cm}{\mbox{a)}}
 \psfig{figure=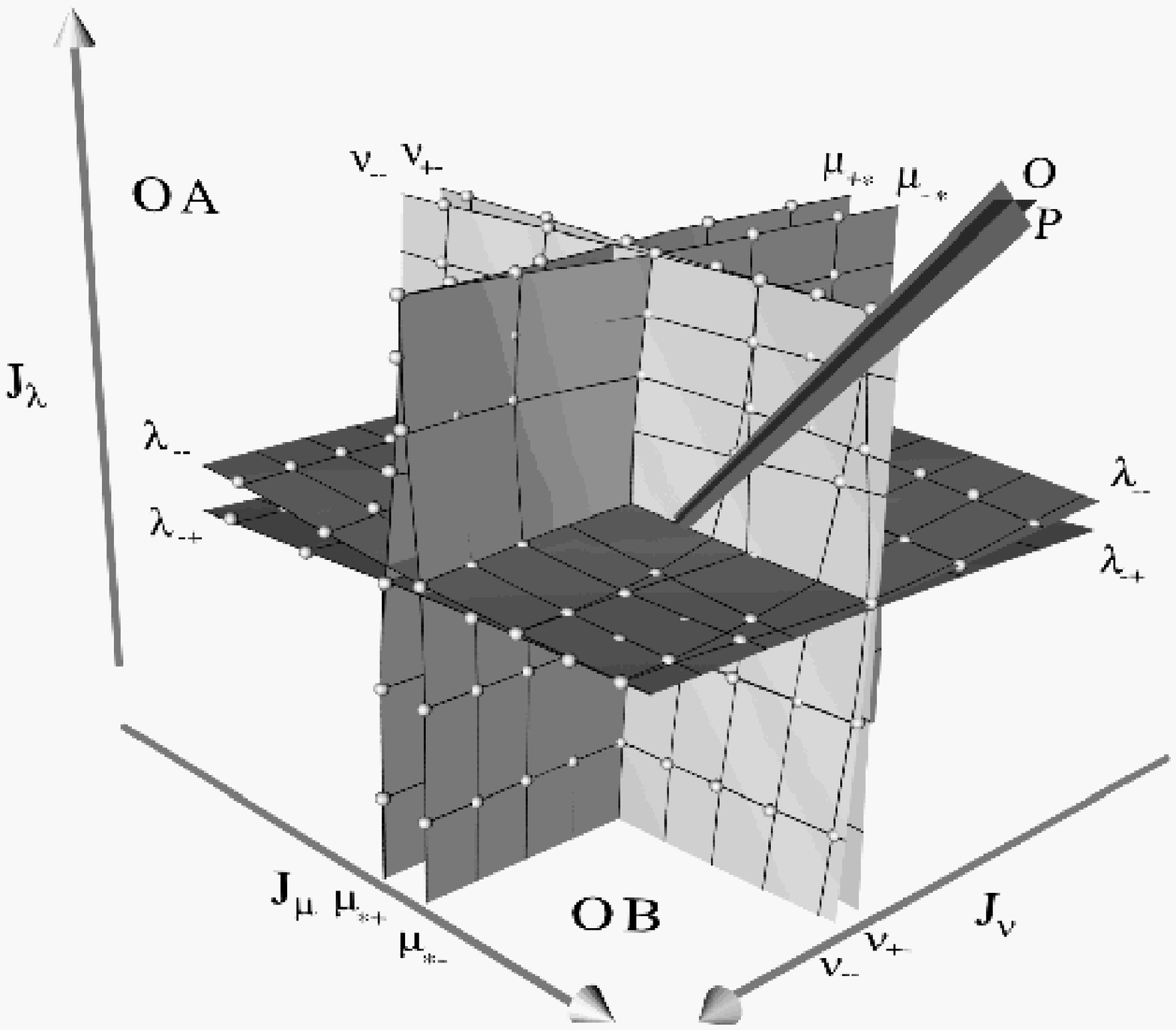,width=12cm}
}
\centerline{
 \raisebox{9.0cm}{\mbox{b)}}
 \psfig{figure=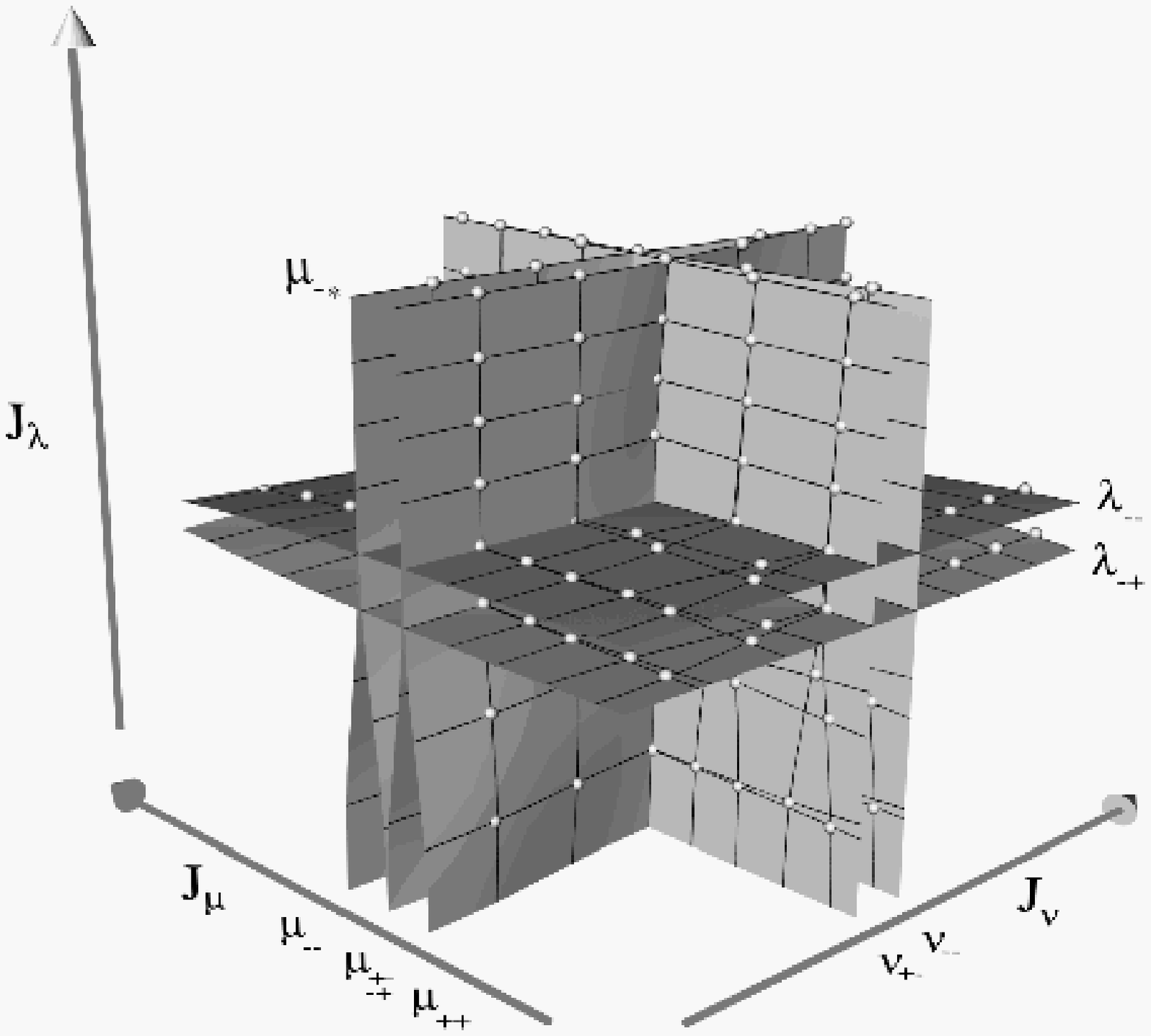,width=12cm}
}}
\def\figmaslovsurfaces{%
a) Transition of the {\sl WKB} lattice upon crossing the intersection
line of the separtrix surfaces in action space. The intersection line
is represented as the ``cross-ray''. 
b) Fig.~\ref{fig:maslovsurfaces}a from behind.
}
\FIGo{fig:maslovsurfaces}{\figmaslovsurfaces}{\FIGmaslovsurfaces}

%% file: degenerate.tex
\section{Degenerate Ellipsoids}
\label{sec:deg}

\noindent So far we treated the billiard in the general triaxial ellipsoid
in its classical, quantum mechanical, and semiclassical aspects.
The triaxial ellipsoid degenerates into simpler systems when
any two or even all of the semiaxes coincide. In the latter case
we obtain the sphere, in the former case  prolate or oblate
ellipsoids which are rotationally symmetric about the longer or
shorter semiaxis, respectively. In this section we want to take a
short look at these degenerate cases where the focus is on the
similarities in the classical, quantum mechanical, and semiclassical
treatment.

The main theme is that the coalescence of semiaxes of the ellipsoid
induces the collision of roots or poles. In the classical
treatment the disappearence of certain types of motions is expressed
by the fact that some roots of a hyperelliptic curve collide and the
genus of the curve drops. In the quantum mechanical treatment
it is the singularities of the Helmholtz equation that coalesce,
which is usually called confluence. Let us look at these transitions in more
detail.

First of all we discuss the ellipsoid itself, without any dynamics.
The general ellipsoid with parameteres $0 < b < a < 1$ has the 
semiaxes $1 > \sqrt{1-b^2} > \sqrt{1-a^2}$.
There are two choices of degenerate cases, either $b=a$ or 
$b=0$. For $b=a$ the longest semiaxes is $1$ and the two
shorter semiaxes coincide, which gives the prolate ellipsoid. 
In the case $b=0$ the shorter semiaxis $\sqrt{1-a^2}$ is singled 
out and we obtain an oblate ellipsoid. 
If $a=b=0$ we obtain the sphere.

The algebraic treatment of the degenerate cases for the classical 
dynamics is quite simple. In the general case the hyperelliptic 
curves ${\cal R}_w$ have four fixed roots at $0 < b < a < 1$ and two movable 
roots (i.e.\ depending on the initial conditions) $0 \le s_1 \le a$ 
and $b \le s_2 \le 1$ with $s_1 \le s_2$. 
Placing the movable roots $s_i$ into the
three intervals marked by the fixed roots we obtain four
possibilities,
corresponding to the four types of motion.  

Let us first consider the prolate case where $b=a$ and the
$\eta$-range has vanished. 
In terms of the ranges for $\xi$ and $\zeta$ only case  \tP\ remains,
because the other three become special cases of it, see
Fig.~\ref{fig:riemannsurffamily}. Only the ordering 
of the roots $0 \le s_1^2 \le a^2 \le s_2^2 \le 1$ is left which means that
 there is only one type of motion in the prolate ellipsoid.
The genus of the hyperelliptic curve has dropped to one and the
remainder of 
the $\eta$-interval is a pole at $z=a^2$, i.e. the integral in
\equ~(\ref{eq:actionintegrals}) is elliptic  and of
the third kind. Integrating around this pole gives the
residue of this pole divided by $2\pi$, which is
\bege \label{eq:Iphiprolate}
      I_\eta = I_\mu =  \pm \sqrt{K^2-L^2-2a^2 E} = L_x.
\ende
This is the angular momentum of the rotational degree of freedom,
which is itself an action because the corresponding angle is cyclic.
Hence even though the $\eta$-interval disappears, the $\eta$-action
of course does not, this is the reason for the 
``appearance'' of the pole.

In the oblate case $b=0$ the  $\zeta$-interval vanishes so that 
in terms of the ranges for $\xi$ and $\eta$ the two cases \tOA\
and \tOB\ remain with the corresponding orderings of the roots $0 \le
s_1^2 \le s_2^2 \le a^2$ and $0 \le s_1^2 \le a^2 \le s_2^2 \le 1$,
respectively. \tO\ becomes a special case of \tOA\ and \tP\ a 
special case of \tOB, see
Fig.~\ref{fig:riemannsurffamily}.
Again the hyperelliptic curve attains a double root, so the 
genus drops and we obtain an elliptic differential of the
third kind. The residue at the pole at $z=b=0$ gives $2\pi I_\nu$,
where $I_\nu$ is the angular momentum about the symmetry
axis; we find 
\bege
        I_\zeta = I_\nu = L = L_z.
\ende
If we finally collapse to the
sphere with  $b=a=0$, there is only the $\xi$-interval left. The
motion in this interval describes the radial dynamics. The
genus of the curve 
has dropped to 0, because now $s_1$ is fixed at zero.
The two angular degrees of freedom are hidden in a pole at 0. 
Considering \equ~(\ref{eq:xtok2EB}) we see that $l=0$ for $a=b=0$
so that in this case we find
\bege
        I = I_\nu+I_\mu = K = |{\bm L}|.
\ende
The fact that there is only one pole while we would like to
obtain two actions can be taken as an indication that the
system is now degenerate, i.e.\ it has more constants of motion 
than degrees of freedom.

Let us now consider the separated Helmholtz equation. The partial 
fraction decomposition of $g/f$ in Section~\ref{sec:qm} shows that we
have regular singular 
points at $\pm a$ and $\pm b$  in the case of 
a general ellipsoid. The solutions with exponents $1/2$ and $0$ gave
the different parities.
In the prolate case we have a confluence of $a$ with $b$ and
of $-a$ with $-b$. 
The equations for $\xi$ and $\zeta$ reduce to the prolate spheroidal
wave equations. 
Scaling the variables according to $\xi=a\tilde{\xi}$ and
$\zeta=a\tilde{\zeta}$ gives them the familiar appearance (see
\cite{AbraSteg65}) 
\bega
\label{eq:prol1}
(\tilde{\xi}^2-1)\psi''+2\tilde{\xi}\psi'-(\lambda
-c^2\tilde{\xi}^2+\frac{m^2}{\tilde{\xi}^2-1})\psi &=& 0,\\
\label{eq:prol2}
(1-\tilde{\zeta}^2)\psi''-2\tilde{\zeta}\psi'+(\lambda
-c^2\tilde{\zeta}^2-\frac{m^2}{1-\tilde{\zeta}^2})\psi &=& 0
\enda
with  parameters
\bege
\label{eq:prolateparam}
\lambda=(4kE-2a^2E)/\hbar^2 ,\quad c = 2a^2E/\hbar^2,\quad m^2 = L_x^2/\hbar^2.
\ende
The variable $\eta$ can be turned into an angle after some scaling
to compensate for the vanishing of the $\eta$-range. The corresponding
differential equation yields the familiar result $m\in \Z$ which means
that each energy eigenvalue is 
twofold degenerate. Equations (\ref{eq:prol1}) and (\ref{eq:prol2})
are identical. The way of
writing them just indicates that they are considered on the different
ranges $\tilde{\zeta} \in [-1,1]$  and $\tilde{\xi}\ge 1$. 
The indicial equations for the regular singular points $\tilde{\xi}=\pm
1$ and $\tilde{\zeta}=\pm 1$ are of course the same and give the
exponents $\pm \alpha=m/2$. 
Note that half the residue of the classical action 
integral over the coalesced singularity gives the
exponent of the indicial equation in the quantum case.

For the oblate case $b=0$ we obtain a confluence of $b$ with 
$-b$ which gives a regular singular point at $0$ for the equations for
$\xi$ and $\eta$.
The two regular singular points at $\pm a$ usually are removed by
transforming $\xi$ and $\eta$ separately according to
$\xi^2=a^2+a^2\tilde{\xi}^2$ and $\eta^2=a^2-a^2\tilde{\eta}^2$. This
gives the familiar pair of oblate spheroidal wave equations (see again
\cite{AbraSteg65})
\bega
\label{eq:obl1}
(\tilde{\xi}^2+1)\psi''+2\tilde{\xi}\psi'-(\lambda -c^2\tilde{\xi}^2
-\frac{m^2}{\tilde{\xi}^2+1})\psi&=&0, \\
\label{eq:obl2}
(1-\tilde{\eta}^2)\psi''
-2\tilde{\eta}\psi'+(\lambda+c^2\tilde{\eta}^2-\frac{m^2}{1-\tilde{\eta}^2})\psi
&=&0
\enda
with $\lambda$ and $c^2$ again defined as in
\equ~(\ref{eq:prolateparam}) but now $m^2=L_z^2/\hbar^2$. Similarly
to the prolate case the variable $\zeta$ can be turned into an angle and
the corresponding equation again gives $m\in\Z$, i.e. the twofold degeneracy
of the energy eigenvalues. 
The indicial equations for the regular singular points $\tilde{\xi}=\pm
i$ and $\tilde{\eta}=\pm 1$ which correspond to the original regular
singular points $\xi=0$ and $\eta=0$ are again the same and again give
the exponents $\alpha=\pm m/2$.
For the sphere let $a=b=0$ which gives $l=0$.  The scaling $\xi=r
\hbar/\sqrt{2E}$  then turns the equation for $\xi$ into 
\bege
\label{eq:bessel}
r^2\psi''+2r\psi' +(r^2-n(n+1))\psi = 0
\ende
with $n(n+1)=|{\bm L}|^2/\hbar^2$.
The variables $\eta$ and
$\zeta$ can be transformed into the azimutal and polar angles of the
sphere and the corresponding equations give the familiar result
$n\in\N_0$ and the $(2n+1)$-fold degeneracy of the energy eigenvalues.
\equ~(\ref{eq:bessel}) obviously is the defining equation for
spherical Bessel functions. It has a regular singular point at $0$
with exponents $n$ and $-(n+1)$. The corresponding solutions are the so
called spherical Bessel functions of first and second kind. The
asymptotics at $0$ picks out the functions of first kind as the
physical solutions for the sphere.

Note that all the equations cited here for the degenerate cases have an
irregular singular point at $\infty$. In the non-degenerate case
$\infty$ is a regular singular point of \equ~(\ref{eq:sephelmholtz}).

Concerning the separation of the separation constants the
non-degenerate Helmholtz equation presents the worst case because it
is not separable in the separation constants. In the
prolate and oblate cases only the constant $m$ can be
separated off, the equations are called partially separable for
separation constants. The billiard in the sphere belongs to the simplest
class of equations which are completely separable for separation
constants. 

For the semiclassical treatment it is again illuminating to have a look at
what happens to the hyperelliptic curve ${\cal R}_w$ in the degenerate
cases.  In the non-degenerate case ${\cal R}_w$ has genus 2 and
therefore has two complex periods giving the penetration integrals $\Theta_\xi$
and $\Theta_\zeta$ in Equations (\ref{eq:thetanu}) and
(\ref{eq:thetalambda}). In the prolate 
and oblate limiting cases one of the handles in
Fig.~\ref{fig:slitriemannsphere} 
vanishes. The genus of the curve drops to 1, i.e. the curve becomes
elliptic. Generally an elliptic curve has one complex period that
gives one penetration integral. In our case it is more useful to think of
the elliptic curves for the prolate and oblate cases as singular limits
of a hyperelliptic curve for the following reasons. During the prolate
limiting process the middle handle in Fig.~\ref{fig:slitriemannsphere}
shrinks. The penetration 
integrals are not only defined for any $b<a$ but even for $a=b$ where
they become infinite. This means that although the curve corresponding
to the prolate billiard motion is elliptic there is no tunneling in
our semiclassical treatment. During the oblate limiting process the
upper handle in Fig.~\ref{fig:slitriemannsphere} shrinks. Again the
penetration integrals are 
even defined for the limiting case $b=0$ where the curve becomes
elliptic. Here $\Theta_\zeta$ diverges 
and $\Theta_\xi$ stays finite. The prolate and the oblate limiting
cases have in common that the penetration integrals connected to the
vanishing handle diverge. But the oblate billiard still has one finite
penetration 
integral that gives the semiclassical description of the tunnelling
between tori corresponding to the two types of motions present here.
This is why the oblate limiting behaviour may be considered as 
the more typical case. The prolate ellipsoidal billiard is peculiar in
this sense. This peculiarity is also reflected by the fact that the
prolate ellipsoidal billiard exhibits quantum monodromy, see
\cite{WD98}. From the point of view of periods of the Riemann surface 
it is important to mention that in the prolate case the sum of the
penetration integrals is finite. 

The degeneracy of the energy eigenlevels can semiclassically be
understood by inspection of the energy surfaces. In the prolate and
oblate cases the energy surface is symmetric with respect to the sign
change of the action corresponding to the conserved angular
momentum \cite{RW95}. Thus the {\sl EBK} quantization condition is each time
fulfilled simultaneously at two
points of the energy surface. The additional degeneracy
of the billiard in the sphere classically manifests itself in the
resonance of the azimutal and polar angular motion. The ratio of the
corresponding frequencies has modulus 1, see  \cite{RW95}. The
energy surface is thus foliated by straight lines what makes the {\sl EBK}
quantization condition being fulfilled not only at one point of the
energy surface but on a whole line of points. This gives the
$(2n+1)$-fold degeneracy of the energy eigenlevels.

%% file: conclu.tex
\newcommand{\old}[1]{}

\section{Conclusions and Outlook}
\label{sec:conclu}

The last section demonstrated the unity of classical, semiclassical and 
quantum mechanical treatment in the complex plane, which is one main aspect 
of this exposition.
Another one is to emphasize the simple and nice
picture of the quantum mechanics of an integrable system as a
discretization of classical action space. 
Away from the separatrix surfaces the discretization gives
regular lattices due to the applicability of the simple \EBK\ quantization
of Liouville-Arnold tori in \equ~(\ref{eq:ebkquantization}). 
It is much harder to give a semiclassical description of the quantum
states whose eigenvalues in action space lie close to the separatrix
surfaces because of the presence of quantum mechanical  tunneling
between tori with different Maslov indices. The tunneling was
incorporated by a uniform \WKB\ quantization scheme. This approach 
is necessary because the application of the simple quantization rule
(\ref{eq:ebkquantization}) close to the separatrix surfaces in action
space can give erroneous additional eigenstates \cite{WWD97}. The
ingredients for the \WKB\ quantization scheme, i.e. the three
classical actions and the two penetration integrals, have a consistent
interpretation as the real and purely imaginary periods of a
single Abelian differential of second kind on a hyperelliptic curve of
genus 2. In this sense quantum mechanics appears as a
``complexification'' of classical mechanics.

For the billiard in the ellipsoid we were able to represent all 
quantum states as a regular \WKB\ lattice in the space of
the slightly modified actions~${\bm J}$ which are twice the
actions~$\tilde{\bm I}$ of the symmetry reduced system.  
This is impossible in the space of the  original actions ${\bm I}$.
The two classical senses of rotations of motion types \tP, \tOA,
and \tOB\ give two different tori in phase space whose actions differ
in sign. Quantum mechanically we cannot distinguish between these tori
and it is therefore impossible to assign different quantum state to
them.  
In \cite{Wiersig98} all systems like the ellipsoidal 
billiard which (after some symmetry reduction) allow for a 
representation of the eigenvalues in action space  are refered to as
``one-component systems''. In order to represent quantum states in
action space it is generally necessary to modify the original
actions. For systems which are no one-component systems the \WKB\
lattice will be much less regular then.

\old{
For the billiard in the ellipsoid 
we were able to describe all the quantum mechanical 
states in one non-overlapping \WKB\ lattice. As was pointed out
in \cite{Wiersig98} this might not be possible in general.
Far away from separatrices the corresponding lattices always exist.
In general the map from phase space to action space 
(modulo discrete symmetries) is not globally invertible, 
such that the corresponding lattices might overlap.
It is not clear if one can always find a smooth transitions
between the different lattices across a separatrix.
In our case for motion types \tP, \tOA\ and \tOB\ 
in the ellipsoidal billiard for every set of valid actions there 
correspond two tori, corresponding to the to senses of
rotational motion. This classical degeneracy 
is reflected by the parities of the separated wave functions.
For ``one component systems'' \cite{Wiersig98} like this
the problem of representing the quantum eigenstates for all
parities in action space can be solved by blowing up the phase 
space of the symmetry reduced system. For the ellipsoidal billiard we
have performed this  via the introduction of the
actions ${\bm J}$ which are twice the actions $\tilde{\bm I}$ of
the symmetry reduced ellipsoidal billiard.
}

\old{
Although  the presentation of the quantum eigenstates of an integrable
system in action space seems very fundamental it works only for a
special class of integrable systems which in \cite{Wiersig98} are
refered to as {\sl one component systems}. In fact a generic system
does not belong to this class, but our ellipsoidal billiard
does. The problem is that generically there correspond
more than one Liouville-Arnold $f$-torus and therefore more than one $f$-tuple
of action variables to a given set of constants of motion. 
As a simple example we mention the one-dimensional motion of a particle
in a non-symmetric  double well potential. For these systems it is not
possible to present the quantum eigenstates in action
space because the map from  the constants of motion to the action
variables is multivalued. Actually this is even the case for motion
types \tP, \tOA\ and \tOB\ in the ellipsoidal billiard, see
Section~\ref{sec:classic}. The map from
the constants of motion to the action variables is here twovalued
because there exist two signs of the action component corresponding to
the rotational degree of freedom. 
The classical degeneracy due to the two senses of rotation of a
rotational degree of freedom is quantum mechanically reflected by the
presence of parities of the separated wave functions. For such systems
the problem of representing the quantum eigenstates for all
parities in action space can be solved by blowing up the phase 
space of the symmetry reduced system. For the ellipsoidal billiard we
have performed this  via the introduction of the
actions ${\bm J}$ which are twotimes the actions $\tilde{\bm I}$ of
the symmetry reduced 
ellipsoidal billiard. Since this can be done for any system whose tori
for a given set of constants of motion 
are simply related by time reversal symmetry  these systems are still
called one component systems.
The representation of the eigenstates in action space for a
system which is not a one component system is still an unsolved
problem.}

For future work on ellipsoidal quantum billiards we want to
mention two directions. On the one hand the ellipsoidal quantum
billiard can be taken as the starting point for computations
of non-integrable quantum billiards resulting from slight distortion
of the ellipsoidal boundary. On the other hand the 
semiclassical analysis in terms of periodic orbits is still to be
worked out. For a chaotic system the Gutzwiller trace formula
gives a semiclassical expression for the quantum density of states as
a summation over isolated periodic orbits \cite{Gutz67,Gutz70,Gutz71}. 
Analogously the quantum density of states of an integrable system can
semiclassically be written as a summation over resonant tori,
i.e. over families of periodic orbits. The Berry-Tabor trace formula
gives the quantum density of states of an integrable system with $f$
degrees of freedom as a summation over resonant $f$-tori
\cite{BerryTabor76}. The main contribution to the density of
states stem from the shortest periodic orbits. Generally the shortest
periodic orbits do not foliate $f$-tori but lower dimensional tori.
Since these contributions are
not included in the generic case discussed in \cite{BerryTabor76} they
demand special considerations \cite{CL91}. In addition to that the
presence of separatrices demands a modification of the Berry-Tabor
trace formula \cite{WWD97}. Both the non-generic
contributions of resonant 2-tori and of isolated periodic orbits, and
the presence of the crossing separatrices complicate the periodic
orbit summation for 
ellipsoidal quantum billiards.
For three-dimensional billiards as models for nuclei
the shortest periodic orbits are important for the explanation of shell
structures. For rotationally symmetric billiards this is considered in 
\cite{Frisk90,Magner97}. For non-symmetric ellipsoids this is still to
be done.